\documentclass[11pt,a4paper]{article}
\usepackage{jheppub}

\usepackage{ dsfont }
\usepackage{amsmath,amssymb,calc}
\usepackage{color}
\usepackage{graphicx}
\def\exd{{\hbox{d}}}
\def\d{\exd}

\makeatletter
\renewcommand\section{\@startsection {section}{1}{\z@}%
                                   {-3.5ex \@plus -1ex \@minus -.2ex}
                                   {2.3ex \@plus.2ex}%
                                   {\normalfont\large\bfseries}}
\renewcommand\subsection{\@startsection{subsection}{2}{\z@}%
                                     {-3.25ex\@plus -1ex \@minus -.2ex}%
                                     {1.5ex \@plus .2ex}%
                                     {\normalfont\bfseries}}
\makeatother

\newcommand{\bea}{\begin{eqnarray}}
\newcommand{\eea}{\end{eqnarray}}
\newcommand{\be}{\begin{equation}}
\newcommand{\ee}{\end{equation}}
\newcommand{\bma}{\begin{pmatrix}}
\newcommand{\ema}{\end{pmatrix}}

\def\barray{\begin{array}}
\def\earray{\end{array}}
\def\be{\begin{equation}}
\def\ee{\end{equation}}
\def\ben{\begin{equation} \nonumber}
\def\een{\end{equation}}
\def\ban{\begin{eqnarray*}}
\def\ean{\end{eqnarray*}}
\def\ba{\begin{eqnarray}}
\def\ea{\end{eqnarray}}

\def\M{\mathcal{M}}

\def\({\left(}
\def\){\right)}
\def\[{\left[}
\def\]{\right]}

\def\tr{{\rm Tr}}
\def\nn{\nonumber}

\def\tr{{\rm Tr}}

\def\One{{\hbox{ 1\kern-.8mm l}}}

\def\dHiggs{\xi}
\def\hvev{Z_0}

\def\X{\vec{X}}

\def\da{\Psi}
\def\dN{\alpha}

\definecolor{darkgreen}{cmyk}{0.85,0.2,1.00,0.2}

\def\be{\begin{equation}}
\def\ee{\end{equation}}
\def\bea{\begin{eqnarray}}
\def\eea{\end{eqnarray}}

\title{Higgsed Gauge-flation}

\author{Peter Adshead}
\author{and Evangelos I. Sfakianakis}
 
\affiliation{Department of Physics, University of Illinois at Urbana-Champaign, Urbana, IL 61801, USA}

\emailAdd{adshead@illinois.edu, esfaki@illinois.edu}

\abstract{
We study a variant of Gauge-flation where the gauge symmetry is spontaneously broken by a Higgs sector. We work in the Stueckelberg limit and demonstrate that the dynamics remain (catastrophically) unstable for cases where the gauge field masses satisfy $\gamma < 2$, where $\gamma = g^2\psi^2/H^2$, $g$ is the gauge coupling, $\psi$ is the gauge field vacuum expectation value, and $H$ is the Hubble rate. We compute the spectrum of density fluctuations and gravitational waves, and show that the model can produce  observationally viable spectra. The background gauge field texture violates parity, resulting in a chiral gravitational wave spectrum. This arises due to an exponential enhancement of one polarization of the spin-2 fluctuation of the gauge field. Higgsed Gauge-flation can produce observable gravitational waves at inflationary energy scales well below the GUT scale.
}

\begin{document}

\maketitle
\flushbottom

\section{Introduction}

Inflation \cite{Guth:1980zm, Linde:1981mu, Albrecht:1982wi} is a remarkably successful paradigm, simultaneously solving fine tuning problems associated with the initial conditions of the standard hot big bang scenario while providing primordial fluctuations with the right amplitude and scale dependence to seed structure formation \cite{Starobinsky:1979ty,Starobinsky:1980te,Lukash:1980iv,Press:1980zz,Mukhanov:1981xt,Guth:1982ec,Hawking:1982cz}.  

With the increasingly exquisite  measurements of the spectrum of temperature and polarization fluctuations in the Cosmic Microwave Background, the basic inflationary paradigm is in good shape \cite{Planck:2013jfk}. The measured fluctuations are adiabatic, Gaussian, and there is evidence at the 5-$\sigma$ level of a red tilt from the CMB alone. While there is currently no evidence for gravitational waves, upcoming experiments such as CMB Stage 4 \cite{Abazajian:2016yjj} will probe tensor-to-scalar ratios as low as $r \sim 10^{-3}$.

In this work we study a massive or Higgsed variation of the model of inflation called Gauge-flation, first proposed in refs.\ \cite{Maleknejad:2011jw, Maleknejad:2011sq}.\footnote{Gauge-flation has also been proposed as a model for dark energy `Gaugessence'  \cite{Mehrabi:2015lfa}} The remarkable aspect of Gauge-flation is that it does not contain scalar fields. Instead the theory utilizes non-Abelian gauge fields in a classical configuration to generate an epoch of accelerated expansion. The Gauge-flation model can be obtained from a related model,  Chromo-Natural inflation \cite{Adshead:2012kp, Adshead:2012qe, Martinec:2012bv}, by integrating out an axion about the minimum of its potential \cite{Adshead:2012qe,SheikhJabbari:2012qf, Maleknejad:2012dt}. Unfortunately, Gauge-flation and Chromo-Natural inflation are ruled out at the level of the fluctuations \cite{Namba:2013kia, Dimastrogiovanni:2012ew, Adshead:2013nka, Adshead:2013qp}. In regions of parameter space that yield acceptable scalar density fluctuations, the tensor-to-scalar ratio is too large; conversely, in the regions where the tensor-to-scalar ratio is acceptable, the scalar spectrum is too red-tilted. In this work, we augment the Gauge-flation model by introducing a Higgs sector which spontaneously breaks the gauge symmetry \cite{Nieto:2016gnp,Adshead:2016omu}. Recently we demonstrated that breaking the gauge symmetry in Chromo-Natural inflation allows that model to generate spectra that are consistent with current data \cite{Adshead:2016omu}, and in this paper we demonstrate that Gauge-flation too can generate acceptable spectra in a broken phase.

While there is certainly no shortage of inflationary models on the market \cite{Martin:2013tda}, and many that fit the data well \cite{Ade:2015lrj}, most rely on  (one or multiple) slowly rolling scalar fields to generate an extended period of nearly exponential expansion. In these models of inflation, the amplitude of the gravitational wave spectrum is set only by the energy scale during inflation. Obtaining a large amplitude gravitational wave spectrum generically requires that inflation occurs at an energy scale near the energy associated with grand unification, and a large tensor-to-scalar ratio requires that the inflaton roll a distance in field space that is comparable to the Planck scale \cite{Lyth:1996im}. As we demonstrate, the remarkable feature of Higgsed Gauge-flation is the generation of observable gravitational waves at much lower energy scales  -- in this model, gravitational waves mix with exponentially enhanced gauge field fluctuations, resulting in their subsequent amplification. This phenomena is also observed in Higgsed Chromo-Natural inflation \cite{Adshead:2016omu} and in models of inflation that have an accompanying spectator Chromo-Natural inflation-like sector \cite{Dimastrogiovanni:2016fuu,Maleknejad:2016qjz,Fujita:2017jwq}

Classical non-Abelian gauge fields lead to striking phenomenology in cosmological settings \cite{Galtsov:2010yqh, Galtsov:2011aa}, most notably chiral gravitational waves \cite{Adshead:2013nka, Maleknejad:2014wsa, Obata:2014loa, Bielefeld:2014nza, Bielefeld:2015daa, Obata:2016tmo, Maleknejad:2016qjz, Caldwell:2016sut,Alexander:2016moy, Obata:2016xcr, Dimastrogiovanni:2016fuu,Maleknejad:2016dve} and the facilitation of gravitational leptogenesis \cite{Maleknejad:2016dci}.  Classical non-Abelian gauge fields have also recently been employed in generalized multi-Proca theories  \cite{Jimenez:2016upj} to build stable inflationary models that do not require gauge invariance \cite{Emami:2016ldl}, and to generate inflation models with Horndeski couplings \cite{Davydov:2015epx,BeltranJimenez:2017cbn}.

Throughout this work, we use natural units where the speed of light and the reduced Planck constant are set to unity, $c = \hbar = 1$.

%
\section{Gauge-flation: Inflation from non-Abelian gauge fields}\label{sec:GFbackground}
%

We consider the theory of Gauge-flation \cite{Maleknejad:2011jw, Maleknejad:2011sq}, which is described by the action
\begin{align}\nn\label{eqn:GFaction}
\mathcal{S} =  \int \exd^{4}x\sqrt{-g}\Bigg[  \frac{M_{\rm Pl}^2}{2} R-\frac{1}{2}\tr\[F_{\mu\nu}F^{\mu\nu}\]+\frac{\kappa}{48}\(\tr\[ F_{\mu\nu}\tilde{ F}^{\mu\nu} \]\)^2\Bigg] ,
\end{align}
and consider a general SU(2) gauge field, $A_\mu$, adopting the conventions of  Peskin and Schroeder \cite{Peskin:1995ev} for its action. In particular, the field-strength tensor and covariant derivative are defined as\footnote{Note that this is opposite to \cite{Adshead:2012kp, Maleknejad:2011jw, Maleknejad:2011sq} who use the opposite sign for the covariant derivative.}
\begin{align}
F_{\mu\nu} = \frac{1}{-ig}\[D_{\mu}, D_\nu\], \quad D_{\mu} = \partial_{\mu} - igA_{\mu},
\end{align}
where $g$ is the gauge field coupling, not to be confused with the determinant of the spacetime metric. We normalize the trace over the SU(2) matrices, which we denote $J_a$, so that
\begin{align}
\tr\[J_a J_b\] = \frac{1}{2}\delta_{ab}, \quad \[J_{a}, J_b\] = i \epsilon_{abc}J_c,
\end{align}
where $\epsilon_{abc}$ are the structure functions. The dual field strength is defined $\tilde{F}^{\mu\nu} = \epsilon^{\mu\nu\alpha\beta}F_{\alpha\beta}/2$, and our convention for the antisymmetric tensor is $\epsilon^{0123} = 1/\sqrt{-g}$, 
while our spacetime metric signature is $(-,+,+,+)$.  Here and throughout, Greek letters  denote spacetime indices, Roman letters from the start of the alphabet denote gauge indices and Roman letters from the middle of the alphabet denote spatial indices. Appendix \ref{app:notation} outlines our remaining conventions and notations.

In addition to the field content of Gauge-flation, we  consider the addition of a symmetry breaking sector proposed in ref.\ \cite{Adshead:2016omu},\footnote{A similar model, `Massive Gauge-flation' was proposed in ref.\ \cite{Nieto:2016gnp} where explicit gauge-symmetry breaking mass terms were added to the action.} which we  write in Stueckelberg form  \cite{Kunimasa:1967zza,Ruegg:2003ps}
\begin{align}\label{eqn:Higgsaction}
\mathcal{S}_{\rm H, eff} =  \int \d^{4}x\sqrt{-g}\[   -g^2\hvev^2\tr\[ A_{\mu} - \frac{i}{g}U^{-1}\partial_\mu U\]^2\]
\end{align}
where 
\begin{align}
U = \exp\[ig \dHiggs\],\quad \xi  =  \dHiggs^a J_a.
\end{align}
The fields $\dHiggs^a$ are the Goldstone modes corresponding to fluctuations of the Higgs along its vacuum manifold. 

%
\subsection{Background solutions}\label{sec:backgroundeqns}
%

The background evolution in Gauge-flation is found by considering the gauge fields in the classical flavor-locked configuration
\begin{align}\label{eqn:gaugevev}
A_{0} = & 0, \quad A_{i} =  \phi\, \delta^{a}_{i} J_{a} = a\psi \, \delta^{a}_{i} J_{a},
\end{align}
where $J_a$ is a generator of SU(2) satisfying the commutation relations
\begin{align}
\[J_a, J_b\] = if_{abc}J_c,
\end{align}
and $f_{abc}$ are the structure functions of SU(2). Note that for SU(2), $f_{ijk} = \epsilon_{ijk}$, where $\epsilon_{ijk}$ is the completely antisymmetric tensor in three-dimensions.

On the background field configuration in eq.\ (\ref{eqn:gaugevev}), the field strength tensor components are,
\begin{align}
F_{0i} = & \phi' \delta^{a}{}_{i}J_a,\quad
F_{ij}  =  g \phi^2 f^{a}_{ij} J_a.
\end{align}
Here and throughout a prime, $\,'\,$, denotes a derivative with respect to conformal time. 
This field configuration results in a stress tensor that is consistent with the symmetries of Friedmann-Robertson-Walker spacetime. 
For these degrees of freedom, the mini-superspace action takes the form (see also ref.\ \cite{Nieto:2016gnp})
\begin{align}\label{eqn:minisupact}
\mathcal{L} =  a^{3}N \Bigg[-3 M_{\rm pl}^2 \frac{\dot a^2}{N^2} + \frac{3}{2}\frac{\dot \phi^2}{N^2}  - \frac{3}{2}g^{2}\frac{\phi^4}{a^4} + \frac{3}{2 N^2}\kappa \frac{ g^2\phi^4\dot{\phi}^2}{a^4}-\frac{3}{2}g^2\hvev^2 \frac{\phi^2}{a^2} \Bigg],
\end{align}
where an overdot here and throughout represents a derivative with respect to cosmic time, and $N = a$ on the background. 

Sheikh-Jabbari and Maleknejad  \cite{Maleknejad:2011jw, Maleknejad:2011sq} demonstrated the existence of inflationary solutions for this system in the absence of the symmetry breaking terms ($Z_0 = 0$). They pointed out that, while the terms in the action arising from the Yang-Mills field have the equation of state of radiation, $p = \rho/3$, where
\begin{align}
\rho_{\rm YM} =\frac{3}{2}\frac{\dot \phi^2}{a^2}  - \frac{3}{2}g^{2}\frac{\phi^4}{a^4},
\end{align}
the term proportional to $\kappa$ has the equation of state of a cosmological constant. That is $p_{\kappa} = -\rho_{\kappa}$, where
\begin{align}
\rho_{\kappa} = & \frac{3\kappa}{2}\frac{g^2 \dot\phi^2\phi^4}{a^6}.
\end{align}
This implies that if  $\rho_{\kappa} \gg \rho_{\rm YM}$, then the background spacetime  undergoes a phase of accelerated expansion.

The addition of the symmetry breaking sector generates additional contributions to both the energy density and the pressure,
\begin{align}
\rho_{\hvev} = \frac{3}{2}g^2\hvev^2 \frac{\phi^2}{a^2} , \quad p_{\hvev} =- \frac{1}{2}g^2\hvev^2 \frac{\phi^2}{a^2} ,
\end{align}
note the equation of state is $w = -1/3$, and thus the presence of the symmetry breaking sector does not affect the conditions for accelerated expansion, which remains  $\rho_{\rm YM} \ll \rho_{\kappa}$ \cite{Nieto:2016gnp}. However, since $\rho_{\hvev}+p_{\hvev} \neq 0$, successful slow roll inflation requires $\rho_{\hvev} \sim \rho_{\rm YM} \ll \rho_{\kappa}$ in order to ensure $\epsilon_{H} \ll 1$ (see eq.\ \eqref{eq:Hdot} below).

The equation of motion for the gauge field vacuum expectation value (vev) that follows from the action at eq.\ \eqref{eqn:minisupact} is
\begin{align}\label{eqn:phieom}
\(1+\kappa \frac{g^2 \phi^4}{a^4}\)\frac{\ddot\phi}{a} + \(1+\kappa \frac{\dot{\phi}^2}{a^2}\)\frac{2g^2\phi^3}{a^3}+\(1-3\kappa\frac{g^2 \phi^4}{a^4}\)\frac{H\dot{\phi}}{a}  + g^2 Z_0^2 {\phi \over a} = 0.
\end{align}
For the remainder of this work we  instead use the variable $\psi=\phi/a$, in terms of which, eq.\ \eqref{eqn:phieom} is
\begin{align}
\label{eq:psibackgroundeom}
\ddot \psi + {3 H \dot \psi  } + \psi \dot H
+   {2\kappa g^2 \psi^3\dot \psi^2 \over (1+\kappa g^2 \psi^4)}+ {\psi (2H^2 + 2g^2 \psi^2 +g^2 Z_0^2 )  \over (1+\kappa g^2 \psi^4)}   = 0 .
\end{align}

The equations of motion for the metric are the Friedmann constraint
\begin{align}\label{eqn:Friedmann}
M_{\rm Pl}^2H^{2} =  &  \frac{1}{2}\frac{\dot\phi^2}{a^2} + \frac{1}{2}g^{2}\frac{\phi^4}{a^4}+ \frac{1}{2}g^{2}\frac{\phi^2}{a^2}\hvev^2 + \frac{1}{2}\kappa \frac{ g^2\phi^4\dot{\phi}^2}{a^6} ,
\end{align}
and
\begin{align}\label{eq:Hdot}
M_{\rm Pl}^2\dot{H} = - \frac{\dot{\phi}^2}{ a^2} - g^2\frac{\phi^4}{a^4} - \frac{1}{2}g^2\frac{\phi^2}{a^2}\hvev^2,
\end{align}
which can be combined to read
\begin{align}
\label{eq:friedmansum}
M_{\rm Pl}^2(\dot H + 2H^2) = {1\over 2} g^2 \psi^2 Z_0^2 + \kappa g^2 \psi^4 (\psi H + \dot \psi)^2.
\end{align}
We introduce the standard Hubble slow roll parameters,
\begin{align}
\epsilon = -\frac{\dot{H}}{H^2}, \quad \eta = -\frac{\ddot{H}}{2H\dot{H}} = \epsilon - \frac{\dot{\epsilon}}{2\epsilon H},
\end{align}
as well as
\begin{align}
\delta = -\frac{\dot{\psi}}{H\psi},
\end{align}
which characterizes the slow-roll of the gauge vev.  The dimensionless mass parameters\footnote{The parameter $\gamma$ is identical to the parameter $m_\psi^2$ defined in ref.\ \cite{Adshead:2016omu}. In this work we use $\gamma$ to be consistent with the nomenclature frequently used for gauge-flation, as for example in ref.\ \cite{Maleknejad:2011jw, Maleknejad:2011sq,Namba:2013kia}. We use $M$ for the contribution to the mass due to the Higgs VEV, rather than $\omega$ as used in ref.\ \cite{Nieto:2016gnp}}
\begin{align}
\label{eq:definegammaM}
\gamma = \frac{g^2\psi^2}{H^2}, \quad M = \frac{g \hvev}{H},
\end{align}
characterize the various contributions to the mass of the gauge field fluctuations in units of the Hubble scale. 

The definition of the slow-roll parameter $\epsilon$ applied to eq.\ \eqref{eq:Hdot} leads to the exact relation \cite{Nieto:2016gnp}
\begin{align}
\epsilon ={ \psi^2 \over M_{\rm Pl}^2} \left (   (1-\delta)^2+\gamma+{M^2\over2} \right ).
\label{eq:epsilonexact}
\end{align}
Alternatively,  using eq.\ \eqref{eq:friedmansum}, $\epsilon$ can be expressed as
\begin{align}
\epsilon =2- \kappa  g^2 \psi^6 (1-\delta)^2  -{1\over 2} \psi^2 M^2.
\label{eq:kappasub}
\end{align}
Equations \eqref{eq:kappasub} and \eqref{eq:epsilonexact}  can be used to express  $\kappa$ and $\psi$ in terms of $\epsilon, \gamma, \delta$ and $M^2$
\begin{align}\label{eqn:kappa}
\kappa  = \frac{2-\epsilon}{g^2 \psi^6 (1-\delta)^2}-{1\over 2} \frac{M^2}{g^2 \psi^4 (1-\delta)^2}, \quad \frac{\psi}{M_{\rm Pl}} = \sqrt{\frac{\epsilon}{(1-\delta)^2+\gamma+\frac{M^2}{2}}} .
\end{align}
Differentiating eq.\ \eqref{eq:kappasub} with respect to cosmic time and using eq.\ \eqref{eq:epsilonexact} gives the exact expression
\begin{align}
\eta = 
\epsilon - 
 \left ( 2 - \epsilon - {   \epsilon \,  {M^2 \over2}  \over     (1-\delta)^2+\gamma  +{M^2\over2}   } \right )
  \left ( {3\delta \over \epsilon }  + {\dot \delta \over \epsilon H (1-\delta) } \right ) + {(\epsilon -\delta ) \, {M^2\over2} \over   (1-\delta)^2+\gamma  +{M^2\over2}  } .
\end{align}
Performing the same differentiation on eq.\ \eqref{eq:epsilonexact}, we arrive at the equivalent exact relation 
\begin{align}
\eta = \frac{\psi^2}{M_{\rm Pl}^2} \((1-\delta)^2+ (1-\delta) {\dot \delta \over \epsilon H} + \gamma{ \delta \over \epsilon}\) +\delta .
\end{align}
Using the exact relations obtained above, the slow-roll parameters can be shown to satisfy the relations \cite{Nieto:2016gnp}
\begin{align}\label{eqn:deltareplace}
\epsilon \simeq \frac{\psi^2}{M_{\rm Pl}^2}\(1+\gamma+\frac{M^2}{2}\),
 \quad
  \eta \simeq \frac{\psi^2}{M_{\rm Pl}^2}, 
\quad
 \delta \simeq \frac{\gamma+M^2}{6\(1+\gamma+\frac{M^2}{2}\)}\epsilon^2,
\end{align}
to lowest non-trivial order. Note that $\eta = {\cal O}(\epsilon)$ and  $\delta ={\cal O}(\epsilon^2)$, which implies that the relative change in $\psi$ during inflation is much smaller compared to the corresponding change in $H$. Therefore, to a very good approximation, $\psi \approx$ constant throughout inflation. 

The Hubble parameter can be re-written using the definition of $\gamma$ and the slow-roll approximation of $\epsilon$ as
\begin{align}
\frac{H^2}{M_{\rm Pl}^2} \approx   {g^2 \epsilon \over \gamma \left(1+\gamma +{M^2\over 2} \right) } \, .
\end{align}
Finally, the total number of $e$-folds of inflation can be conveniently expressed only in terms of initial values as
\begin{align}
\label{eq:Napprox}
N_e \approx    {M_{\rm Pl}^2\over 2\psi_{\rm in} ^2 } \ln\[\frac{1+\gamma_{\rm in}+M_{\rm in}^2/2}{\gamma_{\rm in}+M_{\rm in}^2/2}\] .
\end{align}

We end this section with a comment on the parameters required to fully characterize the background evolution of the system. As in Gauge-flation, the parameter $\kappa$ can be eliminated by rescaling time $t \to t \sqrt\kappa$ and the gauge coupling  $g\to g/\sqrt \kappa$ \cite{Namba:2013kia}. This rescales the value of the Hubble parameter as $H\to H/\sqrt{\kappa}$, and thus the value of $\kappa$ is determined by fixing the overall amplitude of the scalar spectrum to $A_s \simeq 2\times 10^{-9}$. The remaining parameters of the theory are $g$ and $Z_0$.

In what follows, we use $e$-folding number, $N  = -\ln(a/a_0)$, as our time parameter. For convenience, we choose $a_0 = 1$ at $N = 60$ $e$-folds before the end of inflation where necessary. In order to specify a background trajectory, $(\psi(N), \dot{\psi}(N))$, we need to specify the set $\{H_{\rm in}, \psi_{\rm in}, \dot\psi_{\rm in}, \gamma_{\rm in}, M_{\rm in}, g, Z_0\}$ for some initial $N_{\rm in}$. However, note that these seven quantities are not all independent.  We can use the definition of $\delta$ to express $\dot\psi_{\rm in}$ as
\begin{align}\label{eqn:dotpsi}
\dot \psi_{\rm in} =\frac{1}{6} \left(\gamma_{\rm in} +M^2_{\rm in}\right) \left(\gamma_{\rm in} +\frac{M_{\rm in}^2}{2}+1\right) \frac{\psi_{\rm in}^5}{M_{\rm Pl}^4} H_{\rm in}.
\end{align}
Together with the Friedmann constraint eq.\ \eqref{eqn:Friedmann}, eq.\ \eqref{eqn:dotpsi} and the definitions at eq.\ \eqref{eq:definegammaM} provide four relations among the seven variables. This reduces the required parameters to three.  

\begin{figure}[t!]
\centering
\includegraphics[width=\textwidth]{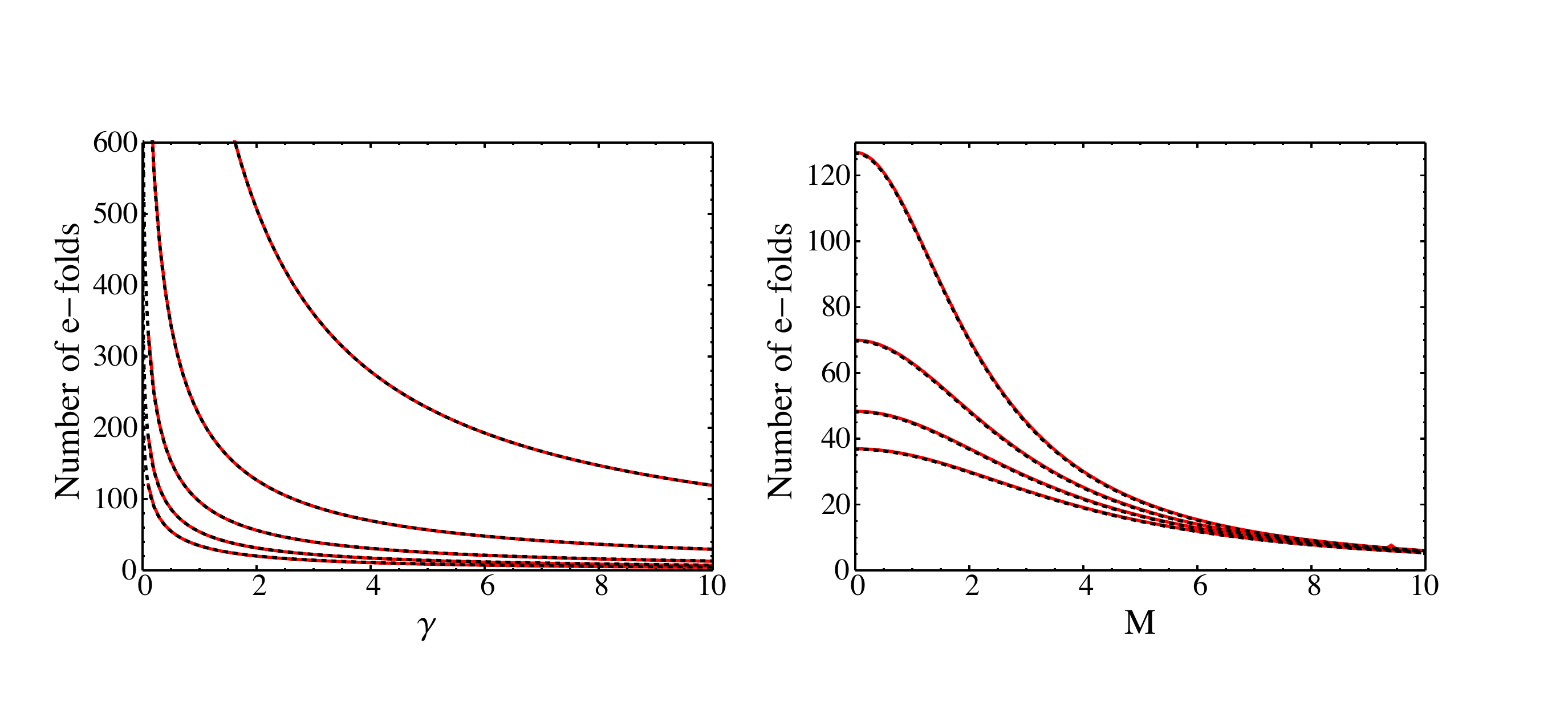}
\caption{
The total number of $e$-folds of inflation using the full numerical evolution of the system (red) and using eq.\ \eqref{eq:Napprox} (black-dotted). Left: The number of $e$-folds is plotted as a function of $\gamma$ for $M=0$ and $\psi/M_{\rm Pl} = 0.02, 0.04,0.06,0.08,0.1$ (top to bottom). Right: The number of $e$-folds is plotted as a function of $M$ with $\psi/M_{\rm Pl}=0.04$ and $\gamma= 2,4,6,8$ (top to bottom). }
\label{fig:efoldscan}
\end{figure}

If we specify $\gamma_{\rm in}$, $\psi_{\rm in}$, and $M_{\rm in}$, eq.\ \eqref{eq:Napprox} determines the length of the inflationary phase. Figure \ref{fig:efoldscan} shows the resulting number of $e$-folds for various parameter combinations. Increasing all three parameters ($\psi_{\rm in}$, $\gamma_{\rm in}$ and $M_{\rm in}$) leads to a decrease in the total number of $e$-folds of inflation. However, there remains a large region of parameter space where sufficient inflation is easily achieved. In the evolution of the perturbations we present below, we choose to specify $\gamma_{\rm in}$ and $M_{\rm in}$ at $N_{\rm in} = 60$ $e$-folds before inflation ends. In this case, eq.\ \eqref{eq:Napprox} specifies $\psi_{\rm in}$. Since eq.\ \eqref{eq:Napprox} is a very good approximation for all $\gamma$, $M$, and $\psi$, with the further (excellent) approximation that $\psi \approx \psi_{\rm in}$, eq.\ \eqref{eq:Napprox} and eq.\ \eqref{eqn:Friedmann} can be solved for the subsequent values of $M$ and $\gamma$ as a function of $N$, the $e$-folding number measured with respect to the end of inflation.

%
\section{Linear perturbations}\label{sec:quadaction}
%

In order to find the spectra of density and gravitational wave fluctuations in Higgsed Gauge-flation, we need to understand how the field and metric fluctuations evolve. In this section we derive the action to quadratic order in small fluctuations about the solutions described above in section \ref{sec:backgroundeqns}. We begin by deriving the action for a the fluctuations of a general SU(2) gauge field about the background field trajectory before we specialize to a two-dimensional representation and introduce a scalar-vector-tensor decomposition of the fluctuations in section \ref{sec:2drep}. Sections \ref{sec:scalars}, \ref{sec:vectors}, and \ref{sec:tensors}, study the scalar, vector, and tensor fluctuations, respectively.

To proceed, we write the metric in ADM form \cite{Arnowitt:1962hi},
\begin{align}\label{eqn:adm}
ds^2 = -N^2 d\tau^2 +  \tilde{h}_{ij}(dx^i+N^i d\tau)(dx^j+N^j d\tau),
\end{align}
where $N$ is the lapse, $N^i$ is the shift vector, and $\tilde{h}_{ij}$ is the metric on the spatial hypersurface. At zeroth order in fluctuations, the FRW metric in conformal time corresponds to $N = a$ and $N^i = 0$ in our conventions. The metric on the hypersurface, $\tilde{h}$, can be decomposed into scalars, vectors and tensors by writing
\begin{align}\label{eqn:spatialmet}
\tilde{h}_{ij} = a^2\left[(1+A)\delta_{ij}+\partial_{i}\partial_j B + \partial_{(i}C_{j)} + \gamma_{ij}\right],
\end{align}
where $\gamma_{ii} = \partial_i \gamma_{ij} = 0$, and $\partial_iC_i = 0$.
The coordinate invariance of general relativity allows to impose four conditions on the fields in eq.\ (\ref{eqn:spatialmet}).  For this work, we choose spatially flat gauge, where the time threading and spatial coordinates are chosen so that $A = B = 0$. The remaining spatial reparametrizations can then be used to set $C_i = 0$, which completely fixes the coordinates. We further write\footnote{Our summation convention is the same as the one above, repeated lower indices are summed with the Kronecker delta, while upper indices paired with lower indices are summed with the metric $\tilde h_{ij}$ and its reciprocal $\tilde h^{ij}$.}
\begin{align}\label{eqn:spatialmet2}
\tilde{h}_{ij} = a^2\[e^{\gamma}\]_{ij} = a^2\left[\delta_{ij} + \gamma_{ij}+\frac{1}{2!}\gamma_{ik}\gamma_{kj}+\ldots\right],
\end{align}
so that $\det[\tilde{h}_{ij}] = a^8$ to all orders in perturbation theory.

Inserting the ADM metric at eq.\ (\ref{eqn:adm}) into the action in eq.\ (\ref{eqn:GFaction}) we find,
\begin{align}\nn
S = & \int \d^4 x  \sqrt{\tilde{h}}\[N R{}^{(3)} + \frac{1}{N}(E^{ij}E_{ij} - E^2)\]+\frac{1}{6}\kappa \int d^{4}x\frac{1}{\sqrt{\tilde h }N}\(\epsilon_{ijk}\tr\[F_{0i}F_{jk}\]\)^2\\ \nn & +
\int \d^{4}x \frac{\sqrt{\tilde h}}{ N}\tr\[(F_{0i}+N^{k}F_{ik})\tilde{h}^{ij}(F_{0j}+N^{l}F_{jl})\]-\frac{1}{2}\int d^{4}x \sqrt{\tilde h}N\tr\[ \tilde{h}^{ik}\tilde{h}^{jl}F_{ij}F_{kl}\]\\\nn &+ \int \d^{4}x\frac{\sqrt{\tilde{h}} }{N}g^2\hvev^2\tr\[ A_{0} - \frac{i}{g}U^{-1}\partial_\tau U+N^i\( A_{i} - \frac{i}{g}U^{-1}\partial_i U\)\]^2\\ & -\int \d^{4}x\sqrt{\tilde{h}}N g^2\hvev^2 \tr\[\[ A_{i} - \frac{i}{g}U^{-1}\partial_i U\]\tilde{h}^{ij}\[ A_{j} - \frac{i}{g}U^{-1}\partial_j U\]\].
\end{align}
In this expression, $E_{ij}$ is related to the extrinsic curvature of the spatial slices
\begin{align}
E_{ij}   = & \frac{1}{2}\left(\partial_{\tau}{h}_{ij} -
\nabla_{i}N_{j}-\nabla_{j}N_{i}\right),\quad 
E  =  E^{i}_{\;i},
\end{align}
and $\nabla_i$ is the covariant derivative constructed from $\tilde{h}$.  Note that spatial indices are raised and lowered using $\tilde h_{ij}$ and its reciprocal $\tilde h^{ij}$.  

We proceed by expanding the lapse and shift about a Friedmann-Robertson-Walker spacetime
\begin{align}
N = & a(1+\dN_{(1)} +\dN_{(2)} +\ldots),\quad 
N^{i} = N^{i}_{(1)}+N^{i}_{(2)} +\ldots,
\end{align}
where $\dN_{(1)}$ and $N^i_{(1)}$, and $\dN_{(2)}$ and $N^i_{(2)}$,  are first and second order in fluctuations, respectively. As is well known, in order to obtain the quadratic action we require only the constraints at linear order, and thus we drop the subscripts in what follows.

In spatially flat gauge, neglecting gravitational waves for a moment, the curvature of the spatial slices vanishes, ${}^{3}R = 0$, and the connection
for $\nabla_{i}$ (the covariant derivative compatible with the metric $\tilde{h}_{ij}$ on the hypersurface) vanishes and thus $\nabla_{i}\rightarrow\partial_{i}$.  The Einstein-Hilbert action to quadratic order in scalar and vector fluctuations is given by
\begin{align}\nn\label{eqn:EHaction}
\delta^2 S_{EH} = & \frac{M_{\rm Pl}^2}{2}\int \d^4 x a^2 \Bigg[-(4aH \partial_{i}N^{i}\dN+6a^2 H^{2}\dN^2)  +4\partial_{(i}N^{j)}\partial_{(i}N^{j)} - \partial_{i}N^{i}\partial_{j}N^{j}
\Bigg].
\end{align}
We denote the fluctuations in the gauge field by 
\begin{align}
\delta\! A_{\mu} = \da_{\mu},
\end{align}
in terms of which the gauge field and Stueckelberg action to quadratic order in field fluctuations, and scalar and vector metric fluctuations is given by
\begin{align}\label{eqn:GFactiongrav}\nn
\delta^2  S_{A} = & \int \d^4 x \[\delta^{2}\!\mathcal{L}_{\rm YM} +{\delta^2}\! \mathcal{L}_{\rm \kappa}+{\delta^2}\! \mathcal{L}_{\rm Higgs}\] - \frac{1}{2}\int \d^{4}x a^4 \dN\(g\frac{\phi^2}{a^4}(2\epsilon_{a ij}\partial_{i}\da_{aj}+4 g\phi\da_{ii} )\)\\ \nn&+ \frac{1}{2}\int \d^{4}x a^2 \Bigg[3\dot{\phi}^2\(  +\kappa g^2  \frac{\phi^4}{a^4}\)\dN^2 -2\frac{\dot{\phi}}{a} (\partial_{\tau}\da_{ii} - \partial_i \da_{i 0} )\dN +(2g^{2}\frac{\phi^4}{a^6} +\frac{1}{2}  g^2 \phi^2 {Z_0}^2 )N_{i}N_{i}\\ \nn & 
+\frac{\dot{\phi}}{a^3}2 N_k((\partial_{i}\da_{ik}-\partial_{k}\da_{ii})+ g \phi \epsilon_{aki}   \da_{ai})+   g\frac{\phi^2}{a^4}2 N_k(  \epsilon_{aik}\partial_\tau \da_{ai} -  \epsilon_{aik}\partial_{i}\da_{a0} -2 g \phi \da_{k0}) \Bigg]\\\nn
& -\kappa \int \d^{4}x\frac{\dN}{a^3}g\dot\phi \phi^2\(g\phi^2 \tr\[\partial_{i}\da_0   J_i \]+g\partial_{\tau}\(\phi^2\tr\[ \da_i J_i\]\)   -\epsilon^{ijk}\partial_{\tau}\phi\tr\[ J_i \partial_{j}\da_{k}\]\) \\ & + \int \d^{4}x a^4\[-g^2 \psi Z_0^2 \dN \da^i{}_{i} - g^2 \psi  Z_0^2
 \delta^a{}_{i}  N^i \da^a_0+\frac{g \phi }{a^2} Z_0^2
   \dN \partial_i \xi^i +\frac{ g \phi }{a^2} Z_0^2
 \delta^a{}_{i} N^i \partial_\tau \xi^a\].
\end{align}
The terms $\delta^{2}\!\mathcal{L}_{\rm YM}$ and  $\delta^2\! \mathcal{L}_{\rm \kappa}$ are given by
\begin{align}\label{eqn:actYM}\nn
\delta^2\mathcal{L}_{\rm YM} = &\tr\[(\partial_{i}\da_{0}-i g\phi\[J_i, \da_0\])^2\] -4ig\partial_{\tau}\phi\tr\[\da_0\[\da_i, J_i\]\]
%
-2\tr\[\da_{0}\partial_{\tau}(\partial_{i}\da_{i}-ig\phi \[J_{i}, \da_i\])\]
\\ \nn& +\tr\[\partial_{\tau}\da_i\partial_{\tau}\da_i\] -\tr\[\partial_j \da_{i}\partial_{j}\da_{i}-\partial_i \da_{j}\partial_{j}\da_{i}\]+2g\phi\epsilon_{ijk}\tr\[\partial_i \da_{j}\Omega_k \] \\ &-g^2\phi^2\tr\[(\Omega_k-\da_{k})\Omega_{k}\],
\end{align}
and we have defined
\begin{align}\label{eqn:omegadef}
\Omega_i = i\epsilon_{ijk}\[J_j,\da_k\],
\end{align}
and 
\begin{align}\label{act:kappa}
\delta^2\mathcal{L}_{\kappa} =& -\frac{\kappa}{2 a^3}g\dot{\phi}\phi^2\partial_{\tau}\tr\[ g\phi \da_i \Omega_{i}-\epsilon^{ijk}\da_i\partial_{j}\da_{k}\]\\\nn &  +\frac{\kappa}{6 a^4}\(g\phi^2 \tr\[\partial_{i}\da_0   J_i \]+g\partial_{\tau}\(\phi^2\tr\[ \da_i J_i\]\)   -\epsilon^{ijk}\partial_{\tau}\phi\tr\[ J_i \partial_{j}\da_{k}\]\)^2.
\end{align}
The Goldstone modes contribute at quadratic order in fluctuations via
\begin{align}
\delta^2\mathcal{L}_{\rm Higgs} = & a^4\Bigg[-\frac{g^2\hvev^2}{2}\bar{g}^{\mu\nu}\(\partial_{\mu}\dHiggs^a + \da^a_\mu\)\(\partial_{\nu}\dHiggs^a + \da^a_\nu\)+\frac{g^2\hvev^2{g\psi}}{a}\epsilon_{bic}\dHiggs^b\partial_{i}\dHiggs^c\Bigg].
\end{align}
The addition of a Higgs sector thus yields an additional mass term for the gauge field fluctuations. Note, however, that retaining gauge-invariance requires us to also add the Goldstone modes $\dHiggs^a$.

Finally, the quadratic Lagrangian density for the transverse-traceless components of the metric, and their interactions with the gauge field fluctuations is given by
\begin{align}\nn\label{eqn:actspin2}
\delta^2\mathcal{L}_{\gamma} = & \frac{a^2M_{\rm Pl}^2}{8} \((\partial_{\tau}\gamma)^{2}-(\partial_i \gamma)^{2}+\frac{2}{M_{\rm Pl}^2}\(\dot\phi^2-g^{2}\frac{\phi^4}{a^2}\)\gamma^2\)\\  & -a^2 \(\frac{\dot{\phi}}{a} \partial_{\tau}\da_{jl} -g\frac{\phi^2}{a^2}(2\epsilon^{a}_{ij}\partial_{[i}\da^a_{l]}+g\phi \da_{jl}) \)\gamma_{jl}  
 -a^2\frac{g^2\phi^2\hvev^2}{4} \gamma^2+a^2 g^2 \hvev^2\phi\gamma_{ij}\da_{ij},
\end{align}
where $\gamma^2 = \gamma_{ij}\gamma_{ij}$.
In order to proceed we need to choose a specific representation for the gauge field. We focus on a two-dimensional representation in what follows for simplicity.

\subsection{Two dimensional representation and scalar-vector-tensor decomposition}\label{sec:2drep}

Specializing to the case of a $N = 2$ dimensional representation of SU(2), the representation matrices are the Pauli matrices, $J_a = \sigma_a/2$, and we can decompose the gauge field fluctuations into scalar, vector, and tensor fluctuations.  In order to make contact with the existing literature, we decompose the gauge field, Goldstone, and metric fluctuations as \cite{Namba:2013kia}
\begin{align}\label{eqn:fields1}
\da^{a}_0 = & a\delta^a_i (\partial_i Y + Y_i),\\
\da^a_{i} = & a((\psi+\delta\psi)\delta^a_{i}+\delta^a_j\partial_i(\M_j+\partial_j \M)+\delta^{a}_k\epsilon_{ikj}(U_j+\partial_j U) +\delta^{a}_{j}t_{ij}),\\
\dHiggs^a = & \delta^{a}_i(\dHiggs_i + \partial_i\dHiggs), \\ \label{eqn:fields4}
N^i = & \partial_i \theta + N^i_V, \\
N = & 1+\alpha,
\end{align}
where $Y$, $\theta$, $\alpha$, $\delta\psi$, $U$, $\dHiggs$, and $\M$ are scalars; $Y_i$, $\M_j$, $\dHiggs_i$, $N^i_V$, and $U_j$ are transverse vectors which satisfy $\partial_iY_i = \partial_j\M_j = \partial_i \dHiggs_i =\partial_i N^i_V  = \partial _jU_j = 0$. Finally, $t_{ia}$ is a transverse and traceless tensor $t_{ii} = \partial_i t_{ia} = \partial_a t_{ia} = 0$. We fix the gauge for the gauge field fluctuations by setting
\begin{align}
U = U_j = 0,
\end{align}
which is equivalent to choosing $\da^a_{i}$ to be symmetric under exchange of $i \leftrightarrow a$. At quadratic order, the Lagrangian separates into separate scalar, vector, and tensor pieces as usual, and in what follows we consider each type of fluctuation separately.

\subsection{Scalar fluctuations}\label{sec:scalars}

After gauge fixing, there are five scalar fluctuation degrees of freedom in this theory,  $\delta\psi$, $\M$, and $Y$, which arise from the gauge sector, $\dHiggs$ arising from the Higgs sector, and $\alpha$ and $\theta$ that arise from the metric perturbations. The quadratic action for these degrees of freedom reads
\begin{align}
\noindent
&\delta^2 S =   \int \frac{\d^3 k}{(2\pi)^3}\d\tau \mathcal{L}_{\rm scalar}\\
&\mathcal{L}_{\rm scalar} = \frac{a^2(1+\kappa g^2 \psi^4)}{6}|3\delta\psi'- k^2\M'|^2+\frac{k^4a^2}{3}|\M'|^2+\frac{a^2k^2}{2}Z_0^2(|\dHiggs'|^2-k^2|\dHiggs|^2) \\ \nn& 
-\frac{a^2}{3}\Bigg\{\frac{k^2}{3}+\frac{g^2\psi^2(1+g^2\kappa \psi^4)}{4 a^2 M_{\rm Pl}^2}(2 \kappa \psi^2(a\psi)'{}^2+a^2 g^2 Z_0^2)+\frac{g^2a^2Z_0^2(3g^2\kappa\psi^4 - 1)}{1+g^2\kappa \psi^4}\\ \nn& +\frac{1}{1+g^2\kappa \psi^4}\[g^2\psi^2(3-g^2\kappa^2\psi^4)(a^2+\kappa\psi^2\frac{a'{}^2}{a^2}+\kappa\psi'{}^2)\]\Bigg\}|3{\delta\psi}- k^2{\M}|^2\\\nn
& -\frac{k^4a^2}{3}(\delta\psi^*\M+\M^*\delta\psi)+\frac{a^2 k^2}{3}\Bigg\{\frac{k^2}{3}-\frac{g^2 \psi^2}{2 M_{\rm Pl}^2}(\kappa \psi^2(a\psi)'{}^2+a^2Z_0^2)-g^2a^2 Z_0^2\\ \nn& -\frac{2g^2\kappa^2 \psi^2}{1+\kappa g^2 \psi^4}\(g^2 a^2 \(\psi^2+\frac{Z_0^2}{2}\)+\frac{a'{}^2}{a^2}\psi^2-\psi'{}^2\)\Bigg\}|\M|^2 -\frac{a^3}{4}g^2Z_0^2 k^2 (\dHiggs^* (\delta\psi-k^2\M)+\text{h.c.})\\\nn
 &+ \frac{1}{2} a^2k^2 \left( a^2 g^2 \left(2 \psi ^2+Z_0^2\right)+k^2 \left(1+\frac{g^2 \kappa  \psi ^4}{3}\right)\right)|Y|^2+\frac{3}{2}\(\left(a \psi\right)'{}^2 \left(1+g^2 \kappa  \psi ^4\right)-2 M_{\rm Pl}^2 a'{}^2\)\dN^2\\\nn
 & + \frac{1}{2} g^2 k^2 a^4\psi ^2 \left( Z_0^2+2\psi^2\right)|\theta|^2+k^2 M_{\rm Pl}^2 a a'(\theta^*\dN+\text{h.c.}) -\frac{1}{2} a^2 g^2 k^2 Z_0^2 \phi (\dN^* \dHiggs+\text{h.c.})\\\nn
& -\frac{a^2k^2}{6}\Bigg[
 \Bigg(\frac{1}{a}\[a'\left(1+g^2 \kappa  \psi ^4\right)+2 g^2 \kappa  \phi ^3 (a\psi)'\](k^2 \M-3 \delta\psi)+2\frac{a'}{a}k^2\M\\ \nn& 
 +a \left(3 a g^2 Z_0^2\dHiggs'+ \left(1+g^2 \kappa  \psi ^4\right)(k^2 \M'-3 \delta\psi')+2k^2 \M'\right)\Bigg)Y+\text{h.c.}\Bigg]\\ \nn& 
 -\frac{a k^2   \left(1+g^2 \kappa  \psi ^4\right)}{2}(a\psi )'(\dN^*Y+\text{h.c.})-\frac{1}{2} a^3 g^2 k^2 \psi  \left( Z_0^2+2 \psi ^2\right)(\theta^*Y+\text{h.c.})\\\nn
 & -\frac{1}{2}
 \left(a'(a\psi)' \left(1+g^2 \kappa  \psi ^4\right)+a g^2 \phi  \left(a^2Z_0^2+2
    \psi ^2+2 \kappa  \phi ^2 \left(a\psi\right)'{}^2\right)\right)\[(3\delta\psi-k^2 \M)\dN^*+\text{h.c.}\]\\\nn
&-\frac{a}{2}(1+g^2\kappa\psi^4)(a\psi)' (\dN^*(3\delta\psi'-k^2\M')+\text{h.c.})   +\frac{1}{2} a k^2\[ \left(a ^2g^2 Z_0^2 \psi  \dHiggs'- 2  (a\psi)'\delta\psi \right)\theta^*+\text{h.c.}\],
\end{align}
where we have integrated by parts, discarded a boundary term, and made use of the background equations of motion. Note that in the limit $Z_0\to 0$, the above action does not quite agree with the corresponding expression in ref.\ \cite{Namba:2013kia}. The difference arises due to a slightly different choice of parametrization of the gravitational constraints. In this work, we have chosen to split the metric using ADM variables.

Note that the fields $Y$, $\alpha$, and $\theta$ appear in the action without time derivatives. As described in detail in ref.\ \cite{Namba:2013kia}, these fields are algebraic constraints, and can be integrated out by solving their linear equations of motion and substituting the solutions back into the action. While this is a straightforward procedure, the result is extremely messy and we do not reproduce it here.

Denoting by $\X = (\delta \psi, \M , \dHiggs)$, we redefine the fields using the transformation
$X_{i} = U_{ij}\Delta_j$, where\footnote{This field redefinition is the redefinition that diagonalizes the kinetic term in the limit where we simply set $\dN\to 0$ and $\theta \to 0$ and only integrate out $Y$. Since the gravitational interactions are only important for momentum $k \lesssim aH$, this transformation also diagonalizes the kinetic term in the limit $k \gg aH$, as required for setting the initial conditions via canonical quantization. In particular, note that the parts of the matrix corresponding to the redefinition of $\delta \psi$ and $\M$ are identical to those of ref.\ \cite{Namba:2013kia}.}
\begin{align}\label{eqn:fieldred}
{\bf U} = \(\begin{matrix} \frac{\sqrt{3+g^2 \kappa  \psi ^4}}{\sqrt{6} a \sqrt{1+g^2 \kappa  \psi ^4}} & 0 & 0\\\frac{\sqrt{\frac{3}{2}} \sqrt{1+g^2 \kappa  \psi ^4}}{a k^2 \sqrt{3+g^2 \kappa  \psi ^4}} & \frac{\sqrt{6 a^2 g^2 \psi ^2+k^2 \left(3+g^2 \kappa  \psi ^4\right)}}{\sqrt{2} a^2 g k^2 \psi  \sqrt{3+g^2 \kappa  \psi ^4}} & 0 \\  0  &\frac{ \sqrt{3 + g^2 \kappa \psi^4}}{
 2 \sqrt{2} a H \sqrt{\gamma}\sqrt{
  6 a^2 H^2\gamma + k^2 (3 + g^2 \kappa \psi^4)}}  & \frac{\sqrt{3 a^2 H^2 (M^2 + 2 \gamma) + 
 k^2 (3 + g^2 \kappa \psi^4)}}{a H k \sqrt{
 6 a^2 H^2\gamma + k^2 (3 + g^2 \kappa \psi^4)}}\end{matrix}\).
\end{align}
After integration by parts and discarding boundary terms, this field redefinition puts the action in the form
\begin{align}
S = \frac{1}{2}\int \frac{\d^3k}{(2\pi)^3}\d\tau\[\Delta^\dagger{}'{\bf T}\Delta' + \Delta^\dagger{}'{\bf K}\Delta-\Delta^\dagger{}{\bf K}\Delta' - \Delta^{\dagger}\boldsymbol{\Omega}^2\Delta\].
\end{align}
In principle it is possible to choose a redefinition, ${\bf U}$, that sets the kinetic matrix, ${\bf T}$, to the identity matrix, however, this requires a much more complicated  transformation which makes the algebra much more involved. As discussed in detail in ref.\ \cite{Namba:2013kia}, this is not necessary to evolve the fluctuations. All that is required for our purposes is that ${\bf T}$ approaches the identity in the limit $k \gg aH$ in order to impose the initial conditions.

While the matrices ${\bf T}$, ${\bf K}$, and ${\bf \Omega}^2$ are obtained in a fairly straightforward  manner as we have described above, they are extremely long, and their full form is not particularly illuminating. In  appendix \ref{app:scalsector}, we present slow-roll expansions of the matrices.

At early times, $k \gg aH$, the symmetric kinetic matrix, ${\bf T}$, is
\begin{align}\begin{array}{llll}
T_{11} \simeq & 1+ \frac{6  (M^2 + 2 \gamma) \epsilon^2 }{(2 + M^2 + 
   2 \gamma)^2}\frac{a^2H^2}{k^2} , & T_{12}\simeq- \frac{2\sqrt{3\gamma}}{2+2\gamma+M^2}\frac{aH}{k} , &T_{13}\simeq  -\frac{\sqrt{6} M \epsilon }{2+2 \gamma +M^2}\frac{aH}{k},\\
T_{22} \simeq & 1+\frac{6  \gamma \epsilon}{(2 + M^2 + 2 \gamma)}\frac{a^2H^2}{k^2}, & T_{23} \simeq  \frac{3 \sqrt{2\gamma }  M \epsilon }{ \left(2+2 \gamma +M^2\right)}\frac{a^2H^2}{k^2}, &T_{33} \simeq 1+\frac{3   M^2 \epsilon}{2 + 2 \gamma + M^2}\frac{a^2H^2}{k^2},
\end{array}
\end{align}
while the anti-symmetric ${\bf K}$ matrix has non-zero entries
\begin{align}
\begin{array}{llll}
K_{12} & \simeq  -\frac{k}{\sqrt{3\gamma}}, &K_{13} \simeq  -\frac{3 \sqrt{\frac{3}{2}} a^2 H^2 M \epsilon }{k \left(2 \gamma+M^2+2\right)}, & K_{23} \simeq \frac{a H M}{\sqrt{2} \sqrt{\gamma }},
\end{array}
\end{align}
and the symmetric $\boldsymbol{\Omega}^2$ matrix has entries
\begin{align}
\Omega_{11}^2 &\simeq \frac{k^2}{3}, \quad \Omega_{12}^2 \simeq \frac{2+2\gamma+M^2}{\sqrt{3\gamma}} aH k ,\quad  \Omega_{13}^2 \simeq   -\sqrt{\frac{2}{3}}M aH k, \\
\Omega_{22}^2 & \simeq  \(1- \frac{2}{\gamma}\)k^2, \quad \Omega_{23}^2 \simeq 3\frac{M}{\sqrt{2\gamma}} a^2 H^2 , \quad \Omega_{33}^2 \simeq k^2.
\end{align}
While, for superhorizon modes $k \ll a H$, 
\begin{align}
\begin{array}{lll}
T_{11} \simeq & 1+ \frac{2+\gamma+M^2}{M^2+\gamma} , 
 & T_{12}\simeq  -\frac{\sqrt{2(2+2\gamma+M^2)}}{(M^2+\gamma)\sqrt{\epsilon}}, \\
T_{13}\simeq & \frac{M (2 + M^2 + 2 \gamma)}{
 \sqrt{3} \sqrt{\gamma} (M^2 + \gamma) \sqrt{
  M^2 + 2 \gamma} \epsilon} \frac{k}{aH}, &
T_{22}\simeq  1+ \frac{2 + M^2 + 2 \gamma}{(M^2 + \gamma) \epsilon}, \\ 
 T_{23} \simeq & -\frac{M (2 + M^2 + 2 \gamma)^{
 3/2}}{\sqrt{6}\sqrt{\gamma} (M^2 + \gamma) \sqrt{
 M^2 + 2 \gamma} \epsilon^{3/2}}\frac{k}{aH} ,
 & T_{33} \simeq  1+ \frac{M^2 (2 + M^2 + 
   2 \gamma)^2}{6  \gamma (M^2 + \gamma) (M^2 + 
   2 \gamma) \epsilon^2}\frac{k^2}{a^2 H^2},
   \end{array}
\end{align}
and
\begin{align}
\begin{array}{ll}
K_{12}\simeq &\frac{2\sqrt{2} aH}{3\sqrt{2+2\gamma+M^2}}\epsilon^{3/2}, \quad K_{13} \simeq -\frac{\sqrt{3} k M \left(2 \gamma +M^2+2\right)}{2 \sqrt{\gamma } \epsilon  \left(\gamma +M^2\right)
   \sqrt{2 \gamma +M^2}}, \quad K_{23}\simeq \frac{\sqrt{\frac{3}{2}} k M \left(2 \gamma +M^2+2\right)^{3/2}}{2 \sqrt{\gamma } \epsilon ^{3/2}
   \left(\gamma +M^2\right) \sqrt{2 \gamma +M^2}},
   \end{array}
\end{align}
and
\begin{align}
\begin{array}{lll}
\Omega_{11}^2 \simeq  & \frac{2a^2 H^2(-2+\gamma+\gamma^2+M^2 +M^4+2\gamma M^2)}{\gamma+M^2} ,  & \Omega^2_{12}\simeq  \frac{2 a^2 H^2 \sqrt{2}  \sqrt{2+2 \gamma +M^2}}{\sqrt{\epsilon } \left(\gamma +M^2\right)}, \\ 
\Omega_{13}^2\simeq  & \frac{2 a H k M \left(2 \gamma +M^2+2\right)}{\sqrt{3} \sqrt{\gamma } \epsilon  \left(\gamma +M^2\right)
   \sqrt{2 \gamma +M^2}},
 & \Omega_{22}^2 \simeq  \frac{2 a^2 H^2 \left(2 \gamma +M^2+2\right)}{\epsilon  \left(\gamma +M^2\right)},\\ \Omega_{23}^2\simeq &  -\frac{\sqrt{\frac{2}{3}} aHkM  \left(2 \gamma +M^2+2\right)^{3/2}}{\sqrt{\gamma } \left(\gamma +M^2\right) \sqrt{2 \gamma +M^2}  \epsilon
   ^{3/2}},  & \Omega_{33}^2\simeq  2 a^2 H^2.
\end{array}
\end{align}
We note that, while these expressions are extremely accurate in the asymptotic regimes, they are not accurate near horizon crossing, $-k\tau \sim 1$. Therefore, in order to solve the equations numerically, we are required to use the full expressions presented in appendix \ref{app:scalsector}.

%
\subsubsection{Initial conditions and quantization}\label{sec:ICandquant}
%

We set the initial conditions for the fields by canonically quantizing them, and using Bunch-Davies  conditions in the asymptotic past. We expand the fields into modes  \cite{Namba:2013kia}
\begin{align}\label{eqn:modeexp}
\Delta_i(\tau, {\bf k})  = \mathcal{Q}_{ij}( \tau, k)a_j({\bf k})+  \mathcal{Q}^*_{ij}(\tau, k)a^\dagger_j(-{\bf k}) , \quad \[a_i({\bf k}), a^{\dagger}_j({\bf k'})\] = \delta^{3}({\bf k} - {\bf k}')\delta_{ij},
\end{align}
where we impose the canonical commutation relation between $\Delta_i$ and its canonically conjugate momentum 
\begin{align}\label{eqn:cancom}
\[\Delta_{i}(\tau, {\bf x}), \pi_j(\tau, {\bf y}) \] = i \delta_{ij}\delta^{3}({\bf x} - {\bf y}), \quad \pi_i  \equiv \frac{\partial \mathcal{L}}{\partial (\partial_{\tau}\Delta_i^{\dagger})}.
\end{align}
We decompose the canonical momenta, $\pi_i$, into the same set of creation/annihilation operators as above,
\begin{align}
\pi_i(\tau, {\bf k}) = \pi_{ij}(\tau, {\bf k}) a_j({\bf k}) +\pi_{ij}^*(\tau, {\bf k})a^{\dagger}_j(-{\bf k}),  \quad \pi_{ij} = (\mathcal{Q}'_{ij}+K_{il} \mathcal{Q}_{lj}).
\end{align}
The relations in eqs.\ (\ref{eqn:modeexp}) and (\ref{eqn:cancom}) can only be simultaneously imposed if the condition
\begin{align}\label{eqn:inicond}
\[\mathcal{Q}\pi^{\dagger} - \mathcal{Q}^* \pi^{T}\]_{ij} = i\delta_{ij}
\end{align}
is obeyed.
As pointed out by ref.\ \cite{Namba:2013kia}, eq.\ (\ref{eqn:inicond}) can be imposed as an initial condition, which then holds at all times if the initial conditions satisfy
\begin{align}
\pi \pi^{\dagger} - \pi^* \pi^{T}  = \mathcal{Q} \mathcal{Q}^{\dagger} - \mathcal{Q}^* \mathcal{Q}^T = 0,
\end{align}
which is equivalent to imposing that the products $\pi \pi^\dagger$ and $\mathcal{Q} \mathcal{Q}^{\dagger}$ are real.

In the limit $x\to \infty$, the fields remain coupled, and  the separate fields cannot be simply quantized independently. That is, $\mathcal{Q}_{ij}$ is not simply proportional to $\delta_{ij}$. Instead, we expand the solutions into normal modes and impose the initial conditions on these solutions to fix the constants. To identify these modes, we use a Wentzel-Kramers-Brillouin (WKB) method. 

Working in the limit $k \gg a H$, and using the expressions for the matrices above, the equations of motion for the fluctuations become
\begin{align}
\mathcal{Q}''+\alpha\mathcal{Q}'+\beta\mathcal{Q} = 0,
\end{align}
where in the limit $k \gg a H$, the matrices $\alpha$ and $\beta$ are given by
\begin{align}
\alpha = \(\begin{matrix} 0 & -\frac{2k}{\sqrt{3\gamma}} & 0\\ \frac{2k}{\sqrt{3\gamma}}  & 0 & 0 \\ 0 & 0 & 0\end{matrix}\), \quad \beta =k^2 \(\begin{matrix} \frac{1}{3} & 0 & 0\\ 0 & \(1-\frac{2}{\gamma}\)& 0 \\ 0 & 0 & 1\end{matrix}\).
\end{align}
Adopting a WKB ansatz for the mode functions
\begin{align}
\vec{\mathcal{Q}}_{j} = \vec{a}_j \exp\[{i\int dx\, \omega(x)}\],
\end{align}
and substituting into the system of equations, neglecting terms of order $\mathcal{O}(\omega'/\omega)$ and $\mathcal{O}(\omega''/\omega)$, we find six solutions for the frequencies
\begin{align}\label{eq:WKBfreqs}
\omega \approx \left\{\pm 1, \pm 1, \pm \frac{\sqrt{\gamma - 2}}{\sqrt{3\gamma}}\right\}.
\end{align}
In order for the system to be stable, all of these instantaneous WKB frequencies must be real. Thus there is an instability in the system for parameters such that  $\gamma = g^2\psi^2/H^2 < 2$, as was found for the original model in ref.\ \cite{Namba:2013kia}. 

The corresponding mode solutions are, up to an irrelevant phase
\begin{align}\nn
\vec{\mathcal{Q}}_{j} = & c_{1j}\vec{a}_{1}e^{i k\tau}+c_{2j} \vec{a}_{2}e^{-ik\tau} +c_{3j} \vec{a}_{3}e^{i k\tau}+c_{4j} \vec{a}_{4}e^{-ik\tau}+c_{5j} \vec{a}_{5}e^{i  \frac{\sqrt{\gamma - 2}}{\sqrt{3\gamma}} k\tau  }+c_{6j} \vec{a}_{6}e^{-i  \frac{\sqrt{\gamma - 2}}{\sqrt{3\gamma}} k\tau}  ,
\end{align}
where   the $c_{ij}$ are constants and the $\vec{a}_{i}$ are the vectors
\begin{align}
\vec{a}_{1} = \vec{a}_2 = &
\left[
\begin{array}{ccc}
0    \\
0 \\
1  \\   
\end{array}
\right],
\quad
\vec{a}_{3} = \vec{a}_4^* = 
\left[
\begin{array}{ccc}
1   \\
-\frac{i\sqrt{\gamma}}{\sqrt{3}}\\
0  \\   
\end{array}
\right],\quad
\vec{a}_{5} = \vec{a}_6^* = 
\left[
\begin{array}{ccc}
1 \\
\frac{i}{\sqrt{\gamma-2}} \\
0   
\end{array}
\right].
\end{align}

Demanding the solutions approach the positive frequency solutions as $x = -k\tau \to \infty$ means we can set $c_{1j} = c_{3j} = c_{5j} =0$. The remaining constants now need to be set by imposing the quantization conditions above. Working in the limit $-k\tau\to\infty$, it is then straightforward to see that a solution that satisfies the initial conditions is
\begin{align}\nn\label{eqn:initialconditions}
  \text{ Goldstone mode:} \quad c_{21} =  & c_{23} = 0, \quad c_{22} = \frac{1}{\sqrt{2k}},\\\nn
 \text{ Regular mode:} \quad c_{42} = &  c_{43}=0 ,\quad c_{41} = \frac{1}{\sqrt{2k}}\sqrt{\frac{3}{{1+\gamma}}},\\
 \text{ Slow mode:} \quad c_{61} =  & c_{62} = 0, \quad c_{63} = \frac{1}{\sqrt{2k}}\frac{(3\gamma)^{1/4}(\gamma-2)^{1/4}}{\sqrt{1+\gamma}}.
\end{align}

\begin{figure}[t!]
\centering
\includegraphics[width=\textwidth]{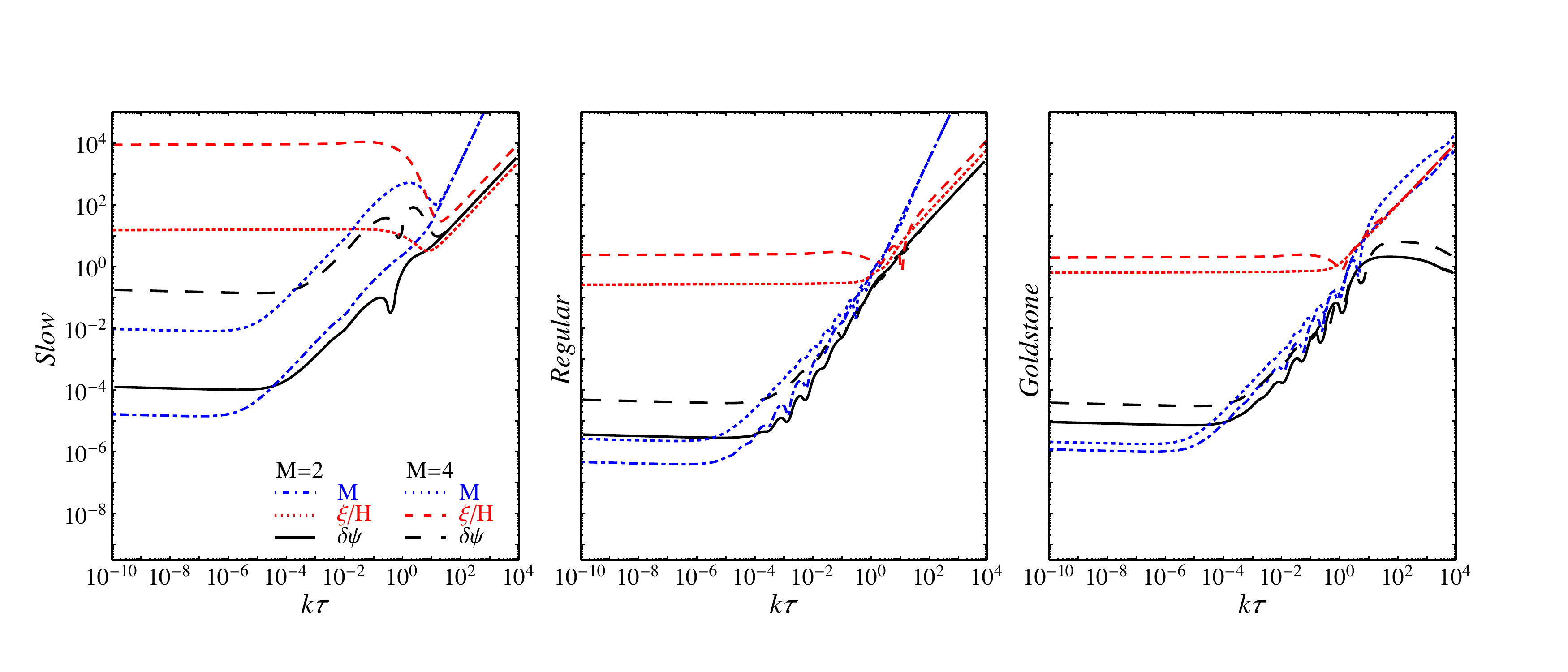}
\caption{
Evolution of scalar fluctuations in Higgsed Gauge-flation. The values of the other parameters here are chosen to be $\epsilon = 0.01$ and $\gamma = 4$. The three panels show the three independent solutions of the equations of motion, corresponding to the three independent initial conditions in eq.\ \eqref{eqn:initialconditions}. Shown are the solutions for  two different values of the Higgs vev, corresponding to $M = 2$ and $M = 4$. }
\label{fig:scalarmodes}
\end{figure}

We show the solutions to all three independent modes in figure \ref{fig:scalarmodes}. Notice that the effect of the Higgs vev and accompanying Goldstone fluctuations boosts the final amplitude of the fluctuations. These dynamics are what allows the model to become consistent with the data -- the scalar curvature fluctuations are boosted, thus lowering the tensor-to-scalar ratio. In numerically solving the system, we initialize the system including $1/k\tau$ corrections to the solutions described above. This allows more efficient and accurate evaluation starting at later times.

\subsubsection{{Superhorizon solutions}}
\label{sec:scalarsuperhorizon}

We can solve the system to a very good approximation in the superhorizon regime, $k\ll aH$, by expanding the ${\bf T}$, ${\bf K}$, and ${\bf \Omega}^2$ matrices in both $k/aH$ and $\epsilon$, keeping only the lowest-order non-trivial terms. We also use the relation $\tau = -(1+\epsilon) / aH$, which is accurate to first order in $\epsilon$.

The matrices at zeroth order in a series in  $k/aH$ and lowest non-trivial order in $\epsilon$ become
\begin{align}
\begin{array}{lll}
T_{11} \simeq & 1+ \frac{2}{M^2+\gamma} , 
 & T_{12}\simeq  -\frac{\sqrt{2(2+2\gamma+M^2)}}{(M^2+\gamma)\sqrt{\epsilon}}, \\
T_{13}\simeq & 0, &
T_{22}\simeq  1+ \frac{2 + M^2 + 2 \gamma}{(M^2 + \gamma) \epsilon}, \\ 
 T_{23} \simeq & 0 ,
 & T_{33} \simeq  1,
   \end{array}
\end{align}
and
\begin{align}
\begin{array}{ll}
K_{12}\simeq &\frac{2\sqrt{2} }{3\sqrt{2+2\gamma+M^2}}\epsilon^{3/2} {(1+\epsilon)\over -\tau}, \quad K_{13} \simeq 0, \quad 
   K_{23}\simeq0   \end{array}.
\end{align}
Finally, ${\bf \Omega}^2$ becomes
\begin{align}
\begin{array}{lll}
\Omega_{11}^2 \simeq  & \frac{2(-2+\gamma+\gamma^2+M^2 +M^4+2\gamma M^2)}{\gamma+M^2} {(1+\epsilon)^2\over \tau^2},  
& \Omega^2_{12}\simeq  \frac{2  \sqrt{2}  \sqrt{2+2 \gamma +M^2}}{\sqrt{\epsilon } \left(\gamma +M^2\right)}{(1+\epsilon)^2\over \tau^2}
, \\ 
\Omega_{13}^2\simeq  & 0,
 & \Omega_{22}^2 \simeq  \frac{2  \left(2 \gamma +M^2+2\right)}{\epsilon  \left(\gamma +M^2\right)} {(1+\epsilon)^2\over \tau^2} ,\\ 
 \Omega_{23}^2\simeq & 0,
   & \Omega_{33}^2\simeq  2 {(1+\epsilon)^2\over \tau^2}.
 \end{array}
\end{align}

\subsubsection*{Gauge-flation}

\begin{figure}[t!]
\centering
\includegraphics[width=\textwidth]{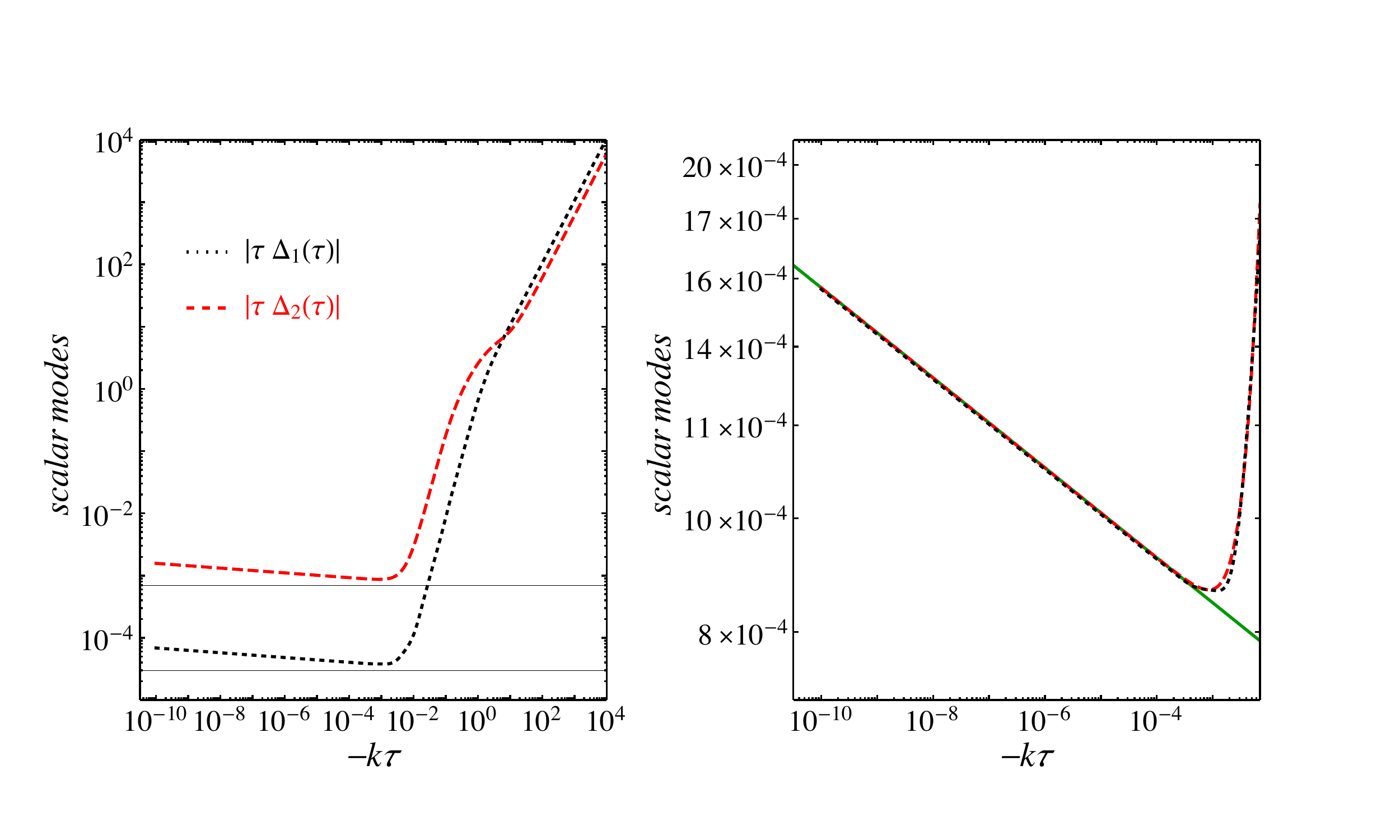}
\caption{
Left: Evolution of scalar fluctuations in Gauge-flation for $\epsilon=0.01$ and $\gamma=5$. 
Right: A zoomed region of the left panel with the modes rescaled according to eq.\ \eqref{eq:scalarmodeslateEigenvectorsM0}. The solid green line corresponds to the analytically calculated growth rate for these parameters. The black-dotted curve has been vertically shifted using the eigenvector given in eq.\ \eqref{eq:scalarmodeslateEigenvectorsM0}, in order to align with the red curve at late times and fit in a single panel.
}
\label{fig:ScalarModesLateM0}
\end{figure}
We study the $M=0$ case---regular Gauge-flation---separately, in order to separate the superhorizon behavior that arises due to the Higgs field from the generic superhorizon evolution. For $M=0$ the $3\times 3$ system of equations becomes $2\times 2$, since the Higgs degree of freedom becomes trivial and decoupled at all times. 

We look for solutions of the form
\begin{align} 
\vec \Delta^{(2)} \sim \vec \Delta^{(2,0)}(-k\tau)^n,
\end{align}
where $\vec \Delta^{(2)} = \{\Delta_1, \Delta_2\}$, and $\vec \Delta^{(2,0)}$ is a constant two-vector. At late times, the eigenvalues, $n$, are
\begin{align}\label{eqn:powers}
n = \left \{  {1\over 2} \left (1-i \sqrt{7+8\gamma} \right )   \pm i \epsilon{ 5+2\gamma \over (1+\gamma) \sqrt{7+8\gamma}}
,\,  
-1 - \left ( 4- {1\over 1+\gamma} \right )\epsilon 
,\,
 2+ \left (4 - {1\over 1+\gamma} \right )\epsilon
\right \}.
\end{align}
The growing mode has $n = -1 - \left ( 4- {1\over 1+\gamma} \right )\epsilon $, leading to the eigenvector
\begin{align}
 \vec \Delta^{(2,0)}  \sim
  \[\begin{matrix} {5+2\gamma \over 2 \sqrt{1+\gamma}(2+\gamma)} \sqrt{ \epsilon} \\ 1\end{matrix}\].
  \label{eq:scalarmodeslateEigenvectorsM0}
 \end{align}
In figure \ref{fig:ScalarModesLateM0} we show the late-time ($k\ll aH$) evolution of the scalar fluctuations. Shown is the numerical solution, as well as the power-law solution from eqs.\ \eqref{eqn:powers} and \eqref{eq:scalarmodeslateEigenvectorsM0} -- our analysis accurately captures both the growth rate and the eigenvectors.

\subsubsection*{Higgsed Gauge-flation}
\begin{figure}[t!]
\centering
\includegraphics[width=\textwidth]{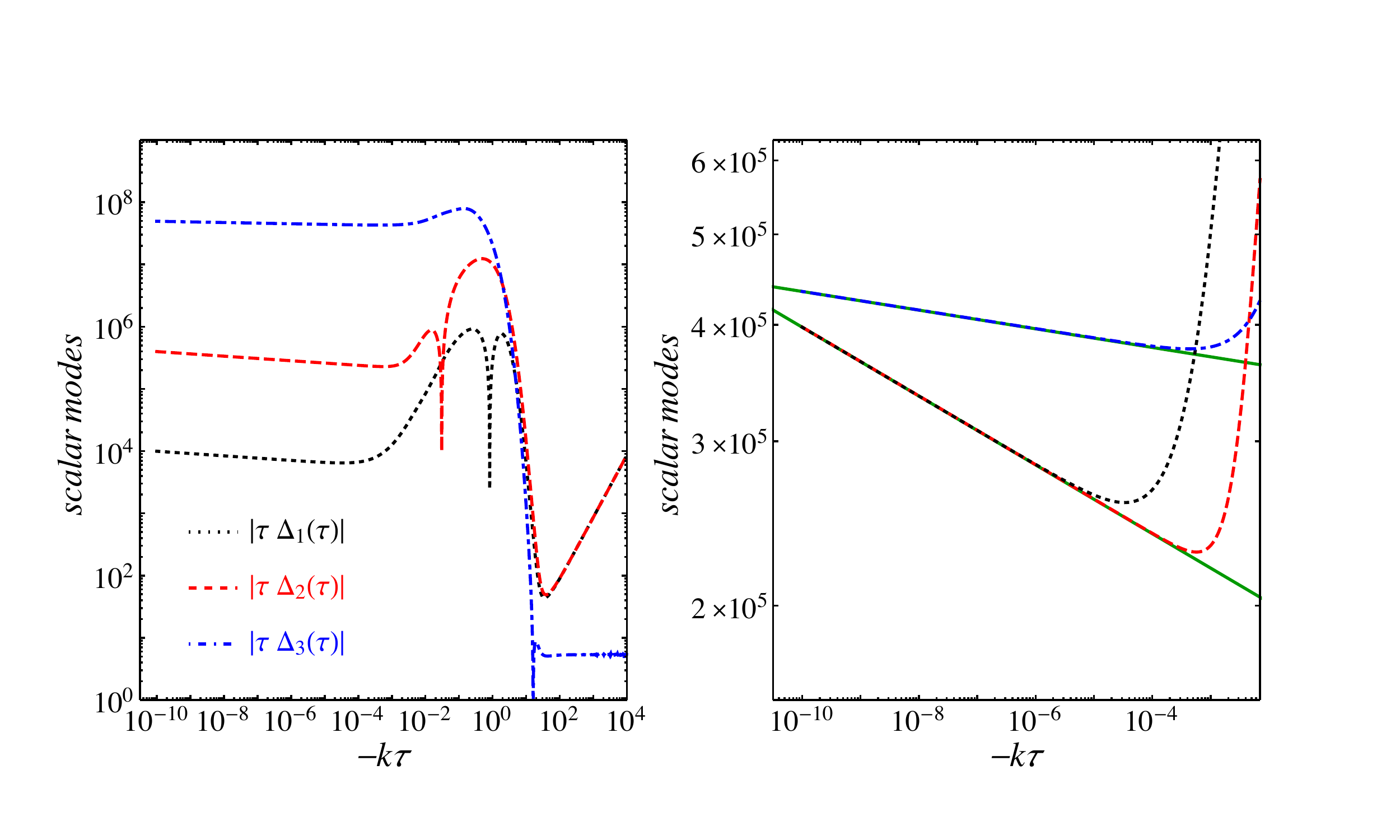}
\caption{
Left: Evolution of scalar fluctuations in Higgsed Gauge-flation for $\epsilon=0.01$, $\gamma=3$ and $M= 5$. 
Right: A zoomed region of the left panel with the modes rescaled according to eq.\ \eqref{eq:scalarmodeslateEigenvectors}. The green solid line corresponds to the analytically calculated growth rate for these parameters. The black-dotted and blue-dot-dashed curves have been vertically shifted, in order to align at late times  with the red curve and fit in a single panel.
}\label{fig:ScalarModesLate}
\end{figure}

We now move to the full Higgsed Gauge-flation case of $M\ne 0$. In this case the Higgs fluctuations are coupled to the other two modes. However, in the superhorizon limit $k/aH \to 0$ the Higgs fluctuation decouples regardless of the mass $M$, splitting the $3\times 3$ system into a $2\times 2$ and $1\times 1$ system. Setting $\xi = \xi_0 (-k\tau)^{n_\xi}$ at late times results in 
\begin{align}
n_\xi = \{ -1-\epsilon , 2+\epsilon \} + {\cal O}\(\epsilon^2\).
\end{align}

The remaining $2\times 2$ system for $\{\Delta_1, \Delta_2\}$ is solved as before, using an ansatz for the solution that scales as $(-k\tau )^n$. There are four solutions, the three non-growing modes are
\begin{align}
n = \left \{2+{\cal O}(\epsilon)  , \, {1\over 2} \pm {i\over 2} \sqrt{1+8\gamma + 8M^2} + {\cal O}(\epsilon )
\right \},
\end{align}
while the late-time growing mode has the exponent
\begin{align}
n=-1 -\epsilon \left ( 4 - {1+M^2/2\over 1+ \gamma +M^2/2 } \right ) ,
\end{align}
with the corresponding eigenvector
\begin{align}
 \vec \Delta^{(2,0)}  \sim
  \[\begin{matrix} 
  {5+2\gamma +2M^2  \over 2 \sqrt{1+\gamma+M^2/2}(2+\gamma+M^2)}     \sqrt{\epsilon} \\ 1\end{matrix}\] .
    \label{eq:scalarmodeslateEigenvectors}
 \end{align}
The excellent agreement of our analytical results with the numerical solution of the corresponding equations is shown in figure \ref{fig:ScalarModesLate}.

It is worth commenting on the superhorizon evolution of the scalar power spectrum ${P}_\zeta$, as defined in appendix \ref{app:Pz}. For the purposes of this section, the exact expression of ${P}_\zeta$ is unimportant and we only consider its parametric dependence for superhorizon modes.
To first order in slow-roll the Hubble parameter evolves as
\begin{align}
H \approx  H_* \left|\frac{\tau}{\tau_*}\right|^\epsilon \, ,
\end{align}
where we take $H_*$ to be the Hubble scale at horizon crossing of a particular mode with comoving wavenumber, $k_*$, i.e.\ $-k_*\tau_* = 1$. Furthermore the effective mass parameters $\gamma$ and $M$ also flow with time. In particular
\begin{align}
M = {g Z_0 \over H } \approx M_* \left|\frac{\tau}{\tau_*}\right|^{-\epsilon}  \, ,
\end{align}
where $M_*$ is the value at horizon crossing. The case of $\gamma$ is in principle more complicated, since both $\psi$ and $H$ evolve in time. However, $\psi = \psi_* |\tau/\tau_*|^\delta$ and since $\delta = {\cal O}\(\epsilon^2\)$ we can regard it as a constant, leading to
\begin{align}\label{eqn:gammainconf}
\gamma = {g^2 \psi^2 \over H^2 } \approx \gamma_* \left|\frac{\tau}{\tau_*}\right|^{-2 \epsilon_*}  \, .
\end{align}
Furthermore, $\eta = {\cal O}(\epsilon)$, which implies that the flow of $\epsilon$ must also be taken into account. To lowest order
\begin{align}\label{eqn:epsinconf}
\epsilon \approx \epsilon_*\left|\frac{\tau}{\tau_*}\right|^{-2\epsilon_* {\gamma +M^2/2 \over 1+\gamma +M^2/2} }.
\end{align}
Numerical evaluation of $\gamma$ and $\epsilon$ show very good agreement with eq.\ \eqref{eqn:gammainconf} and \eqref{eqn:epsinconf}, respectively.

The late-time power spectrum is dominated by the contribution of $\M(\tau)$, therefore we consider only the late-time behavior of its pre-factor. Following the notation of appendix \ref{app:Pz}, the asymptotic behavior is
\begin{align}
\left | (\vec{c}\cdot {\bf U} +\vec{d}\cdot {\bf U }')_2\right |_{k\tau \ll 1} \sim \left|\frac{\tau}{\tau_*}\right| \frac{\sqrt{\frac{2}{3}} \left(2 \gamma+M^2+2\right)}{\epsilon ^2 \left(\gamma+M^2\right)} \sim \left|\frac{\tau}{\tau_*}\right| \epsilon^{-2} \sim\left|\frac{\tau}{\tau_*}\right|^{{1+4\epsilon_* {\gamma +M^2/2 \over 1+\gamma +M^2/2} }} \, ,
\end{align}
where the second-to-last equality holds for $\gamma+ M^2/2\gg 1$. This approximation shows good to excellent agreement with numerical data for all tested values of $\gamma$ and $M$.

\begin{figure}[t!]
\centering
\includegraphics[width=\textwidth]{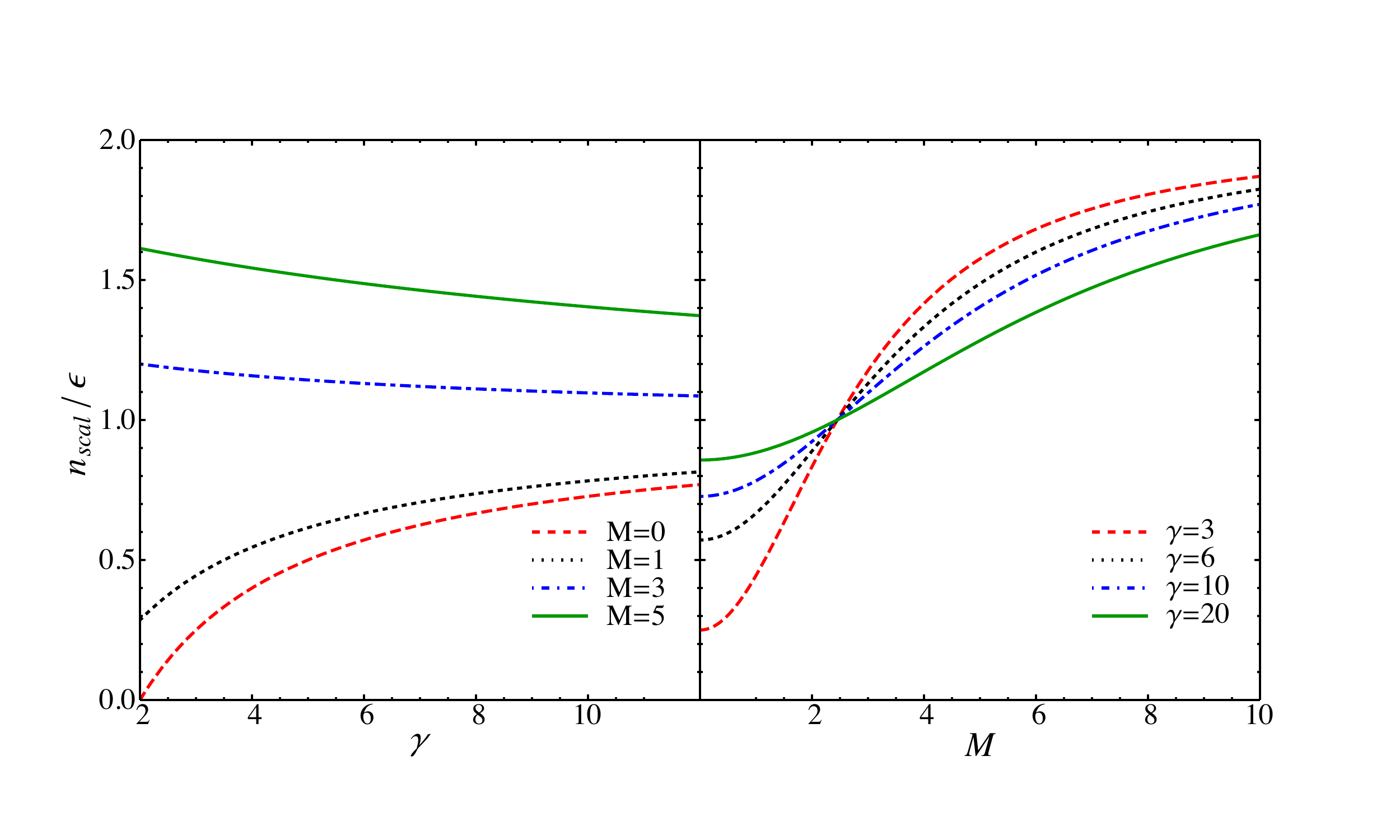}
\caption{
The late-time decay rate of the scalar power spectrum for different values of $\gamma$ and $M$.
}\label{fig:latetimegrowthratescalar}
\end{figure}

The parametric time-dependence of the superhorizon scalar power spectrum is 
\begin{align}
\sqrt{{P}_\zeta} \sim & H \left | (\vec{c}\cdot {\bf U} +\vec{d}\cdot {\bf U }')_2\right | |\M(\tau)| \sim \left|\frac{\tau}{\tau_*}\right|^{n_{\rm scal}}\( \frac{k}{k_*}\)^ {-\epsilon \left(4-{1+M^2/2\over 1+\gamma +M^2/2} \right)},
\end{align}
where
\begin{align}
n_{\rm scal}= & \epsilon \left ( 1 + \frac{M^2-6}{2 \gamma +M^2+2} \right ).
\label{eq:Pz_latetime}
\end{align}
Since $n_{\rm scal}  > 0$  for all values of $\gamma>2$ and $M>0$, the scalar power spectrum (slowly) decays outside the horizon, $|k\tau|\to 0$.
This is different to the case of single-field slow-roll inflation, where the time evolution of the prefactor exactly cancels the time evolution of thee superhorizon field fluctuation , so that the scalar power spectrum is exactly constant at late times. Alternatively, the evolution of the curvature perturbation in Gauge-flation indicates the presence of an isocurvature mode, which is absent in the standard single-field scenario. The late-time decay rate of the scalar power spectrum is plotted in figure \ref{fig:latetimegrowthratescalar}. The consequence of this decay is that the tensor-to-scalar ratio increases during inflation. Hence, we find that the disagreement of Gauge-flation with Planck data is  worse than originally computed in ref.\ \cite{Namba:2013kia}. 

 It is worth noting that the form of the scalar power spectrum given in eq.\ \eqref{eq:Pz_latetime} is missing a wavenumber-dependent prefactor. This prefactor arises from the different amount of enhancement that each mode undergoes due to the evolution of the background; each mode sees a slightly different background. Therefore, an estimate of the spectrum tilt $n_s$ cannot be read-off immediately from this expression, as it can be in the case of simple single-field inflation, where $n_s =1-2\epsilon-\eta$.

\subsection{Vector fluctuations}\label{sec:vectors}

We turn now to the vector fluctuations. Working in fourier space we expand the transverse vector modes into helicity states as
\begin{align}
V_i({\bf x}, \tau) = \sum_{\lambda = \pm}\int \frac{\d^3 k}{(2\pi)^3}V_k^{\lambda}(\tau)\epsilon_i^{\lambda}({\bf k})e^{i{\bf k}\cdot{\bf x}}+{\rm c.c.},
\end{align}
where $V_i$ is any of the modes $N_V^i, Y_i, \M_i$ or $\xi_i$ from above. We have also introduced  the helicity vectors
$\vec{\epsilon}\,{}^{(\pm)} ({\bf k})$ which satisfy the relations ${\bf k}\cdot\vec{\epsilon}\,{}^{(\pm)} ({\bf k}) = 0$, $i{\bf k}\times \vec{\epsilon}\,{}^{(\pm)} ({\bf k}) = \pm k \vec{\epsilon}\,{}^{(\pm)} ({\bf k})$, $\vec{\epsilon}\,{}^{(\pm)} ({\bf k})\cdot \vec{\epsilon}\,{}^{(\pm)} ({\bf k}) = 0$, and $\vec{\epsilon}\,{}^{(\pm)} ({\bf k})\cdot \vec{\epsilon}\,{}^{(\mp)} ({\bf k}) = 1$.

After introducing these modes, we find that the action splits into two non-interacting pieces corresponding to the positive and negative helicity states. Explicitly, the vector action reads
\begin{align}
\delta^2S = \int \frac{\d^3 k}{(2\pi)^3} \d\tau\mathcal{L}_{\rm vector},
\end{align}
where the quadratic Lagrangian density for the vector modes is
\begin{align}\nn
 \mathcal{L}_{\rm vector} = & \frac{a^2 k^2}{4}\M^{\pm}{}' \bar{\M}^{\pm}{}'+a^2H^2 M^2 \xi^{\pm}{}'\bar{\xi}^{\pm}{}'   - \frac{1}{4}a^4 H^2 k^2(2+M^2+\gamma)\M^{\pm}\bar{\M}^{\pm}\\ \nn&-a^2 H^2 M^2 k(k \mp aH\sqrt{\gamma})\xi^{\pm}\bar{\xi}^{\pm}  \pm \frac{ a^3  H^2 k^2 M^2}{2}\( \M^{\pm}\bar{\xi}^{\pm}+\xi^{\pm}\bar{\M}^{\pm}\)\\ \nn
 & +\frac{1}{4}a^3 H (\pm k \sqrt{\gamma} - a H (M^2 + 2\gamma)) \psi (N^{\pm} \bar{Y}^{\pm}+\bar{N}^{\pm} {Y}^{\pm})\\\nn
 &+ \frac{1}{4}a^2 (k^2 \mp 2 a H k \sqrt{\gamma} + a^2 H^2 (M^2 + 2 \gamma)) Y^{\pm}\bar{Y}^{\pm}-  \frac{1}{2}a^3 H^2 M^2 (\xi^{\pm}{}'\bar{Y}^{\pm}+\bar\xi^{\pm}{}'{Y}^{\pm}) \\\nn & - \frac{1}{4}a^2 k (k\mp a H \sqrt{\gamma})( \M^{\pm}{}' \bar{Y}^{\pm}+ \bar\M^{\pm}{}' {Y}^{\pm})-    \frac{1}{4}a^3 H k ( k \mp a H \sqrt{\gamma}\delta) (\M^{\pm}{} \bar{Y}^{\pm} +\bar\M^{\pm}{} {Y}^{\pm} )\\\nn
& +\frac{ a^2}{8}M_{\rm Pl}^2  \(k^2 + 2 a^2 H^2 (M^2 + 2 \gamma ) \frac{\psi^2}{M_{\rm Pl}^2 }\) N^\pm\bar{N}^{\pm} + \frac{ a^3}{2}   H^2\psi M^2 (\xi^{\pm}{}'\bar{N}^{\pm}+ \bar\xi^{\pm}{}'{N}^{\pm})\\
& \mp \frac{ a^3}{4}  k H\psi \sqrt{\gamma}(a H\delta\M^{\pm}+ 
         \M^{\pm}{}')\bar{N}^{\pm} \mp \frac{ a^3}{4}  k H\psi \sqrt{\gamma}(a H\delta\bar\M^{\pm}+ \bar \M^{\pm}{}'){N}^{\pm}.
\end{align}
Note that $Y^{\pm}$ and $N^{\pm}$ appear in the action without time derivatives, and are thus algebraic constraints. We can thus solve their equations of motion, and insert the solutions back into the action. To leading order in slow roll, the shift vector is given by
\begin{align}\label{eqn:shiftsol}
N^{\pm} = 2 a H^2\frac{\psi}{M_{\rm Pl}^2}\frac{ (a k (  a H (M^2 + 2\gamma)\mp k \sqrt{\gamma} ) \M^{\pm} +  a k (M^2 + \gamma) \M^{\pm}{}' - 2 M^2 (k \mp a H  \sqrt{\gamma}) \xi^{\pm}{}')}{ k \(k^2 \mp 2 a H k \sqrt{\gamma} +  a^2 H^2 \(M^2 + 2 \gamma \)\)},
\end{align}
while $Y^{\pm}$ is 
\begin{align}
Y^{\pm} = & \frac{a H  \(k^2 \mp a H k \sqrt{\gamma}\delta + 
      2 a^2 H^2 \(M^2 + 2\gamma\)\frac{\psi^2}{M_{\rm Pl}^2}\) \M^{\pm} }{(k^2 + 2 a H k \sqrt{\gamma}+ 
     a^2 H^2 (M^2 + 2\gamma))}\\ \nn& + \frac{k \(k (k \mp a H  \sqrt{\gamma}) + 
      2 a^2 H^2 (M^2 +\gamma)\frac{ \psi^2}{M_{\rm Pl}^2}\) \M^{\pm}{}' + 
   2 a H^2 M^2 \(k \pm 2 a H \sqrt{\gamma}\frac{\psi^2}{M_{\rm Pl}^2}\) \xi^{\pm}{}'}{k (k^2 \mp 2 a H k \sqrt{\gamma}+ 
     a^2 H^2 (M^2 + 2\gamma ))}.
\end{align}
After substituting back, the resulting action is complicated and not particularly enlightening; we do not reproduce it explicitly here.  

Denoting by $\vec{V}^{\pm} = (\M^{\pm}, \xi^{\pm})$, we redefine the field using the transformation $V^{\pm}_i = R^{\pm}_{ij} W_{j}$, where ${\bf R}$ is the matrix\footnote{Analogously to the scalar case, this transformation---which diagonalizes the kinetic matrix at early times--- is found by neglecting the gravitational constraints.}
\begin{align}
{\bf R} = \(\begin{matrix}\frac{ \sqrt{2k^2 \mp 4 aH k \sqrt{\gamma}+4 a^2 H^2 \gamma}}{a^2 H k \sqrt{\gamma}}& 0\\ \frac{k\mp a H \sqrt{\gamma}}{a H \sqrt{\gamma}\sqrt{2k^2 \mp 4 aH k \sqrt{\gamma}+4 a^2 H^2 \gamma}} & \frac{\sqrt{k^2 \mp  aH k \sqrt{\gamma}+ a^2 H^2(2 \gamma+M^2)}}{aH M \sqrt{2}\sqrt{k^2 \mp  aH k \sqrt{\gamma}+2 a^2 H^2 \gamma}}\end{matrix}\).
\end{align}
After making this transformation, the action takes the form
\begin{align}
\delta^{2}S = \int \frac{\d^3 k}{(2\pi)^3}d\tau \[ {\vec{W}}^{\pm}{}^{\dagger}{}'{\bf T}_{\pm}{\vec{W}}^{\pm}{}' + {\vec{W}}^{\pm}{}^{\dagger}{}'{\bf K}_\pm{\vec{W}}^{\pm} - {\vec{W}}^{\pm}{}^{\dagger}{\bf K}_{\pm}{\vec{W}}^{\pm}{}'  - {\vec{W}}^{\pm}{}^{\dagger}{\bf \Omega}^2_{\pm}{\vec{W}}^{\pm} \].
\end{align}
Again, it is possible to choose a transformation that sets the matrix ${\bf T}_{\pm}$ to the identity for all times, however, this requires a more complicated transformation which significantly complicates the algebra. As discussed above, for our purposes this is not required. All that is required is that ${\bf T}_{\pm}$ approaches the identity for $k \gg aH$, so that we can set the initial conditions, and that ${\bf T}_{\pm}$ is invertible for all times, so that we may smoothly evolve the equations of motion. 

The full matrices ${\bf T}_{\pm}$, ${\bf K}_{\pm}$, and ${\bf \Omega}^2_{\pm}$ are obtained in a straightforward manner, their exact forms are messy and not particularly illuminating. In appendix \ref{app:vecsector} we present slow-roll expansions of these matrices that are valid at all scales. For use in the subsequent sections, we present expansions of the matrices in the limits $k \gg aH$ and $k \ll aH$.

{
On subhorizon scales, $k\gg aH$, the symmetric ${\bf T}_{\pm}$, and antisymmetric ${\bf K}_{\pm}$  matrices are
\begin{align}\label{eqn:vectrans}
{\bf T}_{\pm} = \mathds{1}+ \frac{a^2 H^2}{k^2}\(\begin{matrix} {-2 \gamma \epsilon\over 1+\gamma+M^2/2} & {2 \sqrt{\gamma}M \epsilon\over 1+\gamma+M^2/2}  \\ {2 \sqrt{\gamma}M\epsilon\over 1+\gamma+M^2/2} & {-2 M^2\epsilon\over 1+\gamma+M^2/2}\end{matrix}\), \quad {\bf K}_{\pm} = \frac{a^2 H^2}{k} \(\begin{matrix} 0  & {M\epsilon\over 1+\gamma+M^2/2}  \\ {-M \epsilon\over 1+\gamma+M^2/2} & 0 \end{matrix}\),
\end{align}
and the symmetric ${\bf \Omega}^{2}_{\pm} $ matrix is
\begin{align}
{\bf \Omega}^{2}_{\pm} = \(\begin{matrix} k^2\(1-\frac{1}{\gamma}\) & 0 \\  0 & k^2 \end{matrix}\)\mp aH k \(\begin{matrix} \frac{1+2\gamma+M^2}{\sqrt\gamma} &  0 \\  0  & \sqrt{\gamma}  \end{matrix}\)+ a^2 H^2 \(\begin{matrix} 2\gamma + M^2& -\frac{M(1+2\gamma)}{2\sqrt{\gamma}} \\   -\frac{M(1+2\gamma)}{2\sqrt{\gamma}}  & M^2 - 2 \end{matrix}\).
\end{align}
On superhorizon scales, these matrices are
\begin{align}\label{eqn:superHT}
{\bf T}_{\pm} = & \mathds{1} -{\epsilon\over 1+\gamma+M^2/2} \(\begin{matrix}1 & \frac{M }{ \sqrt{2 \gamma +M^2}}  \\  \frac{M }{ \sqrt{2 \gamma +M^2}} & \frac{M^2}{ \left(2 \gamma +M^2\right)} \end{matrix}\),\\\label{eqn:superHK}
{\bf K}_{\pm} = & a H\frac{ M \epsilon}{(1+\gamma+M^2/2)\sqrt{M^2+2 \gamma }} \(\begin{matrix}  0 & 1  \\ -1 & 0 \end{matrix}\),
\end{align}
and
\begin{align}\label{eqn:superHO}
{\bf \Omega}_{\pm}^2 = a^2 H^2 \(\begin{matrix}  \left(\gamma +M^2+1\right) &  0 \\ 0  & -2 \end{matrix}\)+\epsilon a^2 H^2\(\begin{matrix}   \left(-\frac{8}{2 \gamma +M^2+2}-1\right) & -\frac{2  M  }{\sqrt{2
   \gamma +M^2} \left(2 \gamma +M^2+2\right)}   \\   -\frac{2  M  }{\sqrt{2
   \gamma +M^2} \left(2 \gamma +M^2+2\right)}  &  \frac{  \left(4 \gamma ^2+M^4+4 \gamma 
   \left(M^2+1\right)+6 M^2\right)}{\left(2 \gamma +M^2\right) \left(2 \gamma +M^2+2\right)} \end{matrix}\).
\end{align}
}

\subsubsection{Initial conditions and quantization}

Following the procedure outlined above in section \ref{sec:ICandquant} for the scalar modes, we quantize the vector modes. We begin by expanding the fluctuations into modes
\begin{align}\label{eqn:modeexpvec}
\vec{W}^{\pm}_i(\tau, {\bf k})  = \mathcal{W}_{ij}( \tau, k)a_j({\bf k})+  \mathcal{W}^*_{ij}(\tau, k)a^\dagger_j(-{\bf k}) , \quad \[a_i({\bf k}), a^{\dagger}_j({\bf k'})\] = \delta^{3}({\bf k} - {\bf k}')\delta_{ij}.
\end{align}
Working in the limit $k \gg a H$, and using the expressions for the matrices above, the equations of motion for the fluctuations become
\begin{align}
{\mathcal{W}}^{\pm}{}''+\beta^{\pm}{\mathcal{W}}^{\pm} = 0,
\end{align}
where in the limit $k \gg a H$, the matrix $\beta^{\pm}$ is given by
\begin{align}
\ \beta^{\pm} =k^2 \(\begin{matrix} 1-\frac{1}{\gamma} &0 \\ 0 &1 \end{matrix}\).
\end{align}
Adopting a WKB ansatz for the mode functions
\begin{align}
\vec{\mathcal{W}}_{j} = \vec{b}_j \exp\[{i\int dx\, \omega(x)}\],
\end{align}
and substituting into the system of equations, neglecting terms of order $\mathcal{O}(\omega'/\omega)$ and $\mathcal{O}(\omega''/\omega)$, we find four solutions for the frequencies
\begin{align}\label{eq:WKBfreqsvec}
\omega \approx \left\{ \pm 1, \pm \frac{\sqrt{\gamma - 1}}{\sqrt{\gamma}}\right\}.
\end{align}
In order for the system to be stable, all of these instantaneous WKB frequencies must be real. Therefore, there is an instability in the system for parameters such that  $\gamma = g^2\psi^2/H^2 < 1$. However, since the scalar sector requires $\gamma > 2$, this does not present a further restriction on the model.

The corresponding mode solutions are, up to an irrelevant phase,
\begin{align}\nn
\vec{\mathcal{Q}}_{j} = & c_{1j}\vec{a}_{1}e^{i k\tau}+c_{2j} \vec{a}_{2}e^{-ik\tau} +c_{3j} \vec{a}_{3}e^{i \frac{\sqrt{\gamma - 1}}{\sqrt{\gamma}} \tau}+c_{4j} \vec{a}_{4}e^{-i \frac{\sqrt{\gamma - 1}}{\sqrt{\gamma}} \tau},
\end{align}
where   the $c_{ij}$ are constants and the $\vec{a}_{i}$ are the vectors
\begin{align}
\vec{a}_{1} = \vec{a}_2 = &
\left[
\begin{array}{ccc}
1 \\
0   
\end{array}
\right],
\quad
\vec{a}_{3} = \vec{a}_4^* = 
\left[
\begin{array}{ccc}
0   \\
1  
\end{array}
\right].
\end{align}

\begin{figure}[t!]
\centering
\includegraphics[width=\textwidth]{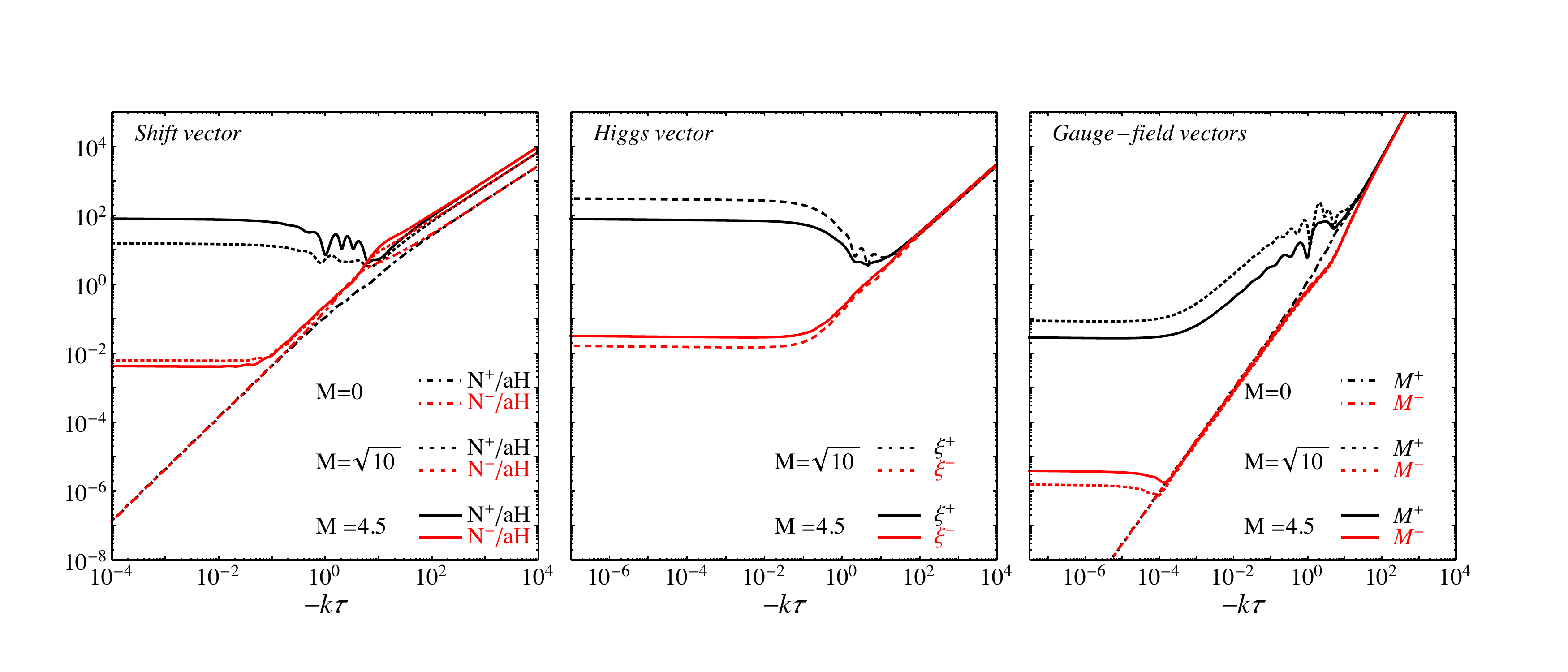}
\caption{
Evolution of vector fluctuations in Higgsed Gauge-flation. Note that in the limit $M\to 0$, the vector modes all decay on superhorizon scales. The values of the other parameters here are chosen to be $\epsilon = 0.01$ and $\gamma = \sqrt{10}$. }
\label{fig:vectormodes}
\end{figure}

Demanding the solutions approach the positive frequency solutions as $x = -k\tau \to \infty$ sets $c_{1j} = c_{3j} =0$. The remaining constants are set by imposing the quantization conditions above. Working in the limit $-k\tau\to\infty$, it is then straightforward to see that a solution that satisfies the initial conditions is
\begin{align}\label{eqn:initialconditionsvecs}
  \text{ Gauge mode:} & \quad c_{21} =  0, \quad c_{22} = \frac{1}{\sqrt{2k}},\\
 \text{ Goldstone mode:} & \quad c_{42}  = 0,\quad c_{41} = \frac{1}{\sqrt{2k\sqrt{\(1-\frac{1}{\gamma}\)}}}.
\end{align}
In figure \ref{fig:vectormodes}, we show the evolution of the Higgs vectors and gauge field vectors, as well the evolution of the shift vector.

\subsubsection{Superhorizon solutions}

We can solve the equations of motion for the vector modes in the superhorizon limit by using the ansatz
\begin{align}
\vec{\mathcal{W}}^\pm_k(\tau)  = \vec{\mathcal{W}}_0 ^\pm(-k\tau)^{n}.
\end{align}
Using the asymptotic matrices in eqs.\ \eqref{eqn:superHT}-\eqref{eqn:superHO},  to first order in $\epsilon$, we find
\begin{align}
n=\left\{ -1-\epsilon, 2+\epsilon, \left ( {1\over 2} \pm {i\over 2}  \sqrt{3+4\gamma + 4M^2} \right ) \pm  i \frac{   \left(4 (\gamma  (\gamma +2)-1)+2 M^4+(6 \gamma +7) M^2\right)}{\left(2 \gamma +M^2+2\right) \sqrt{4 \gamma +4 M^2+3}} \epsilon\right\}.
\end{align}
The growing mode has $n=-1-\epsilon$, and  corresponding eigenvector
\begin{align}
\vec{\mathcal{W}}_0 ^\pm  \sim \(\begin{matrix} -\frac{2 \epsilon ^3 \left(M \left(4 \gamma ^2+4 \gamma +M^4+4 \gamma  M^2+4 M^2\right)\right)}{\sqrt{2 \gamma +M^2} \left(2 \gamma +M^2+2\right)^2 \left(2 \gamma ^2+6 \gamma +M^4+3 \gamma  M^2+3 M^2\right)} \\  1 \end{matrix}\).
\label{eq:vectorgrowingmode}
\end{align}
In the limit $k \ll aH$ 
\begin{align}\label{eqn:shiftsol}
\frac{N^{\pm}}{aH} \sim  2 \frac{\sqrt{2\epsilon}}{M_{\rm Pl} \sqrt{2+2\gamma+M^2}}\(   \M^{\pm} +    \frac{(M^2 + \gamma)}{(M^2 + 2 \gamma)}\frac{1}{aH} \M^{\pm}{}' \pm \frac{2 M^2\sqrt{\gamma}}{M^2 + 2 \gamma}\frac{1}{ak}  \xi^{\pm}{}'\),
\end{align}
and inserting the late-time solution of eq.\ \eqref{eq:vectorgrowingmode} into eq.\ \eqref{eqn:shiftsol} while keeping the lowest order term in $1/a$ we arrive at the solution
\begin{align}
\frac{N^{\pm}}{aH} \sim {\epsilon^3 \sqrt\epsilon \over ak} |k\tau|^{-1-\epsilon} \sim \epsilon^3 \sqrt\epsilon \, ,
\end{align}
where we used the fact that the scale-factor grows like $a\sim |\tau/\tau_*|^{-1-\epsilon}$ during inflation.\footnote{We note here that in this model, we find that the vector part of $g_{0i}$ metric perturbation grows exponentially during the inflationary phase. While vector modes typically decay following inflation,  this  exponentially growing shift potentially invalidates the FRW background during inflation. One can show that the scalar part of the shift vector also generates an exponentially growing $g_{0i}$ perturbation, even in the limit that the Higgs is absent, $M\to 0$.}

\subsection{Tensor fluctuations}\label{sec:tensors}

We now turn to the tensor degrees of freedom. The addition of the Goldstone modes in the Stueckelberg limit does not add any new degrees of freedom to the tensor sector.  At linear order in perturbation theory these fluctuations are gauge invariant under SU(2) transformations and coordinate transformations for the gauge field and metric fluctuations respectively. Furthermore, neither are subject to the Einstein, or Gauss law constraints at this order. 

We expand the tensor modes into a helicity basis in Fourier space
\begin{align}
\gamma_{ij}({\bf x}, \tau) = & \frac{\sqrt{2}}{M_{\rm Pl} a}\sum_{\lambda = \pm}\int \frac{\d^3 k}{(2\pi)^3}\Pi_{ij}^\lambda({\bf k}) \hat{\gamma}^{\lambda}_{\bf k}e^{i {\bf k}\cdot{x}}+\text{c.c.},\\
t_{ij}({\bf x}, \tau) = & \frac{1}{\sqrt{2} a}\sum_{\lambda = \pm}\int \frac{\d^3 k}{(2\pi)^3}\Pi_{ij}^\lambda({\bf k}) \hat{t}^{\lambda}_{\bf k}e^{i {\bf k}\cdot{x}}+\text{c.c.},
\end{align}
where the polarization tensors satisfy (see, e.g.\ ref.\ \cite{Namba:2015gja}) 
\begin{align}
 \Pi_{ij, \pm} ({\bf k}) \equiv \epsilon^{(\pm)}_i ({\bf k})\epsilon^{(\pm)}_j ({\bf k}),
\end{align} 
and $\vec{\epsilon}\,{}^{(\pm)} ({\bf k})$ are the helicity vectors from above. Inserting these decompositions into the action, it splits into two decoupled pieces corresponding to the left-helicity and right-helicity modes. Neglecting boundary terms, the action takes the form
\begin{align}
S_{\pm} = \int \frac{\d^3 k}{(2\pi)^3}d\tau \[\Delta_{\pm}^{\dagger}{}'\Delta_{\pm}' + \Delta_{\pm}^{\dagger}{}'{\bf K}\Delta_{\pm} -  \Delta_{\pm}^{\dagger}{}{\bf K}_{\pm}\Delta_{\pm}' -  \Delta_{\pm}^{\dagger}{}{\bf \Omega}^2_{\pm}\Delta_{\pm} \], \quad \Delta_{\pm} = \(\begin{matrix} \hat{\gamma}^{\pm}_{\bf k} \\ \hat{t}^\pm_{\bf k} \end{matrix}\),
\end{align}
where the antisymmetric matrix ${\bf K}_{\pm}$ has entries
\begin{align}
K_{\pm, 12} = \frac{1}{M_{\rm Pl}}\(\psi'+\frac{a'}{a}\psi\),
\end{align} 
and the symmetric ${\bf \Omega}^2_{\pm}$ matrix has entries
\begin{align}
\Omega^2_{\pm, 11} = & k^2 - \frac{a'{}^2}{a^2} + \frac{g^2 a^2 \psi^2(6\psi^2+5Z_0^2)}{2M_{\rm Pl}^2}-2\frac{(a\psi)'{}^2}{M_{\rm Pl}^2 a^2},\\\nn
\Omega^{2}_{\pm, 22} = & k^2 +a^2 g^2 Z_0^2 \pm k g a \psi\[2+\kappa \frac{g^2 a^4 (2\psi^2+Z_0^2)+2a'{}^2\psi^2-2a^2\psi'^2}{a^4(1+\kappa g^2 \psi^4)}\] \\\nn
& + \frac{ \kappa g^2 \psi^2}{a^2}\[\frac{g^2 a^4 \psi^2(2\psi^2+Z_0^2)+2a'{}^2\psi^2-2a^2\psi'{}^2}{(1+\kappa g^2 \psi^4)}\],\\\nn
\Omega^{2}_{\pm, 12} = &\mp \frac{2 g a \psi^2}{M_{\rm Pl}} - a^2 g^2 \psi Z_0^2+\frac{(a\psi)'}{a M_{\rm Pl}}\frac{a'}{a}-\frac{\kappa g^2 \psi^3}{M_{\rm Pl}}\[\frac{g^2 a^4\psi^2(2\psi^2+Z_0^2) +2a'{}^2\psi^2-2a^2\psi'{}^2}{(1+\kappa g^2\psi^4)}\].
\end{align}
In writing these expressions we have made no slow-roll approximation, however, we have made use of the background equations of motion. Note that we recover the results of ref.\ \cite{Namba:2013kia} in the limit that $Z_0 \to 0$, as expected. We now eliminate as many variables as possible.  Making use of the background equations of motion, and expanding to leading order in slow roll, we obtain
\begin{align}
K_{\pm, 12}  =  & aH\frac{\sqrt{\epsilon}}{\sqrt{1+\gamma+\frac{M^2}{2}}},\\
\Omega^2_{\pm, 11}= & k^2 -2 a^2 H^2 -a^2H^2\frac{\epsilon  \left(2-6  \gamma -5  M^2 \right)}{2 \gamma +M^2+2},\\
\Omega^2_{\pm, 22}= & k^2  \pm aH k\frac{2+4\gamma+M^2}{\sqrt{\gamma}} +2a^2 H^2(1+\gamma+M^2),\\
\Omega^2_{\pm,12} = & \mp 2aH k\sqrt{\frac{2 \gamma\epsilon}{2+2\gamma+M^2}}-a^2 H^2 \sqrt{\frac{2\epsilon}{2+2\gamma+M^2}}(1+2M^2+2\gamma).
\end{align}
Note that in the limit that $a \to 0$, $K_{\pm, 12} \to 0$ and $\Omega^2_{\pm} \to k^2 \mathds{1}$, which indicates that $\gamma^{\pm}$ and $t^{\pm}$ are the canonically normalized fields in the far past.\footnote{Note that the recent work of ref.\ \cite{Maleknejad:2016dve} quantized a different combination of $\gamma_{ij}$ and $t_{ij}$. In that work the kinetic term is not diagonal in the limit $k \gg aH$ due to couplings proportional to $\psi$.}

\begin{figure}[t!]
\centering
\includegraphics[width=\textwidth]{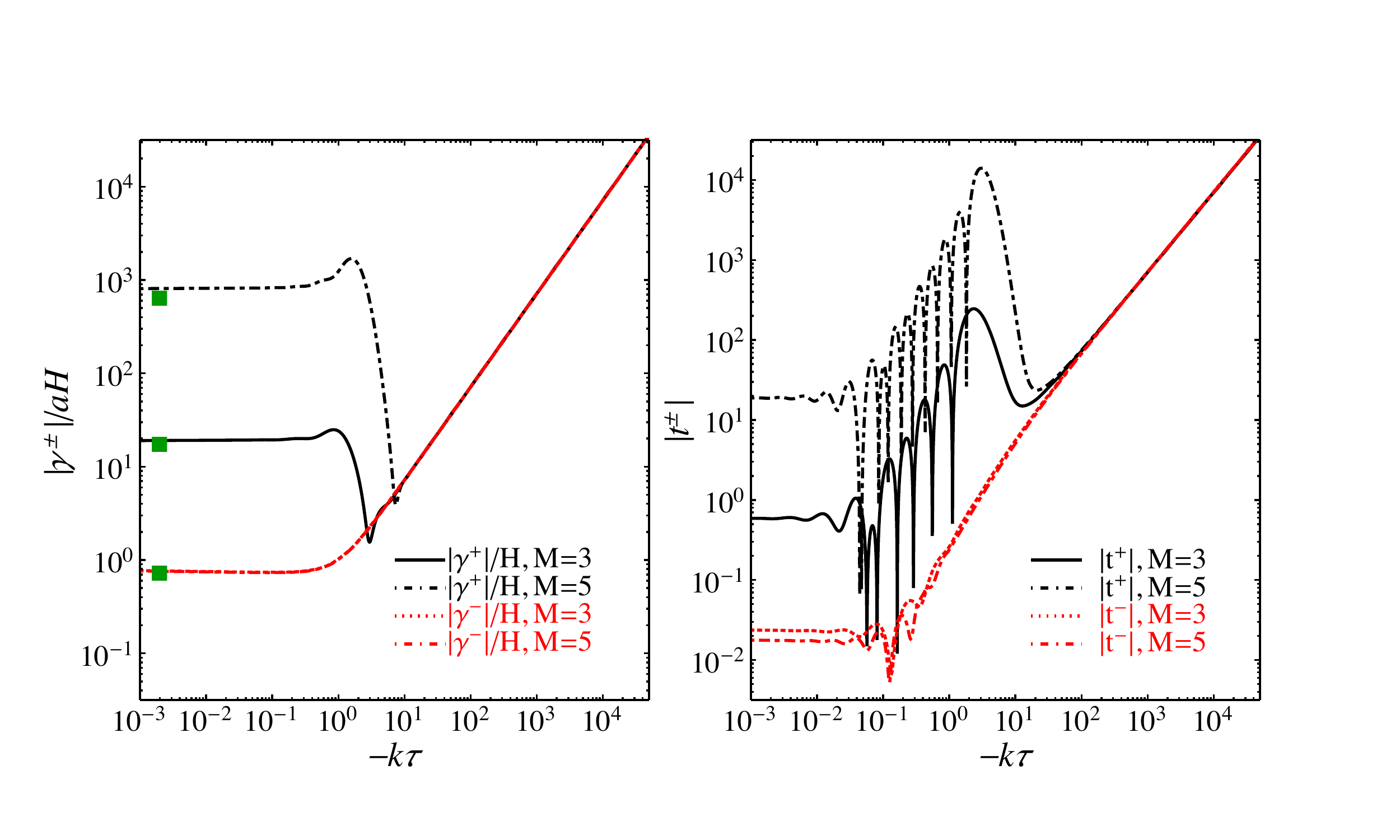}
\caption{
Evolution of tensor fluctuations in Higgsed Gauge-flation.  The parameters are chosen as $\epsilon=0.01$ and $\gamma=5$. 
The green squares show the approximate late-time result as computed in section \ref{sec:TensorApprox}. 
 }
\label{fig:tensormodes}
\end{figure}

The original model of Gauge-flation is ultimately ruled out, because for a range of $k/aH$ one of the helicities $\hat{t}^{\pm}$ experiences tachyonic growth due to its mass becoming negative. This leads to exponentially large gauge tensor modes, which in turn source large gravitational waves. Indeed, we see that the effective mass of $\hat{t}^-$ is negative during the range
\begin{align}
r_* - \Delta r < \frac{k}{a H} <  r_*+\Delta r, \text{ where } r_* \equiv \frac{2+4\gamma+M^2}{2\sqrt{\gamma}}, \quad \Delta r = \frac{\sqrt{\(1+\frac{M^2}{2}\)^2+2\gamma+2\gamma^2}}{\sqrt{\gamma}},
\end{align}
and thus, rather than making the tensors more stable, the effect of the mass in fact makes the instability worse. Examples of the amplification of the tensor modes are shown in figure \ref{fig:tensormodes}.
However, the additional scalar dynamics behave in such a way as to also boost the scalar spectrum, thus  \emph{reducing} the tensor-to-scalar ratio for a fixed value of the Hubble rate.

\subsubsection{Sub-horizon solutions and Born approximation}
\label{sec:TensorApprox}

The rather complicated looking system of equations that results can actually be solved analytically to an excellent approximation, as first pointed out in refs.\ \cite{Adshead:2013qp, Adshead:2013nka}. Note that the off-diagonal terms in  the ${\bf K}$ and ${\bf \Omega}^2$ matrices that couple the gauge field perturbations to the metric fluctuations are slow-roll suppressed by $\sqrt{\epsilon}$. This slow-roll suppression allows us to develop a series solution (in powers of $\sqrt{\epsilon}$) to the equations of motion using the Born approximation. Further, as in refs.\  \cite{Adshead:2013qp, Adshead:2013nka}, the evolution of the gauge modes is dominated by its mass term. The negative helicity mode of the gauge field tensor remains heavy throughout the evolution, and it can be neglected to a good approximation, leaving the negative helicity gravitational wave mode undeflected. We thus focus on the positive helicity modes $\hat{\gamma}^+$ and $\hat{t}^+$.

To leading order in slow roll, neglecting interactions with the gravitational wave sector, the equation of motion for the gauge field reads
\begin{align}\label{eqn:gaugetenfree}
\partial_\tau^2\hat{t}^++\(k^2 - \frac{k}{-\tau}\frac{2+4\gamma+M^2}{\sqrt{\gamma}}+\frac{2(1+\gamma+M^2)}{\tau^2}\)\hat{t}^+ = 0 .
\end{align}
Introducing the variable $z = 2 i k\tau$, and the parameters
\begin{align}
\kappa = \frac{2+4\gamma+M^2}{\sqrt{\gamma}}  = -2i\alpha, \quad \mu = & 2(1+\gamma+M^2) = {1\over 4}-\beta^2,
\end{align}
eq.\ \eqref{eqn:gaugetenfree} is transformed into the Whittaker equation
\begin{align}\label{eqn:gaugeten}
\partial_z^2\hat{t}^{+}+\(-\frac{1}{4}+\frac{\alpha}{z}+\frac{{1\over 4}  -\beta^2}{z^2}\)\hat{t}^{+} = 0.
\end{align}
Equation \eqref{eqn:gaugeten} is solved by the Whittaker functions,\footnote{Note that the basis of solutions to the Whittaker equation given by
\begin{align}\nn
\hat{t}_0^{+}(k, \tau) = A_k W_{\alpha, \beta}(2 i k\tau)+B_k W_{-\alpha, \beta}(-2 i k\tau),
\end{align}
leads to a much simpler set of coefficients $A_k, B_k$. However, we choose to work with the basis in \eqref{eqn:gaugetensol} to make contact with the work in ref.\ \cite{Adshead:2013qp, Adshead:2013nka}.}
\begin{align}\label{eqn:gaugetensol}
\hat{t}_0^{+}(k, \tau)  = &  A_k M_{\alpha, \beta}(2 i k\tau)+B_k W_{\alpha,\beta}(2 i k\tau),
\end{align}
where $M_{\alpha,\beta}(2 i x)$ and $W_{\alpha,\beta}(2 i x)$ are the Whittaker M and W functions. We set the values of the constants $A_k$ and $B_k$, by imposing the Bunch-Davies vacuum conditions in the asymptotic past. That is, we demand that the solutions approach canonically normalized positive frequency free plane waves as $x = -k\tau \to \infty$,
\begin{align}
\hat{t}_0^{+}(k, \tau)  \to \frac{1}{\sqrt{2 k}}e^{i x}.
\end{align}
In this large $x$ limit, the Whittaker functions have asymptotic expansions,
\begin{align}\nn
M_{\alpha,\beta}(2 i x) \to &\frac{(2 i)^{-\alpha }  \Gamma (2 \beta +1) }{\Gamma \left(-\alpha +\beta +\frac{1}{2}\right)}e^{i x - \alpha\ln x} +\frac{i 2^{\alpha } i^{\beta } (-i)^{\alpha -\beta } \Gamma (2
   \beta +1) }{\Gamma \left(\alpha +\beta +\frac{1}{2}\right)}e^{-i x+\alpha\ln x}, \\\label{eqn:asympexp}
W_{\alpha,\beta}(2 i x) \to & (2 i)^{\alpha } e^{-i x+ \alpha \ln x}.
\end{align} 
Using eq.\ \eqref{eqn:asympexp}, we find that the constants are given by
\begin{align}
A_k = & \frac{1}{\sqrt{2k}}\frac{\Gamma \left(-\alpha +\beta +\frac{1}{2}\right)}{(2 i)^{-\alpha }  \Gamma (2 \beta +1) },\\
B_k = &  \frac{1}{\sqrt{2k}}\frac{\Gamma \left(-\alpha +\beta +\frac{1}{2}\right)}{\Gamma \left(\alpha +\beta +\frac{1}{2}\right)}2^{\alpha } i^{\beta+1 } (-i)^{\alpha -\beta }.
\end{align}

To leading order in slow-roll, the positive-helicity tensor modes of the metric obey the equation of motion
\begin{align}\label{eqn:GWeqn}
\partial_x^2\hat{\gamma}^++\(1-\frac{2}{x^2} \)\hat{\gamma}^+ =\frac{\sqrt{\epsilon}}{ \sqrt{1+\gamma+\frac{M^2}{2}}}\(\frac{1}{\sqrt{1+\gamma+\frac{M^2}{2}}}\frac{2}{x}\partial_x-\(\frac{1+2\gamma+2M^2}{x^2}-\frac{\sqrt{\gamma}}{x}\)\)\hat{t}^+,
\end{align}
where $x = -k\tau$. Equation \eqref{eqn:GWeqn} can be solved as a series in $\sqrt{\epsilon}$ using the Born approximation. This solution consists of a homogeneous piece that solves the free equation of motion, and an inhomogeneous piece that is sourced by the gauge field fluctuations
\begin{align}
\hat{\gamma}^+ = \hat{\gamma}_0^+ + \hat{\gamma}^+_{\rm in}, \qquad
\hat{\gamma}^+_0 = \frac{1}{\sqrt{2k}}\(1+\frac{i}{x}\)e^{ix} = u_1(x).
\end{align}
Here, $\hat{\gamma}^+_0$ is the homogeneous solution that matches onto the Bunch-Davies vacuum as $x = -k\tau \to \infty$. To leading order in $\sqrt{\epsilon}$, the inhomogeneous part of the solution is
\begin{align}
\hat{\gamma}^+_{\rm in}(x) = \sqrt{\frac{{\epsilon}}{ {1+\gamma+\frac{M^2}{2}}}}\int^x\!\! dx' \!\!\(\frac{1}{\sqrt{1+\gamma+\frac{M^2}{2}}}\frac{2}{x'}\partial_{x'}-\(\frac{1+2\gamma+2M^2}{x'{}^2}-\frac{\sqrt{\gamma}}{x'}\)\)\!\!G(x, x')t^{+}_0 (x'),
\label{eq:greenintegral}
\end{align}
where
\begin{align}
G(x, x') = \frac{\Im\[u_1(x)u_1^*(x')\]}{W[u_1(x), u^*_1(x)]}\Theta(x-x'),
\end{align}
is the Green's function, and $W[\ldots]$ is the Wronskian.

We can proceed in exact accordance to the analysis found in refs.\ \cite{Adshead:2013nka,Adshead:2016omu} for the physically and mathematically related models of Chromo-Natural and Higgsed Chromo-Natural Inflation. The positive-helicity gravitational wave undergoes exponential amplification and the late-time solution is well-approximated by
\begin{align}
\nonumber
\gamma^+(x) &=  {H x \over M_{\rm Pl}\sqrt{k^3}} u_1(x) 
\\
&+ 2\sqrt 2 {H\over M_{\rm Pl} k} B_k {\sqrt{\epsilon}\over \sqrt{1+\gamma + M^2/2}} \left ( {2\over \sqrt{1+\gamma + M^2/2} }I_1  + \sqrt{\gamma} I_2 -\left (1+2\gamma + 2M^2 \right ) I_3   \right ) \, , 
\end{align}
where the solution is written as the sum of a free and a sourced part, arising from the interaction with the gauge field fluctuations. The functions $I_1, I_2, I_3$ arise from the three distinct integrals of eq.\ \eqref{eq:greenintegral}
\begin{align}\nn
I_1 = & \frac{\left(\mu^2-2 i \mu \kappa+2 \mu-2 \kappa^2\right) \sec \left( \pi \beta \right) \sinh \left(-i\pi\alpha\right) \Gamma \left(\alpha\right)}{2 \mu (\mu+2)} \\ \nn
&-\frac{\pi ^2 \left(\mu^2+2 i \mu \kappa+2 \mu-2 \kappa^2\right) \sec \left( \pi \beta \right) \text{csch}\left(-i\pi\alpha\right)}{2 \mu (\mu+2) \Gamma \left(\alpha+1\right) \Gamma \left(-\alpha-\beta+\frac{1}{2}\right) \Gamma \left(-\alpha+\beta+\frac{1}{2}\right)}~,\\
\nn
I_2 = & \frac{\pi  \sec \left(\pi \beta\right) \Gamma \left(-\alpha\right)}{2 \Gamma \left(-\alpha-\beta+\frac{1}{2}\right)
   \Gamma \left(-\alpha+\beta+\frac{1}{2}\right)}
   -\frac{\pi  \sec \left(\pi \beta\right) \Gamma \left(1-\alpha\right)}{\mu \Gamma \left(-\alpha-\beta+\frac{1}{2}\right) \Gamma \left(-\alpha+\beta+\frac{1}{2}\right)} \\ \nn&
    +\frac{\pi  \mu \sec \left( \pi  \beta\right)-i \pi  \kappa \sec \left( \pi  \beta\right)}{2 \mu \Gamma \left(1-\alpha\right)} ~,
\nn\\
I_3 = &\frac{\pi ^2 (\mu+i \kappa) \text{sec} \left(\pi 
   \beta\right) \text{csch}\left(-i\pi\alpha \right)}{\mu (\mu+2) \Gamma \left(\alpha \right) \Gamma \left(-\alpha-\beta+\frac{1}{2}\right) \Gamma \left(-\alpha+\beta+\frac{1}{2}\right)} + 
    \frac{\pi  (\kappa+i \mu) \text{sec} \left( \pi\beta\right)}{\mu (\mu+2) \Gamma \left(-\alpha\right)}~.
\end{align}
The evolution for the negative helicity gravitational wave is largely independent of the potential parameters and follows closely the free expanding-universe solution 
\begin{align}
\gamma^-(x) = \frac{H}{M_{\rm Pl}} {x \over \sqrt{k^3}} u_1(x), \, \quad
  \quad u_1(x) = \left ( 1+{i\over x} \right ) e^{ix} \, .
\end{align}

\begin{figure}[t!]
\centering
\includegraphics[width=\textwidth]{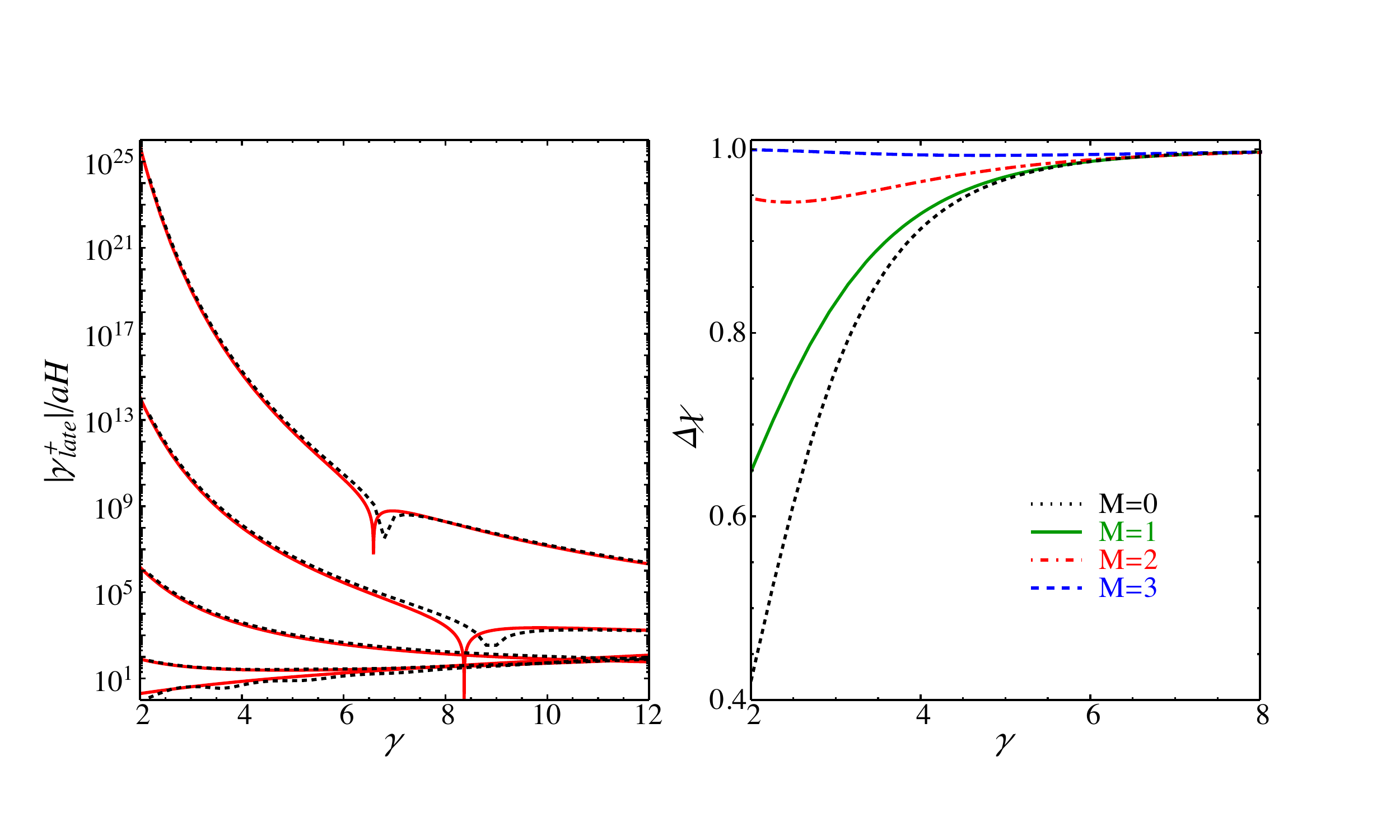}
\caption{
{
Left: Comparison between the late-time amplitude of the positive-helicity gravitational waves computed analytically (red-solid) and numerically (black-dotted) for $M=1,3,5,7,9$ (bottom to top) and $\epsilon=0.01$. Right: The chirality parameter as a function of $\gamma$ for $\epsilon=0.01$ and different values of $M$. }
}
\label{fig:tensorhelicity}
\end{figure}

The total power spectrum at late times is given by the sum of the power in the positive- and negative-helicity modes
\begin{align}
P_T(k) = 2P_{\gamma^+}^2(k)+2P_{\gamma^-}^2(k),
\end{align}
where we define the power in the each mode as
\begin{align}
\langle \gamma_k^\pm (\tau_*) \gamma_{k'}^\pm (\tau*) \rangle = (2\pi)^3 \delta^3(k_k') {2\pi^2\over k^3} P _{\gamma^\pm}(k) \, .
\end{align}
Since the negative-helicity modes follow the free-field result, their power is
\begin{align}
P_{\gamma^-}^2 (k) = {H^2\over 2\pi^2 M_{\rm Pl}^2} \, ,
\end{align}
while the power in the positive-helicity modes can be written as
\begin{align}
P_{\gamma^+}^2 (k) = {H^2\over 2\pi^2 M_{\rm Pl}^2}  + {4k H^2\over \pi^2 M_{\rm Pl}^2}{\epsilon\, |B_k|^2  \over 1+\gamma + M^2/2}  \left  |  	{2\over \sqrt{1+\gamma + M^2/2} }I_1  + \sqrt{\gamma} I_2 -\left (1+2\gamma + 2M^2 \right ) I_3	 \right |^2,
\label{eqn:gammaplusApprox}
\end{align}
because the free and sourced parts are uncorrelated. It is now straightforward to compute the chirality parameter
\begin{align}
\Delta \chi ={ P_{\gamma^+}^2 -  P_{\gamma^-}^2 \over P_{\gamma^+}^2 +  P_{\gamma^-}^2} \, ,
\end{align}
which is plotted in figure \ref{fig:tensorhelicity}.
We see that the gravitational wave spectrum is strongly polarized for most of the parameter-space plotted, especially for larger values of $M$ or $\gamma$.

\subsubsection{Superhorizon solutions}
\label{sec:tensorsuperhorizon}

On superhorizon scales corresponding to $k \ll aH$, we can also solve for the evolution of the system very accurately. In this limit, the parity violating terms become irrelevant, and thus we can ignore the difference between the positive- and negative-helicity modes. In this limit, 
\begin{align}
{\bf K} \sim & \[\begin{matrix} 0 & \frac{\sqrt{\epsilon}}{\sqrt{1+\gamma+\frac{M^2}{2}}} \\ -\frac{\sqrt{\epsilon}}{\sqrt{1+\gamma+\frac{M^2}{2}}}  & 0\end{matrix}\]\frac{(1+\epsilon)}{-\tau},\\ 
{\bf \Omega}^2 \sim & \[\begin{matrix}
 -\(2+\epsilon\frac{2-6\gamma-5M^2}{2+2\gamma+M^2}\) 
 &- \sqrt{\frac{2\epsilon}{2+2\gamma+M^2}}(1+2M^2+2\gamma)
 \\ - \sqrt{\frac{2\epsilon}{2+2\gamma+M^2}}(1+2M^2+2\gamma)
 & 2(1+\gamma+M^2) \end{matrix}\]\frac{(1+\epsilon)^2}{\tau^2}.
\end{align}
We then look for a solution of the form
\begin{align}
\vec{\Delta}_{\pm} \sim \vec{\Delta}_{0} (-k\tau)^{n},
\end{align}
where $\vec{\Delta}_0$ is a constant vector. Substituting into the asymptotic mode equation yields the four solutions for $n$ (two for each independent equation of motion)
\begin{align}
n = \left\{-1-\epsilon, 2+\epsilon, \frac{1}{2}\pm \frac{i}{2}\( \sqrt{7+8M^2+8\gamma}+\frac{8(1+M^2+\gamma)(3+M^2+2\gamma)}{\sqrt{7+8M^2 + 8\gamma}(2+2\gamma+M^2)}\epsilon\)\right\}+\mathcal{O}(\epsilon^2).
\end{align}
Note that the only growing mode is the solution with $n = -1-\epsilon$, which corresponds to the amplitude ratio
\begin{align}\label{eqn:superHgrav}
\vec{\Delta}_{\pm} =\(\begin{matrix} \hat{\gamma}^{\pm}_{\bf k} \\ \hat{t}^\pm_{\bf k} \end{matrix}\) =A_{\pm} \[\begin{matrix} 1 \\ \frac{\sqrt{2\epsilon}}{\sqrt{2+2\gamma+M^2}}\end{matrix}\](-k\tau)^{-1-\epsilon}.
\end{align}
Therefore, with the scale factor given by eq.\ \eqref{eqn:scalefactor}, on superhorizon scales 
the gravitational wave spectrum becomes constant. Despite the appearance of an apparent mass for the graviton, the superhorizon evolution is such that the resulting gravitational waves become constant.\footnote{In the parameterization of ref.\ \cite{Maleknejad:2016dve} this superhorizon solution corresponds to the vanishing of the `genuine tensor perturbation of the gauge field.' In that work, the slow-roll suppressed backreaction of the gravitational wave modes onto the gauge fields was dropped. Generically, this backreaction  sources a growing mode of the gauge field. However, we demonstrate here that the solution corresponding to the adiabatic mode corresponds to a solution where these gauge modes are absent.} While it appears  that the tensor tilt can be read off eq.\ \eqref{eqn:superHgrav} to be given by the standard $n_{t} = -2\epsilon$, in the case at hand, the prefactor $A_{\pm}$ has significant scale dependence, in contrast to standard single-field slow-roll inflation. We demonstrate below that the tensor spectral tilt, $n_T$, can take both negative as well as positive values, corresponding to a red or blue spectrum, respectively.

\section{Phenomenology}\label{sec:results}

In this section we explore the observational consequences of Higgsed Gauge-flation. We numerically compute both  the positive- and negative-helicity gravitational wave amplitudes $\gamma^{\pm}$ as well as the  density fluctuation $\zeta$.\footnote{We verified through explicit computation that our results are unchanged if we instead work with the comoving curvature perturbation in spatially-flat gauge $\mathcal{R} = H\delta u$, and on superhorizon scales $k \ll aH$, $|\zeta| = |\mathcal{R}|$.}

As described above in section \ref{sec:backgroundeqns}, in the slow roll approximation the evolution of the background can be completely specified by the values of $\{\gamma_{\rm in}, M_{\rm in}\}$ $N_{\rm in}$ $e$-folds before the end of inflation. In what follows, we parameterize the resulting spectra in terms of these parameters. The value of the parameter $\kappa$ is then determined by matching the amplitude of the scalar spectrum to the observed value.

In order to solve the equations of motion for the fluctuations numerically, we work with $e$-folding number $N$ as our time variable. For the background, treating $\psi \approx {\rm constant}$, we use the analytical expressions for $\{\gamma(N), M(N), \epsilon(N)\}$ as described above in section \ref{sec:backgroundeqns}.  This method allows for a straightforward computation of the observables closer to the end of inflation, $60$ $e$-folds after horizon-crossing.

As a check for accuracy, we performed the same computation in conformal time $\tau$, simultaneously solving the background and fluctuation equations numerically. We initialize the system close to the slow-roll attractor using the following procedure:
\begin{enumerate}
\item Calculate the value of $\psi$ as ${\psi^2_{\rm in}}={ M_{\rm Pl}^2\over 2N_{\rm in}} \log \left ( {   1+\gamma_{\rm in}+M_{\rm in}^2/2\over \gamma_{\rm in} +M_{\rm in}^2/2    }   \right )$, for chosen values of $\gamma_{\rm in}$ and $M_{\rm in}$ at $N_{\rm in} = 60$ $e$-folds before the end of inflation.
\item The Hubble scale at $60$ $e$-folds is given by $H^2_{\rm in} = 2M_{\rm Pl}^4/ (\gamma_{\rm in} \psi_{\rm in}^2)$.
\item Finally, the initial velocity of the $\psi$ field is $\dot \psi_{\rm in} = - (\gamma_{\rm in}+M_{\rm in}^2) (1+\gamma_{\rm in} +M_{\rm in}^2/2) {\psi_{\rm in}^5 H_{\rm in}\over M_{\rm Pl}^4}$.
\end{enumerate}
These initial conditions start the calculation close to the exact numerical solution for all values of $\gamma$ and $M$. We found excellent agreement between these methods (evolution in conformal time vs $e$-folding number) in both the evolution of each mode-functions, as well as the resulting values of $r$, $n_s$, and $n_T$.

As shown in sections \ref{sec:scalarsuperhorizon} and \ref{sec:tensorsuperhorizon} respectively, the scalar modes evolve outside the horizon, while the tensor modes become constant. This means that the tensor-to-scalar ratio is not constant, but rather evolves from horizon exit until the end of inflation. Even though the evolution is rather weak, since it involves the slow-roll parameter $\epsilon = {\cal O}(0.01)$, a simple order-of-magnitude calculation  shows that the tensor-to-scalar ratio evolves as $( \tau_{\rm comp} / \tau_{\rm end} )^{{\cal O}(\epsilon)} ={\cal O}(10)$, where $\tau_{\rm comp}$ is the conformal time where we compute $r$ and  $\tau_{\rm end}$ is the end of inflation. Thus in the $60$ $e$-folds between horizon-crossing and the end of inflation the tensor-to-scalar ratio can vary by an ${\cal O}(10)$ factor. For this reason, we compute the observables at $5$ $e$-folds before the end of inflation. We did not choose to evolve further, in order to ensure the accuracy of the slow-roll formulas. At $N=5$ $e$-folds before the end of inflation the numerically computed (exact) solution for $\epsilon$ and the slow-roll expression start to deviate. Additionally, at this point $\epsilon \simeq 0.1$, and the $\epsilon$-expansion of the equations of motion begins to break down after this point.

Beyond the tensor-to-scalar ratio, the superhorizon evolution of the scalar power complicates the numerical computation of the spectral index $n_s$. For the standard single-field slow-roll inflationary models,  the power spectrum freezes outside the horizon and the spectral index can be evaluated by comparing $P_\zeta(k)$ for neighboring values of the wavenumber $k \pm \Delta k$
at some constant value of conformal time $\tau$, e-folding number $N$, or $x=-k\tau$, provided the mode has left the horizon.\footnote{In practice we compute the tilt by evaluating the spectrum for near by $k$-values $k \pm \Delta k$ and taking a two-sided numerical derivative. We checked that our results were independent of the precise value of $\Delta k$ used.} In the present case the first two methods (constant $\tau$ or $N$) are equivalent sufficiently far outside the horizon, but the case of taking $x=-k\tau$ constant is not. The value of $n_s$ computed using the two (inequivalent) methods can differ by ${\cal O}(\epsilon) = {\cal O}(0.01)$. We choose to evaluate the spectral index by comparing the power-spectra at a fixed number of $e$-folds before the end of inflation. Even though the spectral index settles to a constant value after around $10$ $e$-folds after horizon-crossing, even for the largest values of $M$ used, we evaluate it at $5$ $e$-folds before the end of inflation, due to the evolution of the tensor-to-scalar ratio as explained above. 

The evolution of the scalar power in the last 5 $e$-folds is a source of uncertainty in our results. One the one hand, the continued decay of the scalar power likely increases the tensor-to-scalar ratio above the values we quote. On the other hand, since all modes evolve identically outside the horizon, the scalar spectral index is frozen and does not evolve during this period.

\subsubsection*{Spectral tilt and tensor-to-scalar ratio}

For the case Gauge-flation, $M=0$, we recover the results of ref.\ \cite{Namba:2013kia} by computing the spectral index at constant $x=-k\tau$, or $3$ $e$-folds after each mode has left the horizon. By computing $n_s$ at constant $N$ we get the same dependence of $n_s$ on $\gamma$, however, the spectral index is larger by about $\Delta n_s= 0.015$ compared to the results of ref.\ \cite{Namba:2013kia}. This is a change of ${\cal O}(\epsilon)$ as expected. However, the tensor-to-scalar ratio is larger by a factor of $10$ when calculated close to the end of inflation ($N=5$) instead of close to horizon crossing ($3$ $e$-folds after horizon-crossing as in ref.\ \cite{Namba:2013kia}), due to the superhorizon decay of the scalar power spectrum, as explained in section \ref{sec:scalars}. This pushes the observables of Gauge-flation even further from the Planck constraints.

\begin{figure}
\centering
\includegraphics[width=\textwidth]{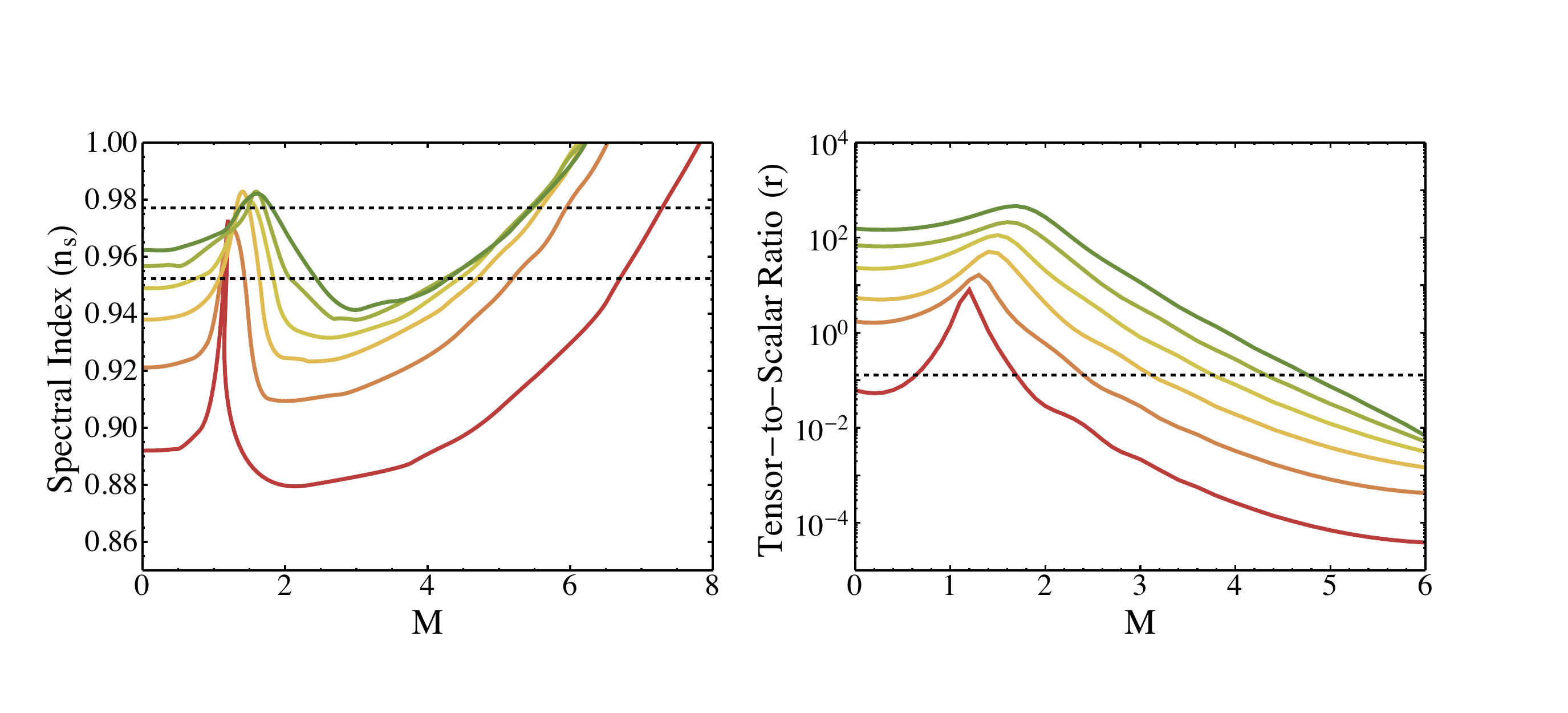}
\caption{
The spectral index $n_s$ (left) and the tensor-to-scalar ratio $r$ (right) as a function of the mass $M$ for $\gamma=4,5,6,7,8,9$, color-coded from red to green. The horizontal black-dotted lines correspond to the Planck 2-$\sigma$ bounds.}
\label{fig:Mscan}
\end{figure}

\begin{figure}
\centering
\includegraphics[width=\textwidth]{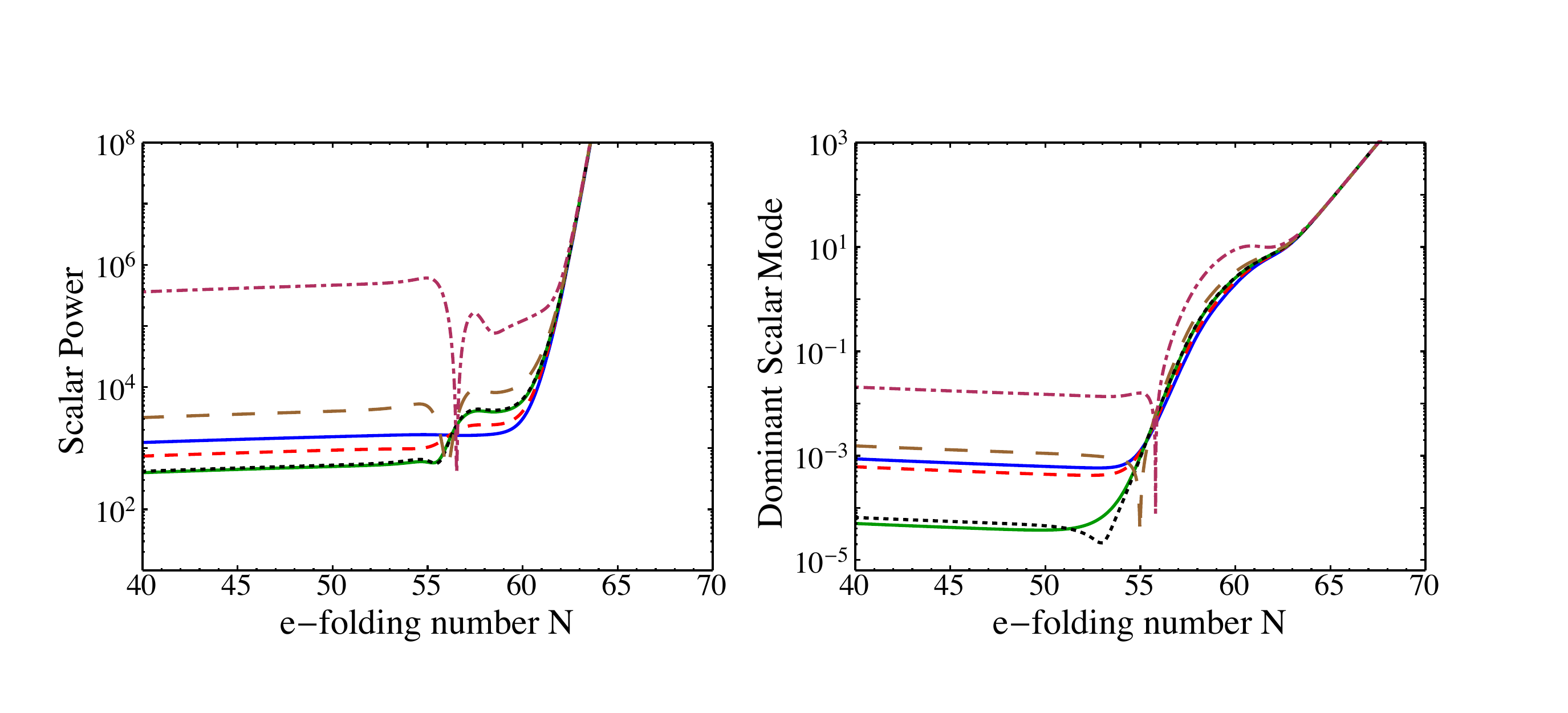}
\caption{
Left: The scalar power for $\gamma=7$ and $M=0,1,1.5,1.55,2,3$ (blue solid, red dashed,  green solid, black dotted, brown long-dashed and maroon dot-dashed respectively). Right: The dominant ``slow'' mode-function $e^{-(N-N_*)}\left |\Delta_2(N) \right |$ for the same parameters and color-coding.}
\label{fig:modefunctionscan}
\end{figure}

\begin{figure}
\centering
\includegraphics[width=\textwidth]{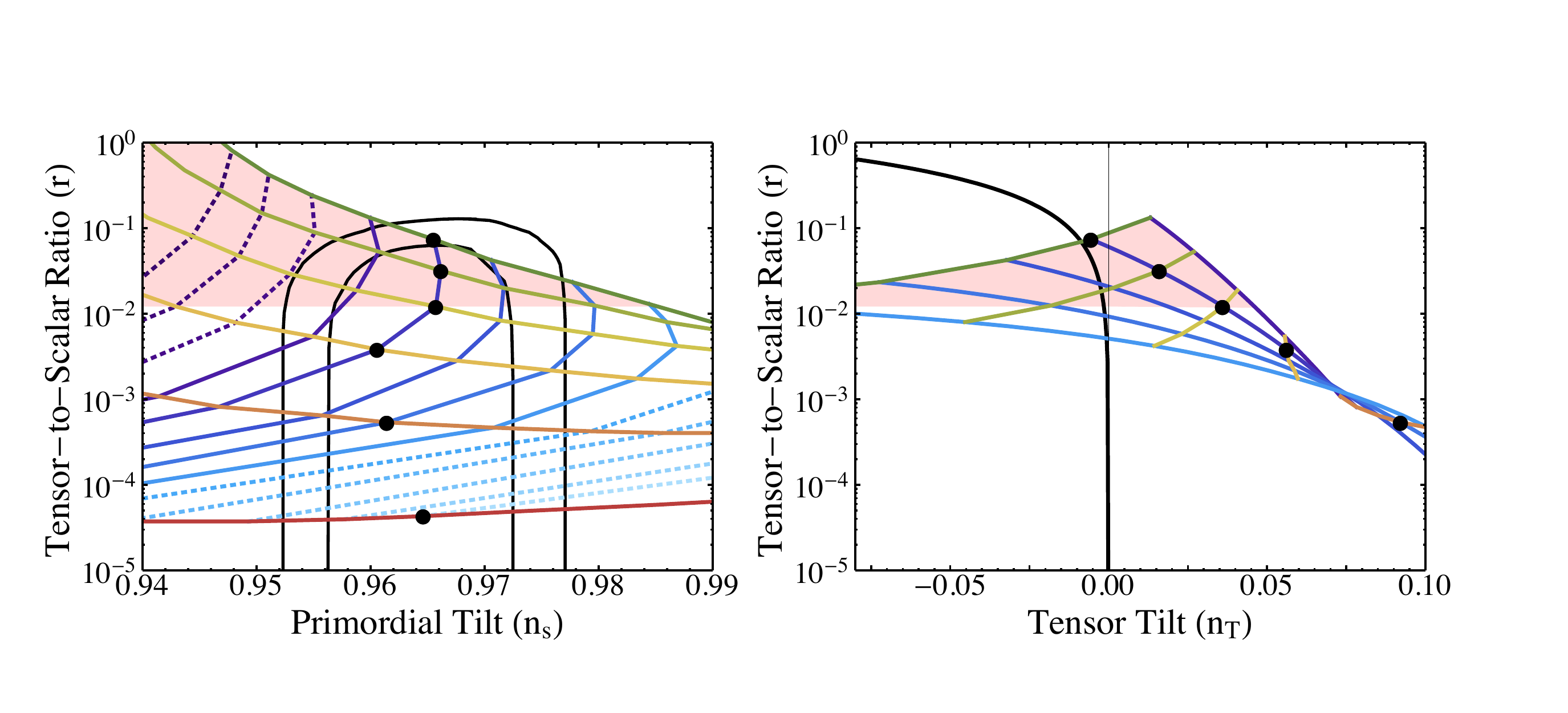}
\caption{
The tensor-to-scalar ratio, $r$, (at $k = 0.002h$ ${\rm Mpc}^{-1}$) as a function of the scalar spectral tilt, $n_s$ (at $k = 0.05h$ ${\rm Mpc}^{-1}$) and the tensor spectral tilt, $n_T$ (at $k = 0.002h$ ${\rm Mpc}^{-1}$)
for models drawn from a grid of values for the parameters $\gamma$ and $M$ measured at $N=60$ $e$-folds before the end of inflation. The mode $k = 0.05 h$ ${\rm Mpc}^{-1}$ is assumed to leave the horizon $60$ $e$-folds before the end of inflation. 
In both panels each rainbow-colored line corresponds to a definite value of $\gamma$ ranging from $\gamma=5$ (red) to $\gamma=9$ (green). Each purple-blue colored line corresponds to a definite value of $M$, ranging from $M=4$ (far left purple-dotted line) to $M=7$ (far right cyan-dotted line). The dots correspond to the values used in Table \ref{tab:paramscan}. The dotted purple-to-blue lines shown on the left panel do not appear on the right one, since they are either largely outside of the Planck-allowed regime (purple-dotted) or lead to a very blue tensor spectrum (cyan-dotted).
 In both panels, the shaded light red region correspond to the $1\%$ limit of the linear regime, as discussed in section \ref{sec:validitylinear}.
}
\label{fig:paramscan}
\end{figure}

As the Higgs vev, and thus  $M$, is increased, both the tensor-to-scalar ratio, $r$, and the scalar spectral index, $n_s$, develop a distinct spike  for any value of $\gamma$, which is however significantly more pronounce for smaller values of $\gamma$. This feature is shown in figure \ref{fig:Mscan}. In order to understand the feature in $r$ and $n_s$, we revisit the behavior of the evolution of the scalar and tensor power spectra for $M\ne 0$. In figure \ref{fig:modefunctionscan} we show the evolution of the scalar power spectrum and the amplitude of mode that contributes dominantly to the scalar spectrum, $\M(-k\tau)$, for constant $\gamma$ and various values of $M$ bracketing the ``spike". 
Figure \ref{fig:modefunctionscan} shows that at fixed $\gamma$ and increasing $M$, the power in the scalar modes initially \emph{decreases} before reaching a minima, and then ultimately increases quickly. In contrast, the power in tensor modes increases monotonically for constant $\gamma$ and increasing $M$. The ``spike" in the spectral parameters is due to the non-monotonic behavior of the scalar spectrum as $M$ is varied; the tensor to scalar ratio reaches its maximum value where the scalar spectrum is minimized. 
Further, across this minima the spectral tilt goes from blue to red. This is counter-intuitive, since $M$ increases during inflation, one might naively expect the opposite. However, as demonstrated below, $\gamma$ and its evolution have a much stronger effect on the tilt of the spectrum (see figure \ref{fig:paramscan}). We note that this behavior was not seen in the related model of Higgsed Chromo-Natural Inflation \cite{Adshead:2016omu}. However, in that work the region of parameter space corresponding to the (Higgsed) Gauge-flation model was not explored.

For large-enough values of $M$, the observables of Higgsed Gauge-flation  pass through the Planck-allowed region in the $n_s$-$r$ plane, as shown in figure \ref{fig:paramscan}. By varying both $\gamma$ and $M$ we can fill-up the whole of the allowed Planck region\footnote{It is worth re-iterating here that we do not compute the evolution of the fluctuations during the last $5$ $e$-folds of inflation. Since we do not expect $n_s$ to vary, these last 5 $e$-folds may shift the curves of figure \ref{fig:paramscan} vertically, hence filing a slightly different part of the Planck-allowed region. } in the $n_s$-$r$ plane for $r \gtrsim 10^{-5}$.

\subsubsection*{Running of the scalar spectral index and inflationary energy scale}

\begin{figure}
\centering
\includegraphics[width=\textwidth]{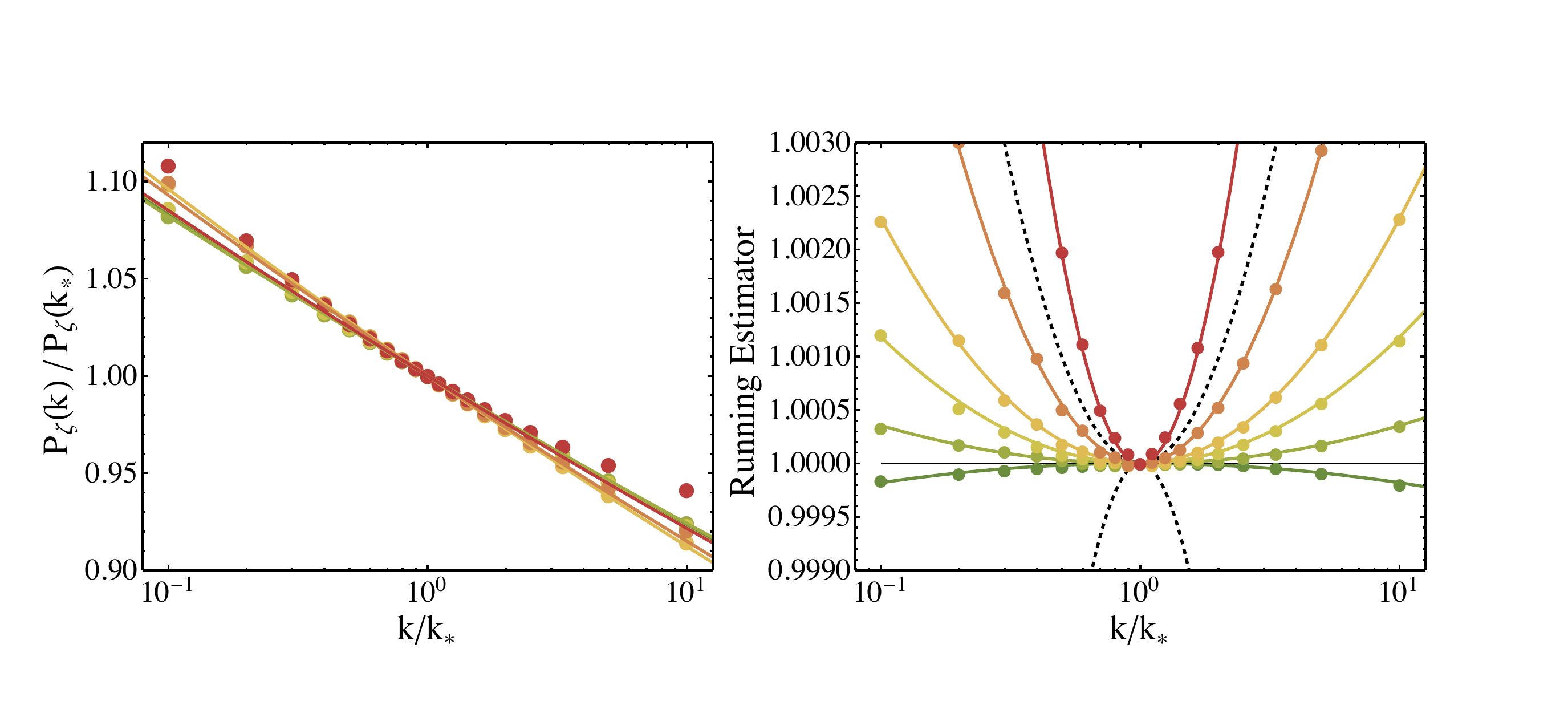}
\caption{
Left: The normalized scalar power spectrum amplitude as a function of wavenumber for the parameters corresponding to the six dots of figure \ref{fig:paramscan}. The lines correspond to power-law fits $P_\zeta(k)\sim k^{n_s-1}$, while the dots show the results of numerical simulations. Right: The ratio of the normalized tensor amplitude (plotted on the left panel) to the form $P_\zeta(k)\sim k^{n_s-1}$. The dots correspond to data points while the lines show the best-fit parabolae. This is a constant line in the case of zero running of the spectral index, hence it can be used as a visual estimator of the magnitude of the running $\alpha$. The black-dotted lines show the latest Planck limits. The color-coding for both panes follows that of figure \ref{fig:paramscan}.
}
\label{fig:running}
\end{figure}

We compute the running of the scalar spectral tilt, $\alpha \equiv {d\,n_s/ d\log k}$, by locally fitting $\log P_\zeta$ as a function of $\log k$ using a second order polynomial around the mode-function $k$ that leaves the horizon $60$ $e$-folds before the end of inflation. The results for the parameters that correspond to the black dots of figure \ref{fig:paramscan} are shown in figure \ref{fig:running}. We see that for all but one of the cases shown the running is positive and also for all but one of the cases shown, the running is within the Planck limits.

Until now, we have rescaled the gauge coupling, $g$, and  cosmic time, $t$,  by $\kappa$, eliminating it from the equations of motion as described above in section \ref{sec:backgroundeqns}. However, the value of $\kappa$ sets the Hubble scale, and the overall energy scale of inflation. We fix $\kappa$ by matching the the amplitude of the observed scalar power spectrum  $P_\zeta \simeq 2.2\times 10^{-9}$ \cite{Ade:2015lrj}. Table \ref{tab:paramscan} shows the required value of $\kappa$ and the corresponding value of the Hubble scale. We note here that the standard inflationary result (see, for example, \cite{Abazajian:2016yjj})
\begin{align}
\Lambda_{\rm inf}\sim \sqrt{H_{\rm inf} M_{\rm Pl}} \sim 1.04\times 10^{16}{\rm GeV}\(\frac{r}{0.01}\)^{1/4},
\end{align}
is  violated.

 \begin{table}[h]
 \begin{small}
 \begin{center}
   \begin{tabular}{ | l | l | l | l | l | l | l | l | l | l | l | l |}
    \hline=
    $\gamma$ & $M$ & $\kappa^{1/4}M_{\rm Pl} 
    $  & $H\,(M_{\rm Pl})
    $  &  $g$  & $\psi\, (M_{\rm Pl})
    $  & $n_s$ &  $r$ & $n_T$ & $\alpha \times 10^{4}$ 
    \\ \hline
     $9$ & $5$ & $1.4\times 10^4 $  & $1.3\times 10^{-7}$  &  $2\times 10^{-5}$  & $0.019$  & $0.965$ & $7.4\times 10^{-2}$ & $-0.006$ & $-0.68$ 
    \\ \hline
    $8$ & $5$ & $2.2\times 10^4 $  & $5.6 \times 10^{-8}$  &  $8\times 10^{-6}$  & $0.020$  & $0.966$ & $3.2\times 10^{-2}$ & $0.016$ & $1.34$ 
    \\ \hline
      $7$ & $5$ & $3.6\times 10^4 $  & $2\times 10^{-8}$  &  $2.6\times 10^{-6}$  & $0.020$  & $0.966$ & $1.2\times 10^{-2}$ & $0.036$ & $4.46$ 
    \\ \hline
     $6$ & $5$ & $7.3 \times 10^4 $  & $5\times 10^{-9}$  &  $6\times 10^{-7}$  & $0.018$  & $0.961$ & $3.8\times 10^{-3}$ & $0.056$ & $8.64$ 
    \\ \hline
    $5$ & $5.5$ & $5.9\times 10^5 $  & $9\times 10^{-11}$  &  $ 1\times 10^{-8}$  & $0.020$  & $0.961$ & $5.4\times 10^{-4}$ & $0.092$ & $22.66$ 
    \\ \hline
    $4$ & $7$ & $4.5 \times 10^8  $  & $2\times10^{-16}$  &  $2.4\times 10^{-14}$  & $0.017$  & $0.965$ & $4.3\times 10^{-5}$ & $0.241$ & $80.46$ 
    \\ \hline
     \end{tabular}
     \end{center}
     \end{small}
 \caption{Potential parameters and observables for the black dots shown in figure. The value of the Hubble scale is given at $N=60$ $e$-folds before the end of inflation. The value at the end of inflation is approximately $8$ times smaller for the parameters in this table.} 
 \label{tab:paramscan}
\end{table}%

\subsection{Validity of the linear theory}\label{sec:validitylinear}

We end this section with a brief discussion of the validity of our analysis. As in the case of Higgsed Chromo-Natural Inflation \cite{Adshead:2016omu}, the introduction of a Higgs sector enhances the tensor mode amplification, rather than suppressing it. Furthermore, the Goldstone mode dynamics contribute constructively to the amplification of the scalar and vector modes. It is thus of paramount importance to examine whether the use of the linear order perturbation theory and of the scalar-vector-tensor decomposition of the fluctuations is justified. Following the analysis of ref.\ \cite{Adshead:2016omu}, we provide an estimate of the non-linearity, rather than attempt a detailed analysis.

The gauge field mode amplification begins around $-k\tau\sim M$ and ceases around the time when the mode exits the horizon at $-k\tau=1$, as shown for scalar, vector and tensor modes in figures \ref{fig:scalarmodes}, \ref{fig:vectormodes}, and   \ref{fig:tensormodes} respectively. In order to keep the linearized analysis under control, it is sufficient for the backround field fluctuations $\delta\! A_\mu$ to be significantly smaller than the classical (background) gauge field value, which is given in eq.\ \eqref{eqn:gaugevev} as $\bar A_\mu = (0,a\psi \delta^c_{i} J_c)$, 
\begin{align}
\label{eq:linearitycondition}
{|\delta\! A_\mu| \over |\bar A_\mu|} \ll 1  \, .
\end{align}
We consider this ratio as a measure of the validity of the linear analysis. We focus on the tensor part of the gauge fluctuations. In order to compute the ratio of eq.\ \eqref{eq:linearitycondition}, we estimate the gauge field fluctuations as
\begin{align}
|\delta\! A| = \sqrt{ \langle (\delta\! A)^2 \rangle} = \sqrt{ \int {\d^3k\over (2\pi)^3} |\delta\! A_k|^2} \sim {M H \over 2\pi} \left |\sqrt{2k} \delta\! A_k \right |  \, ,
\end{align}
where we cut the integral off at the peak of the amplification, which is observed to be near $-k\tau \sim M$. The linearity condition of eq. \eqref{eq:linearitycondition} can thus be rewritten as
\begin{align} 
\label{eq:rescaledgaugeamplidude}
{g\over 2\pi} {M\over \sqrt{\gamma}} \left |\sqrt{2k} \delta\! A_k \right | \ll 1 \, ,
\end{align}
where we used the definition of $\gamma$ to swap $H$ for $g$.

\begin{figure}
\centering
\includegraphics[width=\textwidth]{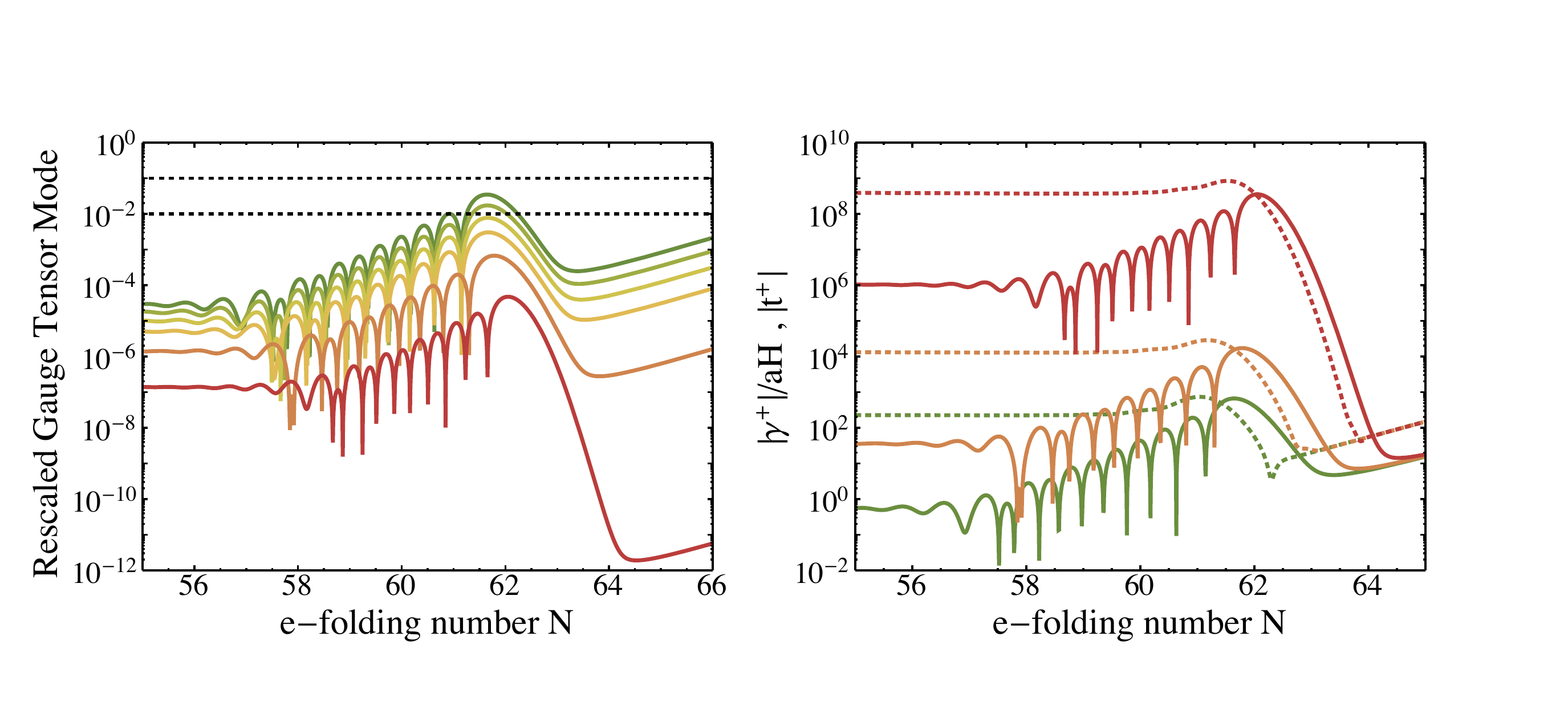}
\caption{
{
Left: The rescaled amplitude of the gauge tensor mode given in eq.\ \eqref{eq:rescaledgaugeamplidude}. The horizontal black-dotted lines correspond to $0.1$ and $0.01$, which indicate the level of the non-linearity. The parameters used follow the black dots of figure \ref{fig:paramscan}, as does the color-coding of the curves.
Right: The gauge (solid) and metric (dotted) tensor modes, using the same color-coding. The gauge modes are multiplied by $0.1$, in order to make the gauge and metric amplitude easier to compare visually. We show only three out of the six curves for clarity.}
}
\label{fig:nonlinearity}
\end{figure}

In figure \ref{fig:nonlinearity}, we plot the left hand side of eq.\ \eqref{eq:rescaledgaugeamplidude}. Note that most of the cases of interest satisfy the linearity criterion at the $1\%$ level or better.
As in the case of Higgsed Chromo-Natural Inflation, the theory remains within the linear regime for lower values of the tensor-to-scalar ratio. A simple estimate of this can be done in a more general way using dimensional arguments.  The background amplitude of the gauge field depends only on the number of $e$-folds, which we take to be $N=60$, and the values of $\gamma$ and $M$. For $\gamma , M={\cal O}(1)$ we get $\psi/M_{\rm Pl} \sim 0.02$. The dominant gravitational wave helicity mode is seeded by the gauge field fluctuation, and we observe that the late time gravitational wave amplitude is directly proportional to the peak gauge field amplitude, $|h_{\ij}|\propto |\delta\! A_\mu|_{\rm Peak}$, with a proportionality factor of ${\cal O}(0.1)$ (see figure \ref{fig:nonlinearity}, right panel). The physical amplitude of the gravitational waves depends on the Hubble scale and can be estimated as $P_T \sim H^2 |h_{\ij}|^2\sim  {\cal O}(0.01)|\delta\! A_\mu|^2$. The linearity condition can thus be re-written as
 \begin{align}
 \label{eq:linearityvalidityHR}
 {|\delta \!A |\over |\bar A |} \sim 10^4 {H \over \sqrt{r}}  \ll 1\, ,
 \end{align}
where we used the fact that $P_T = rP_\zeta$ and $P_\zeta \approx 2.2\times 10^{-9}$ \cite{Ade:2015lrj}. The advantage of this form of the linearity condition is that it involves only the Hubble scale and the tensor-to-scalar ratio. Finally, we emphasize that both eqs.\ \eqref{eq:rescaledgaugeamplidude} and \eqref{eq:linearityvalidityHR} should be used as order-of-magnitude estimates of the non-linearity, rather than a sharp cut-off.

\begin{figure}
\centering
\includegraphics[width=\textwidth]{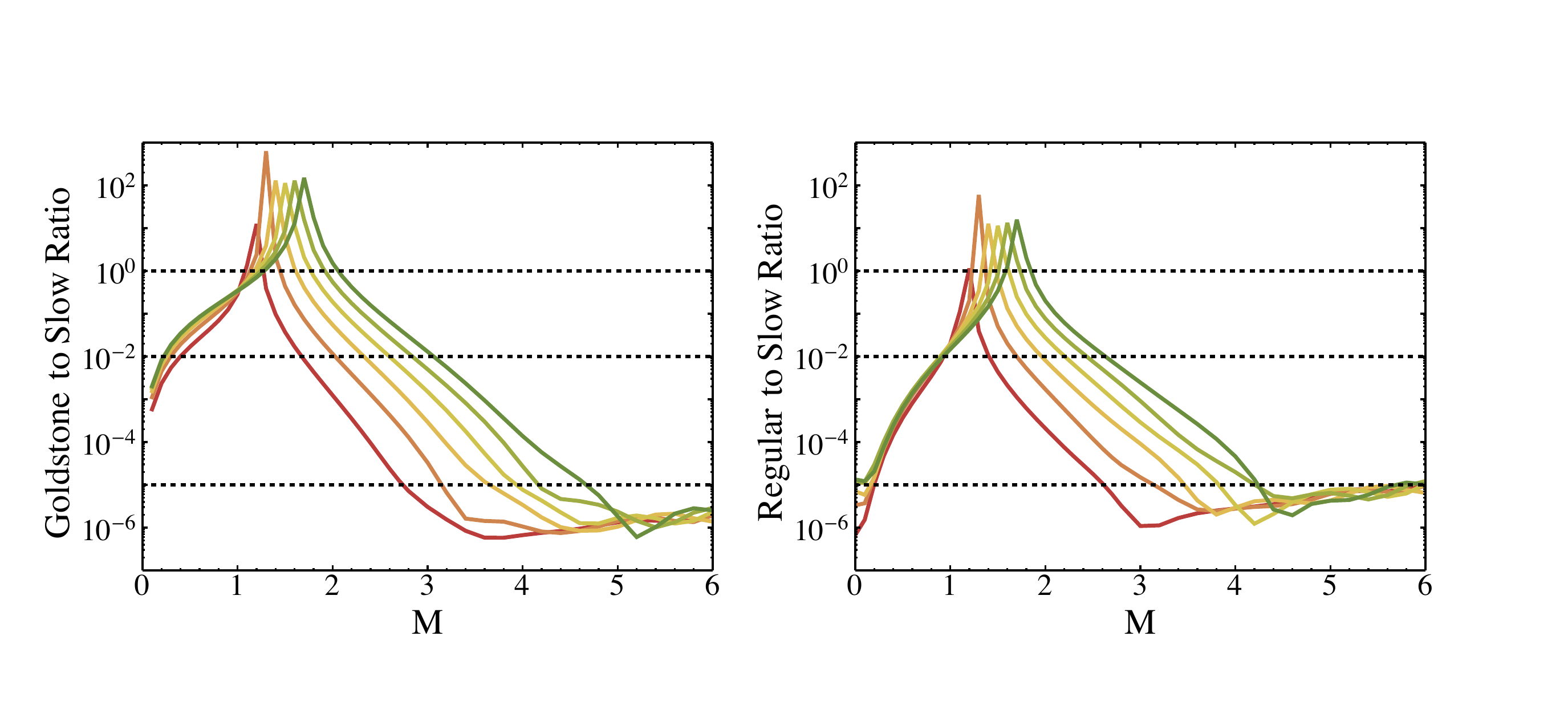}
\caption{
The ratio of the final scalar power spectrum computed using the Higgs and Slow initial conditions (left) and using the Regular and Slow initial conditions (right). The color-coding follows figure \ref{fig:Mscan}. The black-dotted lines correspond to ratio of $1$, $10^{-2}$ and $10^{-5}$.
}
\label{fig:RegHiggsSlowRatio}
\end{figure}

Before concluding this section, we revisit and justify the claim made throughout this work regarding the dominance of the contribution of the `slow' scalar mode compared to the `regular' and `Higgs' modes in the scalar spectrum (see eq.\ \eqref{eqn:initialconditions}). Figure \ref{fig:RegHiggsSlowRatio} shows the ratio of the final scalar power spectrum computed using the three initial conditions, slow, regular and Higgs.  In the region of parameter space where the spectral tilt lies within the Planck bounds, $M\gtrsim 5$ (see figure \ref{fig:Mscan}), the regular and Higgs modes contribute negligibly to the final power spectrum. We are therefore justified in only using the ``slow" mode to initialize the numerical evolution of the fluctuations in this region. Away from this region of parameter space, $M\lesssim 3$, the otherwise subdominant modes can dominate the spectrum, however, in this region of parameter space the model is ruled out by the large tensor-to-scalar ratio, as demonstrated in figure \ref{fig:Mscan}.

\section{Discussion and conclusions}\label{sec:discussion}

In this work we have demonstrated that Gauge-flation can be  made compatible with existing limits from Planck data by introducing an additional mass term for the gauge field fluctuations. We assume that the symmetry is spontaneously broken by a Higgs sector and the resulting Higgs boson is much heavier than the Hubble scale, and is thus irrelevant. We thus work with the theory in the Stueckelberg form, restricting the Higgs sector to the Goldstone modes which fluctuate along the vacuum manifold.

The introduction of a Higgs sector to Gauge-flation significantly changes the phenomenology of the fluctuations. On the one hand, the additional terms in the tensor action have the effect of exacerbating the existing chiral gauge-tensor instability, and boost the overall amplitude of the tensor modes. The resulting chiral spectrum of  gravitational waves becomes constant on superhorizon scales, and is qualitatively similar to the spectrum of gravitational waves from Gauge-flation.  On the other hand, the mass terms and associated scalar Goldstone modes significantly alter the scalar dynamics and resulting spectrum of density fluctuations. The theory remains (catastrophically) unstable for parameters such that $\gamma < 2$. However, as the Higgs vev, $Z_0$, and thus $M$, is increased from zero (with $\gamma$ fixed) the amplitude of the scalar power is initially suppressed before reaching a minima.  This behavior leads to a feature in both the tensor-to-scalar ratio, $r$, and the spectral index, $n_s$, at the points in parameter space where the scalar power reaches its minimum. By increasing the Higgs vev, and thus $M$, beyond this point, the overall amplitude of the resulting scalar density fluctuations grows monotonically, decreasing the tensor-to-scalar ratio $r$, and increasing the scalar spectral index $n_s$. This behavior allows the theory to produce spectra of gravitational waves and density fluctuations that satisfy the latest Planck bounds. The scalar spectrum decays on superhorizon scales in this model due to the presence of an isocurvature mode. We evaluate the tensor-to-scalar ratio and scalar spectral index 5 $e$-folds before the end of inflation to minimize errors, however, the evolution of the spectrum during the last 5 $e$-folds of inflation represents a source of uncertainty in our results.

The exponential enhancement of the tensor modes means that observable gravitational waves may be produced in this model, despite inflation occurring below the GUT scale, and all fields evolving over sub-Planckian distances in field space. The model therefore violates some formulations of the Lyth bound. The production mechanism of gravitational waves is exactly analogous to the cases of Chromo-Natural Inflation, Higgsed Chromo-Natural Inflation, and Gauge-flation. The gravitational waves in these models predominantly arise from linear mixing with the (exponentially amplified) gauge field fluctuations. The form of the gravitational wave spectra produced in this model is therefore significantly altered from the usual form assumed in formulations of the Lyth bound. In contrast to standard inflationary scenarios which uniformly predict red tilted gravitational wave spectra, these gravitational waves can have either red- or blue-tilted spectra on CMB scales, with a strong favoring of a blue tilt, especially for lower values of the tensor-to-scalar ratio $r$. Furthermore, these gravitational waves have the distinct characteristic that they are chirally polarized and, to a very good approximation, consist only of a single helicity. Unfortunately, it seems unlikely that future CMB experiments will be unable to distinguish between unpolarized and chirally polarized gravitational waves \cite{Gluscevic:2010vv, Gerbino:2016mqb}, which would significantly reduce the space of viable inflationary models.

The running of the scalar spectral index is predominately positive and within observational bounds for all but the lowest calculated values of the tensor-to-scalar ratio. Between the linearity constraints for the equations of motion of the fluctuations and the Planck constraints on the running of the spectral index, Higgsed Gauge-flation can fill the whole Planck-allowed region on the $n_s$-$r$ plane for $10^{-4} \lesssim r\lesssim 10^{-2}$,  making this model especially interesting in anticipation of planned Stage-4 CMB experiments. These experiments are aiming to probe tensor-to-scalar ratios as low as $r\sim 10^{-3}$.  The exponential sensitivity of the amplification of both the scalar and tensor power spectra makes some level of fine-tuning necessary to fit observations. In contrast to Gauge-flation, the addition of the Higgs sector causes vector perturbations of the matter sector to freeze out on super-horizon scales, we leave the further study of the consequences of these modes to future work. While we have estimated that the linear theory is under control, we leave the study of non-Gaussian features of this model for future work.


\vskip .5cm

{\bf Acknowledgements:}
We thank Emanuela Dimastrogiovanni, Azadeh Maleknejad, Emil Martinec, Marco Peloso, and Mark Wyman for useful discussions. This work was supported by the United States Department of Energy through grant DE-SC0015655.
EIS gratefully acknowledges support from a Fortner Fellowship at the University of Illinois at Urbana-Champaign.

\appendix

%

\section{Notation and conventions}\label{app:notation}

%

We work with conformal time, which we define to be a negative quantity during inflation
\begin{align}
\tau = \int^{t}_{0}\frac{\d t}{a(t)},
\end{align}
and make use of the near de Sitter expansion to write
\begin{align}\label{eqn:scalefactor}
a \approx \frac{(-\tau)^{-1-\epsilon}}{H},
\end{align}
When we are dealing with fluctuations of the fields, we  work in Fourier space where our convention is
\begin{align}
A({\bf x}) = \int \frac{\d^3 k}{(2\pi)^3}A_{\bf k}e^{-i {\bf k}\cdot{\bf x}},
\end{align}
so that we replace spatial derivatives with $\partial_i A \to -i k_i A_{\bf k}$  
and we  make extensive use of the fact that the fields satisfy a reality condition, which implies $A_{-{\bf k}} = \bar{A}_{\bf k}$. It  proves useful to work with the dimensionless time variable $x = -k\tau$, where $k$ is the Fourier space wavenumber. Throughout we  denote derivatives with respect to cosmic time by an overdot ($\,\dot{}\,$), primes ($\,'\,$)  denote derivatives with respect to conformal time $\tau$, while derivatives with respect to conformal $x$ are kept explicit ($\partial_x$). Our symmetrization and antisymmetrization conventions throughout are
\begin{align}
Z_{[ij]} = & \frac{1}{2}(Z_{ij} - Z_{ji}),\quad 
Z_{(ij)} =  \frac{1}{2}(Z_{ij} + Z_{ji}).
\end{align}

%
\section{Details of the scalar action}\label{app:scalsector}
%

In this appendix we present the details of the scalar action. After eliminating the algebraic constraints from the action, redefining the fields according to eq.\ \eqref{eqn:fieldred}, performing an integration by parts and discarding the boundary terms, the scalar action is put into the form
\begin{align}
S = \frac{1}{2}\int \frac{\d^3k}{(2\pi)^3}\d\tau\[\Delta^\dagger{}'{\bf T}\Delta' + \Delta^\dagger{}'{\bf K}\Delta-\Delta^\dagger{}{\bf K}\Delta' - \Delta^{\dagger}\boldsymbol{\Omega}^2\Delta\]  .
\end{align}
The exact forms of the matrices can be obtained in a straightforward manner, however, they are long and complicated, and not particularly enlightening. We do not present their gory details here.

As in ref.\ \cite{Namba:2013kia}, each entry in the matrices is of the form, or is of the sum of entries of the form
\begin{align}\label{eqn:mtxexp}
\frac{\sum_i c_i k^{\alpha_i}}{\sum_j d_j k^{\alpha_j}}\times\sqrt{\frac{\sum_m \tilde{c}_m k^{\alpha_m}}{\sum_n \tilde{d}_n k^{\alpha_n}}},
\end{align}
where  the sums are finite, and all coefficients $c_i, d_i, \tilde{c}_i, \tilde{d}_i$ are slowly varying functions of time. To perform our numerical evaluations, we expand each of the coefficients in slow roll in the same manner as described in ref.\ \cite{Namba:2013kia}. Specifically, we replace $\kappa$ and $\psi$ using eq.\ \eqref{eqn:kappa}, and then use eq.\ \eqref{eqn:deltareplace} to replace $\delta$. We then expand each term to leading order in $\epsilon \ll 1$. Obtaining the action, as well as expanding each term is performed using {\sc Mathematica}.

To leading order in $\epsilon \ll 1$, the matrices have the entries
\begin{align}
T_{11} \simeq & 1+\frac{6  \epsilon^2 a^2 H^2 \left(2\gamma+M^2\right)}{ k^2 \left(2+2
  \gamma+M^2\right)^2+3 a^2 \epsilon ^2 H^2 \left(\gamma+M^2\right) \left(2\gamma+M^2\right)},\\
   T_{22}\simeq & \frac{6 a^2 \gamma  H^2 \epsilon  \left(2 \gamma +M^2+2\right) \left(3 a^2 H^2 \epsilon  \left(2 \gamma +M^2\right)+k^2 \left(2 \gamma -2 \epsilon  \left(\gamma +M^2-1\right)+M^2+2\right)\right)}{\left(6 a^2 \gamma  H^2 \epsilon +k^2 \left(2 \gamma +M^2+2\right)\right) \left(3 a^2 H^2 \epsilon ^2 \left(\gamma +M^2\right) \left(2 \gamma +M^2\right)+k^2 \left(2 \gamma +M^2+2\right)^2\right)}, \\ 
 T_{12}   \simeq &-\frac{2 \sqrt{3} a H \sqrt{\gamma} \sqrt{
  2 + 2 \gamma + 
   M^2} \epsilon ( k^2 (2 + 2 \gamma + M^2) + 
    3 a^2 H^2 (2 \gamma + M^2) \epsilon)}{
 \sqrt{k^2 (2 + 2 \gamma + M^2) + 
   6 a^2 H^2 \gamma \epsilon} ( k^2 (2 + 2 \gamma + M^2)^2 + 
    3 a^2 H^2 (\gamma + M^2) (2 \gamma + M^2) \epsilon^2)},\\
  T_{33} = & 1 + \frac{3 a^2 H^2 k^2 M^2 (2 + 2 \gamma + 
    M^2)^2\epsilon}{(k^2 (2 + 2 \gamma + M^2) + 
    6 a^2 H^2 \gamma\epsilon) (k^2 (2 + 2 \gamma+ M^2)^2 + 
    3 a^2 H^2 (\gamma+M^2)(2\gamma+M^2)\epsilon^2)},\\
   T_{31} \simeq & \frac{ \sqrt{6} a H k M (2 + 2 \gamma + M^2)\epsilon \sqrt{
   k^2 (2 + 2 \gamma + M^2) + 
   3 a^2 H^2 (2 \gamma + M^2)\epsilon}}{
 \sqrt{k^2 (2 + 2 \gamma + M^2) + 
   6 a^2 H^2 \gamma\epsilon} (k^2 (2 + 2 \gamma + M^2)^2 + 
    3 a^2 H^2 (\gamma+M^2)(2\gamma+M^2)\epsilon^2)},\\
   T_{23} \simeq &-\frac{3 a^2 H^2 k \sqrt{\gamma} M (2 + 2 \gamma + M^2)^{3/2} M \sqrt{
 2 k^2 (2 + 2 \gamma + M^2) + 
  6 a^2 H^2 (2 \gamma + M^2) \epsilon}}{(k^2 (2 + 2 \gamma + M^2) + 
   6 a^2 H^2 \gamma M) \( k^2 (2 + 2 \gamma + M^2)^2 + 
   3 a^2 H^2 (\gamma + M^2) (2 \gamma + M^2)  \epsilon^2\)},
\end{align}
for the ${\bf K}$ matrix, we find
\begin{align}\nn
K_{12}  \simeq &\frac{\sqrt{3} a^2 H^2 k^4 M^2 (2 + 2 \gamma + M^2)^{
 5/2}\epsilon}{2\sqrt{\gamma}(k^2 (2 + 2 \gamma + M^2) + 
   6 a^2 H^2 \gamma\epsilon)^{
 3/2} ( k^2  (2 + 2 \gamma + M^2)^2 + 
   3 a^2 H^2  (\gamma + M^2) (2 \gamma + M^2)\epsilon^2)}\\ \nn&  
   - \frac{(2 + 2 \gamma + M^2)^{5/2}}{{2 \sqrt{3} \sqrt{\gamma}   \sqrt{
   k^2 (2 + 2 \gamma + M^2) + 
    6 a^2 H^2 \gamma\epsilon} \( k^2  (2 + 2 \gamma + M^2)^2 + 
     3 a^2 H^2 (\gamma + M^2) (2 \gamma + M^2)\epsilon^2\)}
}\\  & \times\Bigg(2 k^4  +
  \frac{ 3 a^2 H^2 k^2  (2 \gamma + 
      3 M^2)\epsilon}{(2 + 2 \gamma + M^2)}-
   \frac{24 a^4 H^4 \gamma (\gamma + M^2) (2 \gamma + M^2) \epsilon^4}{(2 + 2 \gamma + M^2)^4}\Bigg)      ,
\end{align}

\begin{align}\nn
K_{13}  \simeq &- \frac{3 a^2 H^2 k M (2 + 2 \gamma + M^2)\epsilon \sqrt{
  k^2 (2 + 2 \gamma + M^2) + 
   3 a^2 H^2 (2 \gamma + M^2)\epsilon}}{
 \sqrt{\frac{2}{3}k^2 (2 + 2 \gamma + M^2) + 
   4 a^2 H^2 \gamma\epsilon} (k^2 (2 + 2 \gamma + M^2)^2 + 
    3 a^2 H^2 (\gamma + M^2) (2 \gamma + M^2)\epsilon^2)}\\
+ &   \frac{3 \sqrt{6}
    a^4 H^4 k^3 M^3 \epsilon^2 (k^2 (2 + M^2 + 2 \gamma) + 
     6 a^2 H^2 \gamma \epsilon)^{-3/2} }{\sqrt{
   4 k^2 (2 + M^2 + 2 \gamma) + 
    3 a^2 H^2 (4 M^2 + 
       8 \gamma) \epsilon} (k^2  + 
     \frac{3 a^2 H^2 (M^2 + \gamma) (M^2 + 2 \gamma) \epsilon^2}{(2 + M^2 + 2 \gamma)^2})},
\end{align}

\begin{align}\nn
K_{23}  \simeq &\frac{a H k M (2 + 2 \gamma + M^2)^{3/2} \sqrt{
     k^2 (2 + 2 \gamma + M^2) + 
     3 a^2 H^2 (2 \gamma + 
         M^2)\epsilon} }{\sqrt{2\gamma} (k^2 (2 + 2 \gamma + M^2) + 
       6 a^2 H^2 \gamma\epsilon)^2 ( k^2 (2 + 2 \gamma + M^2)^2 + 
       3 a^2 H^2 (\gamma + M^2) (2 \gamma + M^2)\epsilon^2)}\\
       & \times\Big\{k^4 (2 + 2 \gamma + M^2)^2 + 
       18 a^2 H^2 k^2 \gamma (2 + 2 \gamma + M^2)\epsilon + 
       54 a^4 H^4 \gamma^2\epsilon^2\Big\},
\end{align}

\begin{align}
\Omega^2_{11} \simeq &   \frac{k^2}{3} +  2a^2 H^2 \(1 +  \gamma + M^2\)  -\frac{2 k^2 \epsilon }{3 \left(2 \gamma +M^2+2\right)} \\ \nn& + \frac{
 2a^2 H^2 (2 + 2 \gamma + 
    M^2)^2\epsilon^2 \(k^4 (2 + 2 \gamma -  M^2) +
    6 a^2 H^2 k^2 (2 \gamma + M^2) - \frac{
    18 a^4 H^4 (\gamma + M^2) (2 \gamma + M^2)^2\epsilon^2}{(2 + 
      2 \gamma + M^2)^2}\)}{(k^2 (2 + 2 \gamma + M^2)^2 + 
   3 a^2 H^2 (\gamma + M^2) (2 \gamma + M^2)\epsilon^2)^2},
\end{align}
\begin{align}
\Omega^2_{22} \simeq & k^2\frac{(\gamma - 2)}{\gamma}-2a^2H^2 \frac{M^2}{\gamma} -\frac{6 k^2 \epsilon }{\gamma  \left(2 \gamma +M^2+2\right)}  \\\nn
& +\frac{a^2 H^2 (2 + 2 \gamma + M^2)^6\epsilon }{(k^2 (2 + 2 \gamma + M^2) + 6 a^2 H^2 \gamma\epsilon)^3(k^2 (2 + 2 \gamma + M^2)^2 + 
    3 a^2 H^2 (\gamma + M^2) (2 \gamma + M^2) \epsilon^2)^2}\\\nn
 \times\Bigg\{ & k^{10}  \(2 \gamma ( \gamma-6) + (4 - 5 \gamma) M^2\)
 +6 a^2 H^2 k^8 (3 \gamma - M^2)+ \frac{36 a^4 H^4 k^6 \gamma  (8 \gamma + 
     M^2) \epsilon}{(2 + 2 \gamma + M^2)}
    \\ \nn&
     +\frac{108 a^6 H^6 k^4 \gamma^2 (14 \gamma + 
     3 M^2) \epsilon^2}{(2 + 2 \gamma + M^2)^2}
    + \frac{1296 a^8 H^8 k^2 \gamma^3 (2 \gamma + M^2) (2 + 2 \gamma + 
     M^2)^2 \epsilon^3}{(2 + 2 \gamma + M^2)^3}
    \\ &
    - \frac{ 3888 a^{10} H^{10} \gamma^3 (\gamma + M^2) (2 \gamma + M^2)^2 \epsilon^5}{(2 + 2 \gamma + M^2)^5}
\Bigg \},
\end{align}

\begin{align}\nn
\Omega^2_{21} \simeq &-\frac{a H  }{4 \sqrt{3\gamma^3} \(k^2  + \frac{6 a^2 H^2 \gamma\epsilon}{(2 + 2 \gamma + M^2)}\)^{
  3/2} \(k^2  + 
   \frac{ 3 a^2 H^2 (\gamma + M^2) (2 \gamma + M^2) \epsilon^2}{(2 + 2 \gamma + M^2)^2}\)^2}\\\nn
    \times\Bigg\{&-k^8 (8 \gamma^2 + 3 M^2 + 4 \gamma (2 + M^2))+\frac{3 k^{10} M^2 (2 + 2 \gamma + M^2)}{
k^2 (2 + 2 \gamma + M^2) + 6 a^2 H^2 \gamma \epsilon}\\ \nn& 
-\frac{6 a^2 H^2 k^6 \gamma  (M^2 + 
   4 \gamma (2 + 4 \gamma + 3 M^2)) \epsilon}{(2 + 2 \gamma + M^2)} -\frac{36 a^4 H^4 k^4 \gamma^2 (8 \gamma^2 + M^2 + 
   2 \gamma (-5 + 4 M^2)) \epsilon^2}{(2 + 2 \gamma + M^2)^2 }\\
 &  +\frac{864 a^6 H^6 k^2 \gamma^3 (2 \gamma + M^2) \epsilon^3}{(2 + 2 \gamma + M^2)^3}   -\frac{2592 a^8 H^8 \gamma^3 (\gamma + M^2) (2 \gamma + M^2)^2\epsilon^5}{(2 + 2 \gamma + M^2)^5 }
  \Bigg\},
\end{align}

\begin{align}
\Omega^2_{33} = & k^2 - 2 a^2 H^2\(1-\frac{\epsilon}{2}\)\\\nn
&-\frac{12 a^2 H^2 k^2 M^2 (2 + 2 \gamma + M^2)^{-1} \epsilon}{\(k^2  + \frac{6 a^2 H^2 \gamma\epsilon}{(2 + 2 \gamma + M^2)}\)^{
  3}\(k^2  +\frac{3 a^2 H^2 (2\gamma+M^2)\epsilon}{(2 + 2 \gamma + M^2)}\) \(k^2  + 
   \frac{ 3 a^2 H^2 (\gamma + M^2) (2 \gamma + M^2) \epsilon^2}{(2 + 2 \gamma + M^2)^2}\)^2}\\\nn
   & \Bigg\{ 
   \frac{k^{10}}{2}+\frac{a^2 H^2 k^8}{4}-\frac{6 a^4 H^4 k^6 \gamma  \epsilon}{ (2 + 2 \gamma + M^2)}
-\frac{9 a^6 H^6 k^4 \gamma (14 \gamma^2 + 
   5 M^2) \epsilon^2}{
2(2 + 2 \gamma+ M^2)^2}\\\nn
   &-\frac{54 a^8 H^8 k^2 \gamma^2 (2 \gamma + M^2)  \epsilon^3}{(2 + 2 \gamma+ M^2)^3}+\frac{81 a^{10} H^{10} \gamma^2 (\gamma^2 + M^2) (2 \gamma^2 + M^2)^2 (6 + 9 \gamma^2 + 
   5 M^2) \epsilon^6}{(2 + 2 \gamma+ M^2)^6}
   \Big\},
\end{align}

\begin{align}
\Omega_{31}^2 = & -\sqrt{\frac{1}{6}}\frac{ a H k M \sqrt{
   k^2 (2 + 2 \gamma + M^2) + 
   3 a^2 H^2 (2 \gamma +  M^2)\epsilon}}{
\sqrt{k^2 (2 + 2 \gamma + M^2) + 6 a^2 H^2 \gamma\epsilon}}\\\nn
& +\sqrt{\frac{2}{3}}\frac{ a H k M (2 + 2 \gamma + M^2)^{-1}\epsilon}{\(k^2 + \frac{6 a^2 H^2 \gamma \epsilon}{ (2 + 2 \gamma + M^2)}\)^{3/2} \sqrt{
 k^2  +   \frac{3 a^2 H^2 (2 \gamma + 
     M^2) \epsilon}{(2 + 2 \gamma + M^2)}} \(k^2  + 
  \frac{ 3 a^2 H^2 (\gamma + M^2) (2 \gamma + M^2) \epsilon^2}{(2 + 2 \gamma + M^2)^2}\)^2}\\\nn
  \times\Bigg\{&
  30 a^2 H^2 k^6  + 2 k^8 + 
 \frac{ 9 a^4 H^4 k^4  (40 \gamma +  9 M^2)\epsilon}{(2 + 2 \gamma + M^2)}+  \frac{540 a^6 H^6 k^2 \gamma (2 \gamma + M^2) \epsilon^2}{(2 + 2 \gamma + M^2)^2}\\ &\nn  + 
 \frac{ 1296 a^8 H^8 \gamma (\gamma + M^2) (2 \gamma + M^2)^2 \epsilon^4}{(2 + 2 \gamma + M^2)^4}
  \Bigg\}\\\nn
  & +\frac{4 \sqrt{6} a^3 H^3 k M (2 + M^2 + 2 \gamma)^{-1}\epsilon}{\(k^2  +\frac{ 6 a^2 H^2 \gamma\epsilon}{(2 + M^2 + 2 \gamma)}\)^{
 5/2} \(4 k^2  + 
   \frac{12 a^2 H^2 (M^2 + 2 \gamma)\epsilon}{(2 + M^2 + 2 \gamma)}\)^{3/2} \(k^2  + 
   \frac{3 a^2 H^2 (M^2 + \gamma) (M^2 + 2 \gamma)\epsilon^2}{(2 + M^2 + 2 \gamma)^2}\)^2}\\\nn
   \times\Bigg\{& -4 k^{10}  - 
 \frac{ 6 a^2 H^2 k^8  (3 M^2 + 
     16 \gamma)\epsilon}{(2 + M^2 + 2 \gamma)}  - 
  \frac{27 a^4 H^4 k^6 (M^4 + 16 M^2 \gamma + 
     32 \gamma^2)\epsilon^2}{(2 + M^2 + 2 \gamma)^2} \\\nn &-
  \frac{108 a^6 H^6 k^4 \gamma (2 + M^2 + 2 \gamma)^3 (5 M^4 + 
     26 M^2 \gamma + 32 \gamma^2)\epsilon^3}{(2 + M^2 + 2 \gamma)^3} \\\nn &- 
  \frac{1296 a^8 H^8 k^2 \gamma^2 (M^2 + 2 \gamma)^2\epsilon^4}{(2 + M^2 + 2 \gamma)^4} - 
  \frac{3888 a^{10} H^{10} \gamma^2 (M^2 + \gamma) (M^2 + 
     2 \gamma)^3\epsilon^6}{(2 + M^2 + 2 \gamma)^6}\Bigg\},
 \end{align}

\begin{align}\nn
\Omega_{32}^2 = &-\frac{12 \sqrt{2} a^2 H^2 k^3 M }{\sqrt{\gamma} \(k^2  + 
   \frac{6 a^2 H^2 \gamma \epsilon}{(2 + M^2 + 2 \gamma)}\)^3 \(4 k^2  + 
   \frac{12 a^2 H^2 (M^2 + 2 \gamma) \epsilon}{(2 + M^2 + 2 \gamma)}\)^{ 3/2} \(k^2  + 
  \frac{ 3 a^2 H^2 (M^2 + \gamma) (M^2 + 2 \gamma) \epsilon^2}{(2 + M^2 + 2 \gamma)^2}\)^2}\\\nn
  \times \Bigg\{ &k^{10}  +
 \frac{a^2 H^2 k^8  (5 M^2 + 
    34 \gamma) \epsilon}{(2 + M^2 + 2 \gamma)} + 
 \frac{3 a^4 H^4 k^6  (2 M^4 + 53 M^2 \gamma + 
    138 \gamma^2) \epsilon^2}{(2 + M^2 + 2 \gamma)^2}\\ \nn&  +
 \frac{36 a^6 H^6 k^4 \gamma  (5 M^4 + 
    39 M^2 \gamma + 66 \gamma^2) \epsilon^3}{(2 + M^2 + 2 \gamma)^3} + 
\frac{ 108 a^8 H^8 k^2 \gamma^2 (8 M^4 + 
    47 M^2 \gamma + 62 \gamma^2) \epsilon^4}{(2 + M^2 + 2 \gamma)^4}\\ &  + \frac{
 1944 a^{10} H^{10} \gamma^3 (M^2 + 2 \gamma)^2 \epsilon^5}{(2 + M^2 + 2 \gamma)^5}+\frac{3888 a^{12} H^{12} \gamma^3 (M^2 + \gamma) (M^2 + 
    2 \gamma)^3 \epsilon^7}{(2 + M^2 + 2 \gamma)^7}\Bigg\}.
\end{align}
Note that, to leading order in slow-roll, we recover the results of \cite{Namba:2013kia} in the limit $M\to 0$.

%
\section{Details of the vector action}\label{app:vecsector}
%

In this appendix, we present the details of the vector action.  After making the transformation in eq.\ \eqref{eqn:vectrans}, the action takes the form
\begin{align}
\delta^{2}S = \int \frac{\d^3 k}{(2\pi)^3}\d\tau \[ {\vec{W}}^{\pm}{}^{\dagger}{}'{\bf T}_{\pm}{\vec{W}}^{\pm}{}' + {\vec{W}}^{\pm}{}^{\dagger}{}'{\bf K}_\pm{\vec{W}}^{\pm} - {\vec{W}}^{\pm}{}^{\dagger}{\bf K}_{\pm}{\vec{W}}^{\pm}{}'  - {\vec{W}}^{\pm}{}^{\dagger}{\bf \Omega}^2_{\pm}{\vec{W}}^{\pm} \].
\end{align}
While it is straightforward to obtain the matrices exactly, they are long and not particularly enlightening. Each entry in the matrices are of the form of eq.\ \eqref{eqn:mtxexp}. To perform the numerical evaluation, we expand each coefficient to leading order in slow-roll in  the same way as described in section \ref{app:scalsector}. We find
\begin{align}
T^{\pm}_{11} = & 1-\frac{2 a^2 \gamma  H^2 \epsilon }{\left(\gamma +\frac{M^2}{2}+1\right) \left(2 a^2 \gamma  H^2\mp2 a \sqrt{\gamma } H k+k^2\right)},\\
T^{\pm}_{22} = & 1-\frac{2 a^2 H^2 M^2 \epsilon  \left(a \sqrt{\gamma } H\pm k\right)^2}{\left(\gamma +\frac{M^2}{2}+1\right) \left(2 a^2 \gamma  H^2\mp 2 a
   \sqrt{\gamma } H k+k^2\right) \left(a^2 H^2 \left(2 \gamma +M^2\right)\mp 2 a \sqrt{\gamma } H k+k^2\right)},\\
T^{\pm}_{21} = &-\frac{4 a^2 \sqrt{\gamma } H^2 M \epsilon  \left(a \sqrt{\gamma } H\mp k\right)}{\left(2 \gamma +M^2+2\right) \left(2 a^2 \gamma  H^2\mp 2 a
   \sqrt{\gamma } H k+k^2\right) \sqrt{a^2 H^2 \left(2 \gamma +M^2\right)\mp 2 a \sqrt{\gamma } H k+k^2}},
\end{align}
and
\begin{align}
K_{12}^{\pm} = & \frac{a^3 \sqrt{\gamma } H^3 k M}{\left(2 a^2 \gamma  H^2-2 a \sqrt{\gamma } H k+k^2\right) \sqrt{a^2 H^2 \left(2 \gamma +M^2\right)\mp 2 a
   \sqrt{\gamma } H k+k^2}}\\\nn
    & \pm  \frac{2 a^2 H^2 M \left(2 a^4 \gamma  H^4 \left(2 \gamma +M^2\right)\mp a^3 \sqrt{\gamma } H^3 k \left(7 \gamma +M^2\right)+a^2 H^2 k^2
   \left(8 \gamma +M^2\right)\mp 4 a \sqrt{\gamma } H k^3+k^4\right)}{\left(2 \gamma +M^2+2\right) \left(2 a^2 \gamma  H^2\mp 2 a \sqrt{\gamma }
   H k+k^2\right) \left(a^2 H^2 \left(2 \gamma +M^2\right)\mp 2 a \sqrt{\gamma } H k+k^2\right)^{3/2}}.
\end{align}
Finally, the entries of the mass matrix take the form
\begin{align}
\Omega^2_{11}{}^{\pm} = &  k^2 \(1 -\frac{1}{\gamma}\) \mp  a H k \frac{(2 + M^2 + 2 \gamma)}{\sqrt{\gamma} }+ a^2 H^2 (M^2 + 2 \gamma ) \\\nn
& -\frac{aH}{3 \sqrt{\gamma}(k^2 \mp 2 a H k \sqrt{\gamma} + 2 a^2 H^2 \gamma)^3 (k^2 \mp 
    2 a H k \sqrt{\gamma} + a^2 H^2 (M^2 + 2 \gamma ))^2}\\\nn
    & \times\bigg(\pm 3 a^2 H^2 k^9 (2 + M^2)\gamma - 24 a^{11} H^{11} (2 + M^2) \gamma^{7/2} (M^2 + 2\gamma)^2
    \\\nn
    & \pm 12 a^{10} H^{10} k (2 + M^2) \gamma^3(M^2 + 2\gamma ) (5 M^2 + 18 \gamma)\pm 6 a^4 H^4 k^7\gamma (M^4 + 57\gamma+ 8 M^2 (1 + 3\gamma))\\\nn
&  -3 a^3 H^3 k^8\sqrt{\gamma} (23 \gamma+ 2 M^2 (1 + 5 \gamma))\\\nn
& \pm 3 a^6 H^6 k^5\gamma(M^6 + 688 \gamma^2 + 
   10 M^4 (1 + 6 \gamma) + 8 M^2 \gamma (24 + 37\gamma))\\\nn
& -6 a^9 H^9 k^2 \gamma^{5/2} (12 M^6 + 332 \gamma^2 + 3 M^4 (9 + 32 \gamma) + 
    4 M^2 \gamma (51 + 40 \gamma))\\ \nn& -
 6 a^5 H^5 k^6 \sqrt{\gamma} (171 \gamma^2 + M^4 (1 + 8 \gamma) +
     M^2 \gamma (35 + 72 \gamma)) \\\nn &\pm
 6 a^8 H^8 k^3 \gamma^2 (8 M^6 + 484 \gamma^2 + 
    8 M^2 \gamma (29 + 28 \gamma) + M^4 (23 + 100 \gamma))\\\nn &  - 
 3 a^7 H^7 k^4 \gamma^{3/2} (6 M^6 + 972 \gamma^2 + 72 M^2 \gamma (5 + 6 \gamma) + 
    M^4 (27 + 136 \gamma))
     \\ \nn&\mp \frac{
 5 k^{11} (M^2 + \gamma) \epsilon^2}{2 + M^2 + 2 \gamma}  + \frac{
 50 a H k^{10} \sqrt{\gamma} (M^2 + \gamma) \epsilon^2}{
 2 + M^2 + 2 \gamma}
    \bigg),
\end{align}
\begin{align}\nn
&\Omega^2_{22}{}^{\pm} =  k^2\\\nn &  - \frac{a^3 H^3 M^2}{(k^2 \mp 2 a H k \sqrt{\gamma}+ 2 a^2 H^2 \gamma)^2 (k^2 \mp 
     2 a H k \sqrt{\gamma} + a^2 H^2 (M^2 + 2 \gamma))^2) + a H ( a H ( M^2-2)\mp k \sqrt{\gamma})}\\\nn
 & \times\Bigg(\pm k^7 \sqrt{\gamma} + 4 a^7 H^7 {\gamma}^2 (M^2 + 2 {\gamma})^2 + 
  a H k^6 (-5 + 8 {\gamma}) \pm
  2 a^2 H^2 k^5 \sqrt{\gamma} (-11 + M^2 + 15 {\gamma}) \\ \nn& 
  +   2 a^3 H^3 k^4 (M^2 (-1 + 6 {\gamma}) + 
     2 {\gamma} (-11 + 17 {\gamma}))\\ \nn&  \pm
  a^4 H^4 k^3 \sqrt{\gamma} (M^4 + 4 {\gamma} (-12 + 25 {\gamma}) + 
     M^2 (-6 + 32 {\gamma})) \\\nn &+ 
  a^5 H^5 k^2 {\gamma} (4 M^4 + 4 {\gamma} (-7 + 24 {\gamma})  + 
     M^2 (-7 + 48 {\gamma}))\\ & 
      \pm 
  2 a^6 H^6 k {\gamma}^{ 3/2} (3 M^4 + 4 {\gamma} (-1 + 7 {\gamma}) + 
     M^2 (-2 + 20 {\gamma})) \Bigg),
\end{align}
\begin{align}\nn
&\Omega^2_{21}{}^{\pm} =\frac{a^2 H^2 M}{2 \sqrt{\gamma} (2 + M^2 + 2 {\gamma}) (k^2 \mp
    2 a H k\sqrt{{\gamma}}+ 2 a^2 H^2 {\gamma})^3 (k^2\mp 
    2 a H k \sqrt{{\gamma}} + a^2 H^2 (M^2 + 2 {\gamma}))^{7/2}}\\\nn
& \times     \Bigg(\pm k^{13} (1 + 2 {\gamma})
     -  a H k^{12} \sqrt{\gamma}(11 + 24 {\gamma}) 
     \pm 2 a^2 H^2 k^{11} (M^2 (2 + 4 {\gamma}) + {\gamma} (35 +  72 {\gamma})) \\\nn
 &     - 4 a^3 H^3 k^{10} \sqrt{\gamma} (M^2 (9 + 20 {\gamma}) + 
     4 {\gamma} (19 + 35 {\gamma})) \\\nn
     & \pm  
  a^4 H^4 k^9 (M^4 (5 + 12 {\gamma}) + 
     120 {\gamma}^2 (8 + 13 {\gamma}) + 8 M^2 {\gamma} (23 + 50 {\gamma})) \\\nn 
     & - 
  a^5 H^5 k^8 \sqrt{\gamma}(M^4 (35 + 96 {\gamma}) + 
     4 M^2 {\gamma} (157 + 320 {\gamma}) + 
     8 {\gamma}^2 (283 + 408 {\gamma})) \\\nn
&\pm   2 a^6 H^6 k^7 (M^6 (1 + 4 {\gamma}) +  96 M^2 {\gamma}^2 (8 + 15 {\gamma}) +   3 M^4 {\gamma} (23 + 64 {\gamma}) +    8 {\gamma}^3 (253 + 328 {\gamma})) \\\nn
  & 
     - 
  2 a^7 H^7 k^6 \sqrt{\gamma} (M^6 (5 + 24 {\gamma}) + 
     16 M^4 {\gamma} (11 + 30 {\gamma}) + 
     64 {\gamma}^3 (43 + 51 {\gamma}) + 
     4 M^2 {\gamma}^2 (343 + 592 {\gamma}))\\\nn
& \pm 2 a^8 H^8 k^5 {\gamma} (M^8 + 360 M^2 {\gamma}^2 (5 + 8 {\gamma}) + 2 M^6 (7 + 36 {\gamma}) + 8 {\gamma}^3 (353 + 390 {\gamma}) +  2 M^4 {\gamma} (157 + 408 {\gamma}))\\\nn& 
     - 4 a^9 H^9 k^4 {\gamma}^{3/2}(2 M^8 + 4 M^6 (3 + 16 {\gamma}) +   15 M^4 {\gamma} (13 + 32 {\gamma}) +    4 {\gamma}^3 (267 + 280 {\gamma}) +   4 M^2 {\gamma}^2 (213 + 320 {\gamma}))\\ \nn
     &  \pm  8 a^{10} H^{10} k^3 {\gamma}^2 (M^2 + 2 {\gamma}) (2 M^6 + M^4 (7 + 32 {\gamma}) + 
     2 {\gamma}^2 (71 + 72 {\gamma}) + 
     M^2 {\gamma} (69 + 128 {\gamma})) \\ \nn
  &   -  8 a^{11} H^{11} k^2 {\gamma}^{ 5/2} (M^2 + 2 {\gamma})^2 (2 M^4 + 
     24 {\gamma} (1 + {\gamma}) + M^2 (5 + 16 {\gamma})) \\&\pm 
  8 a^{12} H^{12} k {\gamma}^3 (M^2 + 2 {\gamma})^3 (2 + M^2 + 
     2 {\gamma})^2 + \frac{
  32 a^{13} H^{13} {\gamma}^{7/2} (M^2 + 2 {\gamma})^3 \epsilon}{(2 + M^2 + 
     2 {\gamma})}\Bigg).
\end{align}

\section{Density fluctuation}
\label{app:Pz}

In this appendix we present the details of the computation of the density fluctuation. We work in spatially flat gauge, where the curvature perturbation is given by
\begin{align}
\zeta = -\frac{H}{\dot{\rho}}\delta\rho = \frac{\delta \rho}{6\[\(\dot{\psi}^2+H\psi\)^2+g^2\psi^4+g^2\psi^2 Z_0^2\]}\simeq \frac{\delta\rho}{3H^2 \psi^2(2+2\gamma+M^2)}.
\end{align}
The perturbation to the energy density is found from the perturbed stress tensor, $\rho = -T^0{}_0$, where
\begin{align}
T_{\mu\nu} =  & 2\tr\[F_{\mu\alpha}F_{\nu\beta}\]g^{\alpha\beta}-\frac{g_{\mu\nu}}{2}\tr\[F_{\alpha\beta}F^{\alpha\beta}\] \\\nn&  -\frac{1}{12}\kappa \tr\[F_{\alpha\beta}F^{\alpha\beta}\]\tr\[(F_{\mu}{}^{\alpha}\tilde{F}_{\nu\alpha}+F_{\nu}{}^{\alpha}\tilde F_{\mu\alpha})\]+g_{\mu\nu}\frac{\kappa}{32}\tr\[F_{\alpha\beta}\tilde{F}^{\alpha\beta}\]^2\\\nn
&    +2g^2\hvev^2\tr\[ (A_{\mu} - \frac{i}{g}U^{-1}\partial_\mu U)(A_{\nu} - \frac{i}{g}U^{-1}\partial_\nu U)\] - g_{\mu\nu}g^2\hvev^2\tr\[ (A_{\mu} - \frac{i}{g}U^{-1}\partial_\mu U)^2\].
\end{align}
Inserting the field configuration in eqs.\ \eqref{eqn:fields1}-\eqref{eqn:fields4}, we find 
\begin{align}\nn
\delta\rho = & 3\frac{(a\psi)'}{a^3}(1+\kappa g^2\psi^4)\delta\psi'- k^2\frac{(a\psi)'}{a^3}(1+\kappa g^2 \psi^4)M'\\\nn
& + 3\[g^2\psi(2 \psi^2+Z_0^2)+(1+3\kappa g^2 \psi^4)\frac{a'{}^2}{a^4}\psi+(1+5\kappa g^2\psi^4)\frac{a' \psi'}{a^3}+2\kappa g^2 \psi^3\frac{\psi'^2}{a^2}\]\delta\psi\\\nn
& -k^2\[g^2\psi(2 \psi^2+Z_0^2)+(1+3\kappa g^2 \psi^4)\frac{a'{}^2}{a^4}\psi+(1+5\kappa g^2\psi^4)\frac{a' \psi'}{a^3}+2\kappa g^2 \psi^3\frac{\psi'^2}{a^2}\] M\\
&+ \frac{g^2 Z_0^2 \psi }{a} k^2 \dHiggs+k^2\frac{(a\psi)'}{a^3}(1+\kappa g^2 \psi^4)Y-3\frac{(a\psi)'{}^2}{a^4}(1+\kappa g^2 \psi^4)\dN.
\end{align}
Next, we insert the solutions for the constraints, and expand to leading order in slow roll, to find
\begin{align}
 \delta\rho \simeq & \frac{\sqrt{2} H^2 M_{\rm Pl} \sqrt{\epsilon } (2+2 \gamma +M^2)^{3/2}}{k^2
   \left(2+2 \gamma +M^2\right)^2+ 3
   a^2 H^2 \epsilon ^2 \left(\gamma +M^2\right) \left(2 \gamma +M^2\right)}\\\nn  \times \Bigg\{ &3   (9 a^2 H^2 (M^2 + 2 \gamma) + 
     k^2 (6 + M^2 + 2 \gamma)) \delta\psi + 
  3 (2 k^2 + 
     3 a^2 H^2 (M^2 + 2 \gamma)) \frac{\delta\psi'}{aH}\\\nn
 & -   k^2 (9 a^2 H^2 (M^2 + 2 \gamma) + 
     k^2 (4 + M^2 + 2 \gamma)) \M   - 
 3 a H k^2 (2\gamma+M^2) \M'\\ \nn& + 
  2\frac{k^2  M^2 (k^2 (2 + M^2 + 2 \gamma) + 
     3 a^2 H^2 (M^2 + 2 \gamma) \epsilon) }{aH(2+2 \gamma +M^2)}\dHiggs +  6  H k^2 M^2 \dHiggs')\Bigg\}.
\end{align}
Note that, as expected, we recover the results of ref.\ \cite{Namba:2013kia} in the limit $M\to 0$. The curvature is then of the form
\begin{align}
\zeta = c_1 \delta\psi+c_2 \M+c_3 \dHiggs+ d_1 \delta\psi'+d_2 \M' +d_3 \dHiggs',
\end{align}
where
\begin{align}
c_1 \simeq & \frac{(2+2 \gamma +M^2)^{3/2}}{\sqrt{2}M_{\rm Pl}\sqrt{\epsilon}}  \frac{(9 a^2 H^2 (M^2 + 2 \gamma) + 
     k^2 (6 + M^2 + 2 \gamma))}{k^2
   \left(2+2 \gamma +M^2\right)^2+ 3
   a^2 H^2 \epsilon ^2 \left(\gamma +M^2\right) \left(2 \gamma +M^2\right)},\\
c_2 \simeq & - k^2\frac{(2+2 \gamma +M^2)^{3/2}}{3\sqrt{2}M_{\rm Pl}\sqrt{\epsilon}}  \frac{(9 a^2 H^2 (M^2 + 2 \gamma) + 
     k^2 (4 + M^2 + 2 \gamma))}{k^2
   \left(2+2 \gamma +M^2\right)^2+ 3
   a^2 H^2 \epsilon ^2 \left(\gamma +M^2\right) \left(2 \gamma +M^2\right)}, \\
c_{3} \simeq& 2 \frac{\sqrt{(2+2 \gamma +M^2)}}{3 \sqrt{2} a HM_{\rm Pl} \sqrt{\epsilon}}\frac{k^2 M^2 (k^2 (2 + M^2 + 2 \gamma) + 
     3 a^2 H^2 (M^2 + 2 \gamma) \epsilon)}{k^2
   \left(2+2 \gamma +M^2\right)^2+ 3
   a^2 H^2 \epsilon ^2 \left(\gamma +M^2\right) \left(2 \gamma +M^2\right)} ,
\end{align}
and
\begin{align}
d_1 \simeq & \frac{ (2+2 \gamma +M^2)^{3/2}}{\sqrt{2}{aH}M_{\rm Pl} \sqrt{\epsilon }}\frac{(2 k^2 + 
     3 a^2 H^2 (M^2 + 2 \gamma))}{k^2
   \left(2+2 \gamma +M^2\right)^2+ 3
   a^2 H^2 \epsilon ^2 \left(\gamma +M^2\right) \left(2 \gamma +M^2\right)},\\
d_2 \simeq &- \frac{ (2+2 \gamma +M^2)^{3/2}}{\sqrt{2}M_{\rm Pl}  \sqrt{\epsilon } }\frac{a H k^2 (2\gamma+M^2)}{k^2
   \left(2+2 \gamma +M^2\right)^2+ 3
   a^2 H^2 \epsilon ^2 \left(\gamma +M^2\right) \left(2 \gamma +M^2\right)},\\
d_{3} \simeq&  \frac{  (2+2 \gamma +M^2)^{3/2}}{\sqrt{2}M_{\rm Pl} \sqrt{\epsilon }}\frac{2H k^2 M^2}{k^2
   \left(2+2 \gamma +M^2\right)^2+ 3
   a^2 H^2 \epsilon ^2 \left(\gamma +M^2\right) \left(2 \gamma +M^2\right)}.
\end{align}
Our numerical solutions are not in terms of the variables $\vec{X} = \{\delta\psi, \M, \dHiggs\}$, but are in terms of $\Delta$, which are related to the $\vec{X}$ via the redefinition at eq.\ \eqref{eqn:fieldred}. The power spectrum is then given by
\begin{align}
P_{\zeta} = \frac{k^3}{2\pi^2}\sum_{i = 1}^3\left|(\vec{c}\cdot {\bf U} +\vec{d}\cdot {\bf U }')\cdot\vec{\mathcal{Q}}_i+\vec{d}\cdot {\bf U}\cdot \vec{\mathcal{Q}}'_i  \right|^2,
\label{eq:powerspectrum}
\end{align}
and the sum runs over the independent solutions of the equations of motion. That is, the solutions initialized on the independent initial conditions in eq.\ \eqref{eqn:initialconditions}. Explicitly, 
\begin{align}
& (\vec{c}\cdot {\bf U} +\vec{d}\cdot {\bf U }')_1 =  \frac{\sqrt{\epsilon } \sqrt{2+2 \gamma +M^2} \left(6 a^2 H^2 \left(2 \gamma
   +M^2\right)+k^2 \left(4+2 \gamma +M^2\right)\right)}{a\sqrt{3} \left(k^2 \left(2
   \gamma +M^2+2\right)^2+3 a^2 H^2
   \epsilon ^2 \left(\gamma +M^2\right) \left(2 \gamma +M^2\right)\right)},\\
& (\vec{c}\cdot {\bf U} +\vec{d}\cdot {\bf U }')_2 = \\ & 
-\frac{\left(\frac{36 a^4 \gamma  H^4 \left(2 \gamma
   +M^2\right)}{(2 \gamma +M^2+2)}+3 a^2 H^2 k^2 \left(2 \gamma +M^2\right) +k^4\left(2 \gamma +M^2+4\right)\right)}{6
   a^2H \sqrt{\gamma } \sqrt{\epsilon } \sqrt{6 a^2 \gamma  H^2 \epsilon +k^2 \left(2
   \gamma +M^2+2\right)} \left(k^2+\frac{3 a^2 H^2 \epsilon ^2 \left(\gamma +M^2\right) \left(2
   \gamma +M^2\right)}{\left(2 \gamma +M^2+2\right)^2} \right)},\\
& (\vec{c}\cdot {\bf U} +\vec{d}\cdot {\bf U }')_3 =   \frac{\sqrt{2} k^3 M \left(2 \gamma +M^2+2\right)^{5/2} }{3 a^2H
   \sqrt{\epsilon }  \sqrt{3 a^2 H^2 \epsilon  \left(2 \gamma +M^2\right)+k^2
   \left(2 \gamma +M^2+2\right)} }\\\nn
   & \times\frac{\left(k^2 \left(2 \gamma
   +M^2+2\right) \left(k^2-3 a^2 H^2\right)-36 a^4 \gamma  H^4 \epsilon \right)}{\left(6 a^2 \gamma  H^2 \epsilon +k^2 \left(2 \gamma
   +M^2+2\right)\right)^{3/2}\left( k^2 \left(2 \gamma +M^2+2\right)^2+3 a^2 H^2 \epsilon ^2 \left(\gamma +M^2\right)
   \left(2 \gamma +M^2\right)\right)},
\end{align}
and
\begin{align}
& (\vec{d}\cdot {\bf U })_1 = \frac{\sqrt{2 \gamma +M^2+2} \left(k^2
   \left(2 \gamma +M^2 +2\right)+3 a^2 H^2 \epsilon  \left(2 \gamma +M^2\right)\right)}{\sqrt{3} a^2 H \sqrt{\epsilon }
   \left(
   k^2 \left(2 \gamma +M^2+2\right)^2+3 a^2 H^2 \epsilon ^2 \left(\gamma +M^2\right) \left(2 \gamma +M^2\right)\right)},\\
& (\vec{d}\cdot {\bf U })_2 =    -\frac{\left(2 \gamma +M^2\right) \left(2+2\gamma +M^2  \right) \sqrt{k^2 \left(2+2 \gamma
   +M^2\right)+6 a^2 \gamma  H^2 \epsilon}}{2 \sqrt{\gamma } \sqrt{\epsilon } a \left(k^2 \left(2+2 \gamma
   +M^2\right)^2+3 a^2H^2 \epsilon ^2
   \left(\gamma +M^2\right) \left(2 \gamma +M^2\right)\right)},\\\nn
& (\vec{d}\cdot {\bf U })_3 =    \frac{k M \left(2 \gamma +M^2+2\right)^{3/2} }{\sqrt{\epsilon }a \sqrt{k^2 \left(2 \gamma +M^2+2\right)+6 a^2
   \gamma  H^2 \epsilon } } \\ & \hskip 1.75cm\times \frac{\sqrt{2 k^2 \left(2 \gamma +M^2+2\right)+6 a^2 H^2 \epsilon  \left(2 \gamma
   +M^2\right)}}{\left(k^2 \left(2 \gamma
   +M^2+2\right)^2+3 a^2 H^2 \epsilon ^2
   \left(\gamma +M^2\right) \left(2 \gamma +M^2\right)\right)}.
\end{align}


\bibliographystyle{JHEP}
\bibliography{HGF}


\end{document}